\documentclass[11pt]{article} 

\usepackage{jheppub}
\usepackage[utf8]{inputenc} 

\usepackage{graphicx} 
\graphicspath{{Figures/}}
\usepackage{changes}
\usepackage{xcolor}
\usepackage{amsmath}
\usepackage{hyperref}
\hypersetup{
	colorlinks,
	citecolor=blue,
	filecolor=black,
	linkcolor=blue,
	urlcolor=blue
}

\usepackage{array} 
\usepackage{verbatim} 

\title{
Non-vanishing zero-temperature normal density in holographic superfluids}
\author{Blaise Gout\'eraux}
\emailAdd{blaise.gouteraux@polytechnique.edu}
\author{and Eric Mefford}
\emailAdd{eric.mefford@polytechnique.edu}

\preprint{CPHT-051.072020}
\affiliation{CPHT, CNRS, \'Ecole polytechnique, Institut Polytechnique de Paris, Route de Saclay, 91128 PALAISEAU, France}

\abstract{The low energy and finite temperature excitations of a $d+1$-dimensional system exhibiting superfluidity are well described by a hydrodynamic model with two fluid flows: a normal flow and a superfluid flow. In the vicinity of a quantum critical point, thermodynamics and transport in the system are expected to be controlled by the critical exponents and by the spectrum of irrelevant deformations away from the quantum critical point. Here, using gauge-gravity duality, we present the low temperature dependence of thermodynamic and charge transport coefficients at first order in the hydrodynamic derivative expansion in terms of the critical exponents. Special attention will be paid to the behavior of the charge density of the normal flow in systems with emergent infrared conformal and Lifshitz symmetries, parameterized by a Lifshitz dynamical exponent $z>1$. When $1\leq z<d+2$, we recover ($z=1$) and extend ($z>1$) previous results obtained by relativistic effective field theory techniques. Instead, when $z>d+2$, we show that the normal charge density becomes non-vanishing at zero temperature. An extended appendix generalizes these results to systems that violate hyperscaling as well as systems with generalized photon masses. Our results clarify previous work in the holographic literature and have relevance to recent experimental measurements of the superfluid density on cuprate superconductors.}
\begin{document}

\maketitle


\newpage

\section{Introduction and results}

The spontaneous breaking of a U(1) symmetry and the associated phenomenon of superfluidity is one of most studied subjects of contemporary Physics. Superfluidity characterizes the low temperature behavior of systems ranging from the two isotopes of Helium, cold atoms and conventional superconductors. For these systems, microscopic descriptions are available. More generally, their low energy dynamics are captured by a universal effective theory (EFT) originally due to Landau and Tisza based on an extension of hydrodynamics \cite{Landau,Tisza} and soon after extended to relativistic superfluids \cite{Khalatnikov1, Khalatnikov2, Khalatnikov3, Khalatnikov4, Israel1, Israel2} -- see also \cite{Son:2000ht,Bhattacharya:2011eea,Bhattacharya:2011tra,Bhattacharyya:2012xi} for more recent treatments.
$\\$

Hydrodynamics posits that slow deviations about local thermodynamic equilibrium can be captured by a small set of conservation equations following from the symmetries of the system, together with constitutive relations for the spatial currents associated to the conserved densities. For a superfluid, these equations need to be supplemented by a `Josephson' relation, which follows from gauge invariance and relates the time derivative of the Goldstone following from the spontaneously broken U(1) symmetry to the chemical potential of the system. This equation is derived from the realization that the Goldstone is canonically conjugate to the charge density of the system \cite{Son:2000ht}.
$\\$

In the hydrodynamic theory, the dynamics at finite temperature can be thought of as the superposition of two types of flows, a normal flow and a superfluid flow. The normal flow is dissipative and is carried by a fraction of the total charge density $\rho$, the normal density $\rho_n$. On the other hand, the superfluid flow is dissipationless and is responsible for the phenomenon of superfluidity. It is carried by the superfluid fraction $\rho_s$ of the total density, such that the total density $\rho=\rho_n+\rho_s$.
$\\$

In all the systems mentioned above, the normal density vanishes at zero temperature $\rho_n^{(0)}\equiv \rho_n(T=0)=0$, leaving only the superfluid component of the flow, $\rho^{(0)}=\rho_s^{(0)}$. This is consistent both with microscopic calculations, \cite{Leggett,Schmitt:2014eka} and relativistic superfluid effective field theories (EFT) \cite{Son:2002zn,Nicolis:2011cs,Delacretaz:2019brr}. Thus the zero temperature EFT only needs to account for a single, linearly-dispersing degree of freedom, the Goldstone mode.\footnote{\label{footnote1} The Goldstone is also referred to as `superfluid phonon' in the literature, in relation to the extra sound mode it sources compared to normal fluids.} Under these assumptions, the following expression has been derived for normal density at leading order in a small temperature expansion (see e.g. \cite{Schmitt:2014eka})
\begin{equation}
\label{rhonlowTeftSchmitt}
\rho_n\simeq\frac{s T}{\mu c_{ir}^2}+\dots
\end{equation}
where $s$ is the entropy density and $c_{ir}^2$ the effective lightcone velocity. The dots stand for subleading corrections in temperature and $c_{ir}$. In general, the superfluid EFT yields $\rho_n\sim T^{d+1}$, \cite{Delacretaz:2019brr}, in agreement with the result \eqref{rhonlowTeftSchmitt} for ${}^4$Helium, see e.g. \cite{Schmitt:2014eka}. In this work, we will derive an improved expression \eqref{rhonlowTeftSchmitt} for the low temperature behavior of the normal density in relativistic superfluid phases, which holds for finite $c_{ir}$.
$\\$

The result from the superfluid EFT \eqref{rhonlowTeftSchmitt} is in line with the general expectation \cite{Leggett,Leggett2} that $\rho_n^{(0)}=0$. One of the arguments in favor of $\rho_n^{(0)}=0$ provided in \cite{Leggett,Leggett2} is based on linearized hydrodynamics. In \cite{Gouteraux:2019kuy}, we have shown that this argument does not correctly account for the source of superfluid velocity fluctuations, and that one cannot conclude that $\rho_n^{(0)}=0$ from these equations. This is not entirely surprising, as the specific temperature dependence of various thermodynamic quantities is an input to the hydrodynamic equations, which only allow to obtain the general expression for the collective modes and retarded Green's functions at late times and long wavelengths. The temperature dependence of thermodynamic quantities and transport coefficients must instead be computed given a specific microscopic theory. Another argument was given in \cite{Leggett2}. However it assumes the existence of coherent (quasi-)particles, which need not be true in strongly-coupled systems.
$\\$

Indeed, the vanishing of the normal density at zero temperature has recently been questioned by measurements of the superfluid density in overdoped high $T_c$ superconductors \cite{Bozovic1}. There, it was found that the superfluid density was anomalously low at low temperatures, and scaled linearly with temperature $\rho_s\simeq\rho_s^{(0)}+\# T$. Later, measurements of the optical conductivity in the superconducting phase at low temperature revealed a very modest depression of the spectral weight inside the low frequency Drude peak \cite{Bozovic2}, in tension with the expectation that $\rho^{(0)}=\rho_s^{(0)}$. In a BCS superconductor, the normal density vanishes towards zero temperature exponentially quickly, see e.g. \cite{Leggett}. Thus, as $T<T_c$, the normal spectral weight in the Drude peak is strongly suppressed and is transferred to the superfluid delta function. A legitimate concern is whether these expectations can be spoiled by disorder effects. However, these systems are expected to be quite clean, as evidenced by their very low residual resistivities. Moreover, disorder should induce a crossover from a linear to quadratic temperature dependence of the superfluid density at the lowest temperatures, which is not observed in the data \cite{Bozovic1}. This led \cite{Bozovic1, Bozovic2} to argue that the anomalously low superfluid density could not originate from pair breaking effects due to disorder, although counter-arguments have been presented in \cite{LeeHone1, LeeHone2, LeeHone3} in the context of the `dirty BCS' theory.
$\\$

In the absence of a well-posed microscopic theory of strongly-coupled superfluids, gauge-gravity duality offers an attractive framework to investigate the low temperature behavior of the normal density in superfluids. A pioneering achievement was the construction of holographic systems spontaneously breaking a U(1) symmetry, \cite{Gubser:2008px,Hartnoll:2008vx,Hartnoll:2008kx}. While the original constructions were `bottom-up' and relied on including a simple charged, complex scalar field in the bulk, holographic superfluids were subsequently studied in top-down string theory models \cite{Ammon:2008fc,Ammon:2009fe,Gauntlett:2009dn,Gauntlett:2009bh,Arean:2010wu,Arean:2011gz,Bobev:2011rv,Donos:2012yu,Adam:2012mw}. A number of studies also verified that the low energy dynamics of these systems matched various aspects of relativistic superfluid hydrodynamics, \cite{Herzog:2008he,Herzog:2009md,Sonner:2010yx,Herzog:2011ec,Bhattacharya:2011eea,Bhattacharya:2011tra}.
$\\$

In a recent paper, \cite{Gouteraux:2019kuy}, we demonstrated that $\rho_n^{(0)}$ need not vanish for certain holographic superfluids in the vicinity of a quantum critical phase. The main purpose of this work is to extend the results of that paper to general critical phases using gauge-gravity duality methods, establish criteria for $\rho_n^{(0)}\neq 0$ and work out the sub-leading low-temperature dependence on the critical exponents characterizing the phase. In particular, we find that the temperature dependence is not always given by \eqref{rhonlowTeft}.
$\\$

Hints that $\rho_n^{(0)}\neq0$ in holographic superfluids have previously been reported, see e.g. figure 7 of \cite{Herzog:2009md}, though the reason for this was unclear at the time. Here, we explain and generalize those results.  Whether or not $\rho_n^{(0)}$ vanishes depends on the nature of the zero temperature superfluid groundstate and on the spectrum of irrelevant deformations in its vicinity. In the main text of this work, we derive the normal density for a general translation invariant superfluid. We then apply this to Lifshitz superfluid groundstates which have been constructed in previous literature \cite{Gubser:2009cg,Horowitz:2009ij}. In Appendix \ref{scalecovariant}, we extend this analysis to include groundstates which feature hyperscaling violation \cite{Gouteraux:2012yr,Adam:2012mw,Gouteraux:2013oca} as well as cases with novel superfluid actions.\footnote{An analytic top-down solution featuring hyperscaling violation in the normal state can be found in \cite{Ammon:2012je}.} 
$\\$

While the analysis involves a somewhat subtle competition between various deformations, we can illustrate the general idea in two particularly simple and representative examples. In \cite{Gubser:2009cg}, it was shown that for a quartic scalar potential, the IR groundstate has an emergent conformal symmetry and is just another copy of Anti de Sitter. The existence of this groundstate relies on a certain deformation sourced by the gauge field being irrelevant in the IR, which restores the isotropy between time and space. We find that in this case $\rho_n^{(0)}$ vanishes as in \eqref{rhonlowTeft}:
\begin{equation}
\label{rhonlowTeft}
\rho_n\simeq\frac{s T}{\mu c_{ir}^2}\left(1-c_{ir}^2\right)+\dots
\end{equation}
where the dots denote subleading temperature dependence. \eqref{rhonlowTeft} reduces to \eqref{rhonlowTeftSchmitt} in the limit of small $c_{ir}\ll1$, as expected. \eqref{rhonlowTeft} can also be derived within the relativistic superfluid EFT, \cite{Delacretaz:2019brr,Delacretaz:2020nit}.
$\\$

On the other hand, for different choices of the parameters in the scalar potential, the deformation sourced by the gauge field becomes relevant and breaks the isotropy between time and space. This leads to a groundstate invariant under Lifshitz scale transformations with a non-trivial dynamical exponent $z>1$. In this case, dimensional analysis tells us that, for a Lifshitz-invariant state, the entropy density $s\sim T^{d/z}$ and the effective IR velocity $c_{ir}\sim T^{1-1/z}$. Naively extrapolating \eqref{rhonlowTeft} to the Lifshitz case implies that $\rho_n\sim T^{\frac{d+2-z}{z}}$. For sufficiently low values of $z<d+2$, $\rho_n$ does vanish at zero temperature and a more detailed calculation (see main text) reveals that it is still given by \eqref{rhonlowTeft} at low temperatures, with an appropriate definition of $c_{ir}$. 
$\\$

However, for $z> d+2$, the same \eqref{rhonlowTeft} naively predicts that $\rho_n^{(0)}$ diverges. The actual calculation (see main text) shows that it tends instead to a nonzero, finite value. This feature explains in particular the results of \cite{Herzog:2009md}. 
$\\$

{Our results shed new light on the relation between charge densities in the boundary theory and in the bulk, various aspects of which have been explored in previous literature \cite{Hartnoll:2011pp,Hartnoll:2011fn,Hartnoll:2012wm,Hartnoll:2012ux,Anantua:2012nj,Adam:2012mw,Gouteraux:2012yr,Gouteraux:2016arz,Martin:2019sxc,Martin:2019ulo}. In some of these works, semi-local geometries with $z=+\infty$ were singled out as they allow for Fermi surface-like features in their spectral functions. In such cases, it would be natural to expect that these fermionic-like, uncondensed degrees of freedom lead to a non-vanishing normal density at zero temperature, as in superfluid systems where Bogoliubov Fermi surfaces are formed, see e.g. \cite{Autti_2020}. 
$\\$

In the case at hand, one might have expected a priori that the normal density should be controlled by the charge behind the horizon $\rho_{in}$, and the superfluid density by the charge carried by the condensate in the bulk (which can be thought as a proxy for the density of condensed electrons). We find no such relation, and also do not find that $z=+\infty$ geometries play any special role. Instead, in conformal or Lifshitz IR phases with $z<d+2$, while $\rho_n^{(0)}$ and $\rho_{in}^{(0)}$ both vanish, they do so with a different temperature dependence. In Lifshitz phases with $z>d+2$, $\rho_n^{(0)}$ is non-vanishing as we noted above, but we still find that $\rho_{in}^{(0)}$ vanishes. We are also able to find (see appendix \ref{convergenceappendix}) phases similar to ${}^4$He, where $\rho_n^{(0)}$ vanishes but $\rho_{in}^{(0)}$ does not, as well as phases where $\rho_n^{(0)}$ and $\rho_{in}^{(0)}$ (the zero temperature limit of the charge behind the horizon) are both non-vanishing, see appendix \ref{scalecovariant}. From these results, we conclude that strongly-coupled holographic superfluids do not necessitate any fermionic-like degrees of freedom to exhibit a non-vanishing normal density at zero temperature, and provide an alternative mechanism to Bogoliubov Fermi surfaces.}

$\\$
In the remainder of this paper, we give the details of our holographic setup and of the quantum critical superfluid groundstates we are interested in section \ref{holosetup}. Then we explain how we compute the normal density and give a general criterion for its vanishing at zero temperature in section \ref{section:transport}. We illustrate our criterion on various numerical examples in section \ref{section:numerics}. We comment on the relevance of our results to previous literature in section \ref{section:connections}. Finally, we present our conclusion and discuss our results in a broader context in section \ref{section:discussion}. An extended appendix is devoted to deriving the hydrodynamic equations (appendix \ref{hydrosection}) as well as details of a technical proof (appendix \ref{app:importantintegral}), extending our results to theories with a hyperscaling violating exponent $\theta$ (appendix \ref{scalecovariant}), the case of weakly broken translations (appendix \ref{translationbreaking}), semi-local quantum critical geometries with Lifshitz exponent $z=+\infty$ (appendix \ref{semilocal}), and geometries with non-vanishing horizon charge density but vanishing normal density (appendix \ref{convergenceappendix}).
$\\$

An abridged version of our results can be found in \cite{Gouteraux:2019kuy}.

\section{Holographic superfluid model, generalities \label{holosetup}}
The class of holographic superfluids that interests us can be described by the following bulk action, \cite{Gubser:2008px,Hartnoll:2008vx,Hartnoll:2008kx,Gubser:2009cg,Horowitz:2009ij},
\begin{align}
\label{bulkaction}
S = \frac{1}{16\pi G} \int d^{d+2}x \sqrt{-g}\left\{R - \frac{1}{4}F_{MN}F^{MN} - |D\eta|^2 - V(|\eta|)\right\}.
\end{align}
Here $A_M$ is a $U(1)$ gauge field with field strength, $F_{MN} = \partial_MA_N-\partial_NA_M$. We choose to source only a time component of the gauge potential which breaks the charge-conjugation symmetry of the boundary fluid. The complex scalar field $\eta$ has charge $q$ under the $U(1)$ symmetry and covariant derivative $D_{M} \equiv \partial_{M}-iqA_{M}$. For the rest of the paper, we work in units in where $16\pi G = 1$.
$\\$

A general ansatz for the metric and matter fields consistent with translation and rotation symmetry that only breaks the charge-conjugation symmetry is 
\begin{align}
\label{generalmetric}
ds^2 = -D(r)dt^2 + B(r)dr^2 + C(r)d\vec{x}_d^2,\quad A = A_t(r) dt,\quad \eta = \eta^* = \eta(r).
\end{align}

The equations of motion arising from this ansatz are given in Appendix \ref{scalecovariant}.\footnote{There, we include an extra neutral scalar $\psi$. For the main text, $\psi = 0$ and $Z_F=M_F=1$.} Importantly, these equations can be combined into the conservation equation
\begin{align}
\label{eq:conservation}
\frac{d}{dr}\left\{\sqrt{\frac{C^d}{BD}}\left[C\left(\frac{D}{C}\right)' - A_t A_t'\right]\right\} = 0,
\end{align}
which follows from an invariance under certain rescalings in the bulk, \cite{Gubser:2009cg}, or more recently as the Noether charge following from one of the bulk Killing vectors, \cite{Gouteraux:2018wfe}.
In the UV, we are interested in solutions that have an asymptotic form ($r\to 0^+$),
\begin{align}
\label{UVmetric}
ds^2 \to \frac{dr^2 -dt^2+d\vec{x}_d^2}{r^2}.
\end{align}
For simplicity, we will chose a potential that satisfies
 \begin{align}
 V(|\eta|) = - d(d+1) - d\; |\eta|^2 +...
  \end{align}
 This choice of potential enforces that the matter fields have an asymptotic fall-off
 \begin{align}
 \eta&\simeq \eta^{(1)}r + \eta^{(2)}r^d+...,\quad
 A_t\simeq \mu - \frac{\rho}{(d-1)}r^{d-1}+...,\quad B \sim r^{-2}+...\nonumber\\
 D &\simeq r^{-2} - \frac{\left(\eta^{(1)}\right)^2}{2d} - \frac{\epsilon}{d+1}\;r^{d-1}+...,\quad
 C \simeq r^{-2} -  \frac{\left(\eta^{(1)}\right)^2}{2d} + \frac{P}{d+1}\;r^{d-1} +...\;.
 \end{align}
 
 Here the dots indicate terms subleading in the limit $r\to0^+$. $\mu$ is the chemical potential of the system and $\rho$ the total charge density. $P$ is the pressure and $\epsilon$ the energy density.
As is well-known, in the conventional quantization scheme, $\eta^{(1)}$ and $\mu$ act as sources for the expectation values $\eta^{(2)}$ and $\rho$. In particular, the choice $\mu \neq 0$ gives rise to a finite charge density and breaks the background charge-conjugation symmetry. With the choice $\eta^{(1)} = 0$, a non-zero $\eta^{(2)}$ indicates spontaneous breaking of the $U(1)$ symmetry and characterizes a holographic superfluid. 
$\\$

 The bulk action (\ref{bulkaction}) leads to an RG flow from a conformally invariant UV fixed point to a quantum critical phase in the IR, characterized by a Lifshitz dynamical exponent $z\geq 1$, \cite{Gubser:2009cg}. If $z=1$, the RG flow connects two scale-invariant Anti de Sitter spacetimes, described in section \ref{AdS4section}, whereas if $z>1$, the IR is described by a Lifshitz geometry,  see section \ref{Lifshitzsection}.

  \subsection{IR geometries with emergent conformal invariance \label{AdS4section}}

The condensation of the a scalar condensate in the action (\ref{bulkaction}) leads to an RG flow either to an emergent conformally invariant IR ($z=1$) or to a Lifshitz symmetric IR ($z>1$) depending on the IR behavior of the complex scalar and the relevance of the Maxwell field. Specifically, if the scalar field $\eta$ minimizes the scalar potential in the IR, $\frac{\partial V(\eta_0,\eta_0)}{\partial\eta^*} =\frac{\partial V(\eta_0,\eta_0)}{\partial\eta} = 0$, the RG flow is a domain wall solution interpolating between two conformally invariant fixed points with $z=1$ \cite{Gubser:2009cg}. We may write the metric in the IR as
\begin{align}
\label{AdSIRmetric}
ds^2 = -\frac{L^{2}}{\hat{r}^{2}}L_t^2\, dt^2 + \tilde{L}^2\,\frac{d\hat{r}^2}{\hat{r}^2} + \frac{L^2}{\hat{r}^2}L_x^2\,d\vec{x}_d^2.
\end{align}
$\hat r$ is an appropriately chosen IR coordinate that does not extend all the way to the UV region. Instead, $\hat r\gg L$, where $L$ is a scale that defines the region of spacetime where the metric takes the form \eqref{AdSIRmetric}. $\tilde L$ is the IR AdS radius, given by
\begin{align}
\tilde{L}^2\,V(\eta_0,\eta_0) = -d(d+1)\,.
\end{align}

In this solution, the gauge field is irrelevant. In particular, it does not support the IR geometry and we need to work out its leading behavior by solving the Maxwell equation in the background \eqref{AdSIRmetric}. By backreacting the solution on the metric and iterating the procedure, we can develop a consistent perturbative series that approximates the IR solution. We find
\begin{align}
\label{AtIRAdS}
A_t &\simeq L_t c_A \left(\frac{\hat{r}}{L}\right)^{\tilde{\Delta}_{A_0}-1}\left(1+\#c_A^2\left(\frac{\hat{r}}{L}\right)^{2\tilde{\Delta}_{A_0}}+...\right),\nonumber\\
 \tilde{\Delta}_{A_0} &= d - \nu_A, \quad \nu_A = \frac{d-1}{2}\left(1+\sqrt{1+\frac{8}{(d-1)^2}\tilde{L}^2q^2\eta_0^2}\;\;\right).
\end{align}
For the perturbation to be irrelevant, $\tilde \Delta_{A_0}<0$. Clearly this requires a non-vanishing condensate, $\eta_0\neq 0$, and so, in the context of the RG flow, both $A_t$ and $\eta$ drive the system away from the IR fixed point towards the UV. 
$\\$
There is also a deformation associated with the charged scalar,
\begin{align}
\label{scaleinvariantscalarperturbation}
&\quad\quad\quad\quad\eta \simeq \eta_0 + c_\eta \left(\frac{\hat{r}}{L}\right)^{\tilde{\Delta}_\eta}\left[1+\#c_\eta^2\left(\frac{\hat{r}}{L}\right)^{2\tilde{\Delta}_\eta}+...\right]\nonumber\\
\tilde{\Delta}_\eta &= \frac{d+1}{2}\left(1-\sqrt{1+\frac{4}{(d+1)^2}\tilde{L}^2 M^2_{IR}}\;\;\right),\quad M^2_{IR} = \frac{\partial^2 V}{\partial\eta\partial\eta^*} + \frac{\partial^2 V}{\partial\eta^{*2}}\biggr|_{\eta=\eta^*=\eta_0}
\end{align}
We note that, because $\eta_0$ minimizes the scalar potential, $M^2_{IR}$ is always positive and real, and hence $\tilde{\Delta}_{\eta}<0$ so that the perturbation is always irrelevant. Both of these perturbations contribute to the RG flow from the IR fixed point to the conformally invariant UV. We write this schematically as
\begin{align}
ds^2 \simeq ds_0^2\left[1+\#c_A^2\left(\frac{\hat{r}}{L}\right)^{2\tilde{\Delta}_{A_0}}+\#c_\eta^2\left(\frac{\hat{r}}{L}\right)^{2\tilde{\Delta}_\eta}+...\right].
\end{align}
We also note that this agrees with \cite{Gubser:2009cg} for the case $d=2$ when the scalar potential is quartic.

\subsection{Lifshitz symmetric IR geometries \label{Lifshitzsection}}
While $\tilde{\Delta}_\eta<0$ always, the perturbation associated with $A_t$ can become relevant when $\tilde \Delta_{A_0}>0$ \cite{Gubser:2009cg}.
When this occurs, the Maxwell field strongly backreacts on the IR geometry and forces the scalar field to minimize the effective potential, $\frac{\partial V_{eff}(\eta_0,\eta_0)}{\partial\eta^*} =\frac{ \partial V(\eta_0,\eta_0)}{\partial\eta^*}+2q^2A_t^2\eta_0=0$. In this case, the RG flow is from a conformally invariant UV fixed point to a Lifshitz symmetric quantum critical phase in the IR with $z>1$ that is determined in terms of $q$ and parameters in $V$ :
\begin{align}
\label{IRmetric}
ds^2 = -\frac{L^{2z}}{\hat{r}^{2z}}L_t^2 dt^2 + \tilde{L}^2\frac{d\hat{r}^2}{\hat{r}^2} + \frac{L^2}{\hat{r}^2}L_x^2d\vec{x}_d^2\,.
\end{align}
We can rescale $\hat{r} \to (L/\tilde{L})\hat{r}$ so that we see only the scale $\tilde{L}$ appears in the IR metric. The dynamical critical exponent is
\begin{align}
z = \frac{2}{d}\;q^2\tilde{L}^2\eta_0^2.
\end{align}
A consistent solution also requires
\begin{align}
V(\eta_0,\eta_0) = -\frac{d^2+(d-1)z+z^2}{\tilde{L}^2},\quad \frac{\partial V}{\partial \eta^*}(\eta_0,\eta_0) =\frac{d\; (z-1)}{\tilde{L}^2\eta_0}.
\end{align}
These criteria can be used to find $q, \eta_0$ and $\tilde{L}$ for a given value of $z$. The gauge field now behaves as
\begin{align}
\label{IRAtz}
A = L_t\sqrt{2-\frac{2}{z}}\left(\frac{\hat{r}}{L}\right)^{-z}dt.
\end{align}
A boundary case to the superfluid Lifshitz geometries is $z\to\infty$, leading to an AdS$_2\times$R$^d$ superfluid geometry (also sometimes called a semi-local quantum critical geometry \cite{Iqbal:2011in}). Under a rescaling, 
\begin{align}
\left(\frac{\hat{r}}{L}\right) \to \left(\frac{\tilde{r}}{L}\right)^{-1/z}
\end{align}
and taking the $z\to \infty$ limit, the metric (\ref{IRmetric}) becomes
\begin{align}
ds^2 = \left(\frac{\tilde{r}}{L}\right)^{-2}(-L_t^2dt^2+d\tilde{r}^2)+L_x^2d\vec{x}_d^2, \quad A_t = \sqrt{2}L_t \left(\frac{\tilde{r}}{L}\right)^{-1}
\end{align}
To find a consistent holographic RG flow to such a phase with a superfluid requires a modified action. We detail this in Appendix \ref{semilocal}.
$\\$

As before, we investigate the irrelevant deformations about the IR Lifshitz fixed point (we leave the $z\to\infty$ details to Appendix \ref{semilocal}). For $d=2$ and a quartic potential, this was done in \cite{Gubser:2009cg}. We are interested in a particular set of deformations which connect the IR fixed point to the UV. Writing $\vec{X} = \{D,B,C,A_t\}$,
\begin{align}
\eta = \eta^* \simeq \eta_0 + c_\eta\left(\frac{L}{\hat{r}}\right)^{\nu_\eta}+...\,,\quad\vec{X} \simeq\vec{X}_{c_\eta=0}\left[1+c_{\vec{X}}\left(\frac{L}{\hat{r}}\right)^{\nu_\eta}+...\right]\,.
\end{align}
We omit writing the details of $c_i$ since they are not particularly enlightening. The exponent is
\begin{align}
\nu_\eta(\sigma_1,\sigma_2) = -\frac{d+z}{2} + \frac{\sigma_1}{2\eta_0}\sqrt{D_1+2\sigma_2\sqrt{D_2}}
\end{align}
where
\begin{align}
D_1 &= 2d(1-z)+ \left(5z^2+2M_{IR}^2\tilde{L}^2 + d(4+d) - 2(2+d)z\right)\eta_0^2\nonumber\\
D_2 &= \left[\left(z^2-(d+1)z+d-\frac{M^2_{IR}\tilde{L}^2}{2}\right)\eta_0^2+\frac{d(z-1)}{2}\right]^2-4d\left[z^2+(d-2)z-(d-1)\right]\eta_0^2
\end{align}
and $\sigma_i = \pm 1$. $M_{IR}$ is defined as in (\ref{scaleinvariantscalarperturbation}). It is worth noting that these deformations depend on the form of the scalar potential and can be complex for certain choices of parameters. Nevertheless, there should always exist a parameter range with real deformations for any choice of $z$. In this paper, we will fix the form of the scalar potential and choose $z$ such that the deformations are real. It is clear that $\nu_\eta(-1,\sigma_2)$ will always be a relevant perturbation, so this deformation does not participate in the RG flow. The least irrelevant perturbation will have $\tilde{\nu}_\eta \equiv \nu_{\eta}(+1,-1)$. For the semi-local phase, the irrelevant deformations will depend on our choice of action. This is detailed in Appendix \ref{semilocal}.

\subsection{Nonzero temperature}
The solutions just described always possess a relevant deformation of the IR critical phase that introduces a temperature. In the language of the previous section, the finite temperature deformation of the background fields has a radial dependence $\nu_\eta = d+z$. Unfortunately, since $\eta_0\neq 0$, we cannot write the finite temperature in closed form. This is in contrast to the solutions we discuss in Appendix \ref{scalecovariant} whose finite temperature metric can be written in closed form for $\eta=0$. There, we see that when we set the hyperscaling violation parameter $\theta=0$, the finite temperature deformation also carries the same radial dependence, captured by an emblackening factor in the metric: $g_{tt} = g_{tt,0}[1-(\hat{r}/\hat{r}_h)^{d+z}]$. Semi-classical gravity tells us to interpret the area of the black hole horizon as the entropy density of the boundary superfluid and regularity of the Euclidean solution at the horizon gives us the superfluid temperature. For solutions that we can write in closed form, we find
\begin{align}
\label{scalingTsrh}
T=c_T \left(\frac{\hat{r}_h}{L}\right)^{-z}, \quad s = 4\pi L_x^d \left(\frac{\hat{r}}{L}\right)^{-d}
\end{align}
where $c_T$ is some constant which can be found in terms of the parameters in the IR metric. In particular, we find that $s \sim T^{d/z}$ which is the expected temperature dependence of the entropy following from dimensional analysis. For the Lifshitz solutions we describe in the main text, though we have no closed form expression for $c_T$, to leading order, $T$ carries the same dependence on $\hat{r}_h$. Furthermore, the expression for $s$ does not change.
$\\$

Despite the lack of a closed form expression for the finite temperature metric, sufficiently close to the horizon we may write the metric as
\begin{align}
\label{nearhorizonmetric}
ds^2 \to -4\pi T(r_h-r)dt^2 + \frac{dr^2}{4\pi T(r_h-r)} + \frac{s}{4\pi}d\vec{x}_d^2\,.
\end{align}
where $r_h\sim \hat{r}_h^{-1/z}$. Furthermore, $A_t$ must vanish at the horizon, so in the near horizon region we may write
\begin{align}
A_t = A_h(r_h-r)+...
\end{align}
 From this expansion, it is clear that by integrating (\ref{eq:conservation}) we obtain the identity
\begin{align}
\label{eq:conservationsT}
\sqrt{\frac{C^d}{BD}}\left[C\left(\frac{D}{C}\right)' - A_t A_t'\right]=-s T\,.
\end{align}
Moreover, by evaluating the left hand side at the UV boundary, we recover the Smarr relation
\begin{equation}
\epsilon+P=s T+\mu\rho\,.
\end{equation}

It is important to note that the deformation that leads to a temperature also contributes to a variation in the charge density which likewise contributes universal deformations of the normal and superfluid densities. 
This deformation can be the leading deformation but for certain critical phases we will see that the universal deformation is subleading to non-universal deformations that are sourced by the condensate.
$\\$

\subsection{Horizon fluxes}
Next, because horizons have a finite, large $N_c^2$ entropy, they are naturally associated with deconfined degrees of freedom in the system, or in our hydrodynamic interpretation, degrees of freedom that have dissipated into the thermal bath, see \cite{Hartnoll:2011fn}. We interpret the low-temperature horizon flux as a measure of the charge of these degrees of freedom. Upon condensation, we will see that this charge may be less than the total charge of the system, indicating the presence of charged degrees of freedom that are isolated from the thermal bath. These charged degrees of freedom sit instead in the condensate. They can be thought of as the analogue of the condensed degrees of freedom. When there is no horizon flux at $T=0$, all the charge is in the condensed degrees of freedom. This phase has previously been referred to as ``cohesive'' following the condensed matter literature; however, we feel that fully condensed is a more faithful reflection of the holographic picture. Any residual horizon flux at $T=0$ will indicate a phase that is not fully condensed. In the holographic literature, this has been called ``fractionalized.'' Here, we will refer to this as partially condensed if there is a condensate or uncondensed if there is no condensate. As an example, an AdS Reissner-Nordstrom black hole is an uncondensed phase (our semi-local geometries share the same IR). In the main text, we only consider fully condensed phases, though in Appendix \ref{scalecovariant} we will also consider partially condensed phases (see also \cite{Gouteraux:2019kuy}).
$\\$

The horizon flux is defined as
\begin{align}
\rho_{in}(r_h) \equiv -\frac{C^{d/2}}{\sqrt{BD}}A_t'\biggr|_{r=r_h}
\end{align}
While we do not have closed form expressions for the finite temperature fields, we can use the near horizon metric to find
\begin{align}
\rho_{in}(r_h) = C_h^{d/2}A_h
\end{align}
where we have defined $C_h = C(r_h)$. Given that $A_t = A_h(r_h-r)+...$ near the horizon, and that $\lim_{T\to0}A_{t}'(r) = A_{t,0}'(r)$, we can approximate $A_h$ by the zero temperature metric at the finite temperature horizon $-A_{t,0}'(r_h)$. Then, for scale invariant geometries, we have
\begin{align}
\label{rhoinscaleinvariant}
\rho_{in} = c_{in}T^{d-\tilde{\Delta}_{A_0}}+...
\end{align}
while for the Lifshitz solutions, we have
\begin{align}
\label{rhoinscalecovariant}
\rho_{in} = c_{in}T^{-d/z}+...
\end{align}
In these expressions, $c_{in}$ is a temperature-independent prefactor which depends on the precise form of the finite temperature metric through the relation between the temperature and $\hat{r}_h$, \eqref{scalingTsrh}. In both the scale invariant and Lifshitz cases, $\lim_{T\to 0}\rho_{in} = 0$ and hence are both fully condensed. On the other hand, semi-local quantum critical phases have
\begin{align}
\rho_{in} = c_{in}T^{0} + ...
\end{align}
and hence are partially condensed. The subleading temperature behavior depends on the precise form of the action.

\section{Transport \label{section:transport}}

The long-wavelength, low frequency fluctuations of our holographic system are well-described by the two-fluid hydrodynamical model of Landau and Tisza \cite{Landau,Tisza}. As we detail in Appendix \ref{hydrosection}, for a slowly fluctuating electric field in the $\hat{x}$ direction, 
the longitudinal conductivity\footnote{In the absence of a background superfluid velocity, the transverse conductivity does not feel the effects of the superfluid and is written $\sigma_\perp(\omega) = \frac{i}{\omega}\left(G_{J_{\hat{y}}J_{\hat{y}}}^R(\omega,0)-G_{J_{\hat{y}}J_{\hat{y}}}^R(\omega=0,k\to0)\right) = \left(\frac{i}{\omega}+\delta(\omega)\right)\frac{\rho_n^2}{\mu\rho_n + sT} + \sigma_0$, where the second term is a static susceptibility that needs to be subtracted out. In this term, the $\omega=0$ limit is taken first, and the zero wavector limit second, where $k$ is the component of the wavevector along the $\hat x$ direction.} can be written as
\begin{align}
\sigma(\omega) = \frac{i}{\omega}G_{J_{\hat{x}}J_{\hat{x}}}^R(\omega,0) = \left(\frac{i}{\omega}+\delta(\omega)\right)\left[\frac{\rho_n^2}{\mu\rho_n + sT} + \frac{\rho_s}{\mu}\right] + \sigma_0
\end{align}
$\\$

Here, $\sigma_0$ is the incoherent conductivity which describes the dissipative part of transport, which we discuss in section \ref{section:incoherentconductivity}. The other terms stem from translation invariance and spontaneous breaking of the $U(1)$ global symmetry in the boundary, and we discuss them in section \ref{section:normalsuperfluiddensities}. Before doing so, it is useful to review the general strategy to compute the longitudinal conductivity holographically.
$\\$

To do this, we solve for the linearized fluctuations sourced by $\delta A_{\hat{x}} = a_{\hat{x}}(r)e^{-i\omega t}$. The only other field sourced by this fluctuation is $\delta g_{t\hat{x}}$. The two equations of motion are
\begin{align}f
\label{conductivityequations}
\frac{1}{C^{d/2-1}}\frac{d}{dr}\left[C^{d/2-1} \sqrt{\frac{D}{B}}a_{\hat{x}}'\right]
- \sqrt{\frac{B}{D}}\left(2q^2D\eta^2 + \frac{(A_t')^2}{B} - \omega^2\right)a_{\hat{x}} &=0\nonumber\\
\frac{d}{dr}\left[\frac{g_{t\hat{x}}}{C}\right] + \frac{a_{\hat{x}} A_t'}{C} &=0
\end{align}
The UV expansion of the fluctuations are (where as before $r\to 0^+$ is the boundary) 
\begin{align}
\label{conductivityUVexpansions}
a_{\hat{x}}(r) = a_{\hat{x}}^{(0)} + \frac{a_{\hat{x}}^{(1)}}{d-1}r^{d-1}+...\nonumber\\
g_{t\hat{x}}(r) = u^2 g_{t\hat{x}}^{(0)} - \frac{\langle T_{t\hat{x}}\rangle}{(d+1)}r^{d-1} +...
\end{align}
Following standard holographic renormalization, $a_{\hat{x}}^{(0)} = E_{\hat{x}}/(i\omega)$ and if there is no temperature gradient $g_{t\hat{x}}^{(0)} = 0$. Then
\begin{align}
\label{holographicconductivity}
\sigma(\omega) = \frac{a_{\hat{x}}^{(1)}}{i\omega a_{\hat{x}}^{(0)}}.
\end{align}

\subsection{Incoherent conductivity \label{section:incoherentconductivity}}
The incoherent conductivity measures the transport of charged operators that do not have overlap with the momentum \cite{Davison:2015bea,Davison:2015taa,Davison:2018ofp,Davison:2018nxm}. As a consequence, it only carries diffusive excitations, hence the name `incoherent'. In particular, even in translation invariant phases, like the ones we discuss here, its DC limit is non-divergent. It can be defined as:
\begin{align}
\label{DefIncConductivity}
\sigma_0 \equiv \lim_{\omega\to 0} \text{Re} [\sigma(\omega)]\,
\end{align}
and as we now review, can be obtained through a near-horizon analysis, \cite{Lucas:2015vna,Davison:2015taa}. Since (\ref{conductivityequations}) is a second order ordinary differential equation, it admits two independent solutions. It is helpful to distinguish between the solution regular at the horizon and the singular solution. Given knowledge of the regular solution, $a^{reg}_{\hat{x}}(r)$, to the zero frequency limit of \eqref{conductivityequations}, the singular solution can be obtained using the Wronskian method,
\begin{align}
\label{BdySingularSol}
a^{sing}_{\hat{x}}(r) = a^{reg}_{\hat{x}}(r)\int_0^r \left[C^{d/2-1}\sqrt{\frac{B}{D}}a^{reg}_{\hat{x}}(r')^2\right]^{-1} dr'.
\end{align}
Near the horizon, we use the fact that at finite temperature $a^{reg}_{\hat{x}}(r_h) \neq 0$ and we verify that
\begin{align}
a^{sing}_{\hat{x}}(r) \to - \frac{1}{4\pi Ta^{reg}_{\hat{x}}(r_h)}\ln(r-r_h)+\text{finite}
\end{align}
is indeed singular as $r\to r_h$.
$\\$

Finally, we note that the general solution to \eqref{conductivityequations} must satisfy ingoing boundary conditions in order to correspond to the calculation of retarded Green's functions \cite{Son:2002sd},
\begin{align}
\partial_ra_{\hat{x}}(r) \to -\frac{i\omega}{4\pi(r-r_h)}a_{\hat{x}}(r).
\end{align}
Expanding $a_{\hat{x}}(r)$ slightly away from the horizon, we find 
\begin{align}
\label{NearHorizonIngoingBC}
a_{\hat{x}}(r) \approx a^{reg}_{\hat{x}}(r_h)\left[1-\frac{i\omega}{4\pi T}\ln(r-r_h)+...\right],
\end{align}
and so must be a specific combination of the regular and singular solution.
The subleading pieces in this expansion come from the smooth parts of $a_{\hat{x}}(r)$ near the horizon and do not contribute to $\sigma_0$. This reflects the fact that dissipation in the dual field theory is intimately connected with the presence of an event horizon in the bulk geometry.
$\\$

At small frequencies, one can expand the general solution $a_{x}(r)$ to \eqref{conductivityequations} as
\begin{align}
a_{\hat{x}}(r) = a^{reg}_{\hat{x}}(r) + i\omega \tilde{a}_{\hat{x}}(r) +\mathcal{O}(\omega^2).
\label{smallomegaexpansion}
\end{align}
The first term must be the regular solution $a^{reg}_{\hat{x}}(r)$ to \eqref{conductivityequations} with $\omega\to 0$. 
The second term is identified by requiring that the frequency dependence of \eqref{smallomegaexpansion} and \eqref{NearHorizonIngoingBC} match near the horizon
\begin{align}
\tilde{a}_{\hat{x}}(r) = (a^{reg}_{\hat{x}}(r_h))^2 a^{sing}_{\hat{x}}(r).
\end{align}
Away from the horizon there may be extra terms, but these do not contribute to the leading low frequency behaviour of the conductivity. Hence the solution to \eqref{conductivityequations} valid in the entire spacetime in the limit of low frequencies is
\begin{align}
a_{\hat{x}}(r) = a^{reg}_{\hat{x}}(r) + i\omega  (a^{reg}_{\hat{x}}(r_h))^2 a^{sing}_{\hat{x}}(r)+\mathcal{O}(\omega^2).
\label{smallomegaexpansion2}
\end{align}
We can now expand this expression near the boundary
\begin{align}
a_{\hat{x}}(r\to 0^+) \to a^{reg,(0)}_{\hat{x}} + \frac{a^{reg,(1)}_{\hat{x}}}{(d-1)}r^{d-1} + \frac{i\omega  a^{reg,(0)}_{\hat{x}}  (a^{reg}_{\hat{x}}(r_h))^2}{(d-1)}r^{d-1} +\mathcal{O}(\omega^2),
\label{UVexpansion3}
\end{align}
where we used that $a^{sing}_{\hat{x}}(r)$ vanishes at the boundary \eqref{BdySingularSol}.
$\\$

Finally, we use that $a^{reg,(1)}_{\hat{x}}$ must be real together with the definitions \eqref{holographicconductivity}, \eqref{DefIncConductivity}, to deduce that the incoherent conductivity is given by
\begin{align}
\sigma_0\equiv\lim_{\omega\to 0} Re[\sigma] =  (a^{reg}_{\hat{x}}(r_h))^2.
\label{sigmaQ}
\end{align}

The incoherent conductivity is always given by this horizon quantity. However, as we have seen, we generally do not know $a^{reg}_{\hat{x}}(r_h)$. For small temperatures, however, we show in section \ref{section:normalsuperfluiddensities} that $a^{reg}_{\hat{x}}(r)\approx\frac{a_{\hat{x}}^{(0)}}{\mu}\;A_{t,0}(r)$ plus terms which are subleading in the temperature. The zero temperature gauge field $\;A_{t,0}$ does not vanish on the (finite temperature) horizon, and we have
\begin{align}
\label{LowTscalingSigmaQ}
\sigma_{0} = \frac{(a_{\hat{x}}^{(0)})^2}{\mu^2}\;A_{t,0}(r_h)^2 \sim \begin{cases} \#\left(\frac{T}{\mu}\right)^{2-2\tilde{\Delta}_{A_0}}&z=1\,\\
\# \left(\frac{T}{\mu}\right)^{2} &z>1\,.\end{cases}
\end{align}

\subsection{Normal and superfluid densities \label{section:normalsuperfluiddensities}}
The normal and superfluid densities can be found from looking at 
\begin{align}
\label{rhonZdefinition}
Z \equiv \lim_{\omega\to 0}\omega \text{Im}[\sigma(\omega)] = \frac{\rho_n^2}{sT+\mu\rho_n} + \frac{\rho_s}{\mu} \quad\Rightarrow\quad \rho_n = \frac{sT(\rho-\mu Z)}{sT-\mu(\rho-\mu Z)}
\end{align}
where we used $\rho = \rho_n + \rho_s$. Importantly, $\lim_{T\to 0}Z =  \frac{\rho}{\mu}$. It is a separate question whether or not $\lim_{T\to 0}\rho_n = 0$.
$\\$

In a recent paper \cite{Gouteraux:2019kuy}, we showed that there are no inconsistencies in the two-fluid hydrodynamic model when $\rho_n^{(0)}\neq 0$ and illustrated this behavior with a holographic example. Here we establish the criteria for this behavior in quantum critical phases with scale-invariant ($z=1$) or Lifshitz ($z>1$) symmetries for the specific action (\ref{bulkaction}). In Appendix \ref{scalecovariant}, we will generalize this result to more general quantum critical phases with nonzero hyperscaling violation exponent $\theta\neq 0$ as well as novel superfluid actions in Appendix \ref{convergenceappendix}. For the purpose of the main text, we find that
\begin{align}
\label{CriterionNonVanishingrhon0}
d + 2 - z < 0 \Rightarrow \rho_n^{(0)}\neq 0.
\end{align}
As explained in the previous section \ref{section:incoherentconductivity}, the equation for $a_{\hat{x}}$ has two independent solutions, one of which is regular at the black hole horizon and another of which is singular. At low frequencies, it is easily seen that the singular part does not contribute to the imaginary conductivity, though it does contribute to the dissipative part, $\sigma_0$. Thus, to find the pole in the imaginary conductivity (and from there the normal density through \eqref{rhonZdefinition}), one needs only find the regular solution to (\ref{conductivityequations}) at $\omega=0$,
\begin{align}
Z = -\frac{a_{\hat{x}}^{reg,(1)}}{a_{\hat{x}}^{reg,(0)}}\,.
\end{align}
Since the normal density is a thermodynamic property of the system, it is not surprising that it is enough to work at $\omega=0$ to determine it, as was done in e.g. \cite{Herzog:2009md,Sonner:2010yx,Herzog:2011ec,Bhattacharya:2011eea,Bhattacharya:2011tra}.
$\\$

We now explain in detail how we arrive at \eqref{LowTscalingSigmaQ} and \eqref{CriterionNonVanishingrhon0}.
$\\$

Due to the presence of the condensate, (\ref{conductivityequations}) cannot be solved exactly. However, it does allow for a perturbative solution at small temperature. It will be useful throughout the derivation to keep in mind a few approximations we will make. The first is that everywhere in the derivation, we will use the radial coordinate $r$ that aligns with Appendix \ref{scalecovariant}. In particular, $r\to 0$ corresponds to $u\to\infty$ for the UV coordinate defined in \eqref{UVmetric}. In addition, $r\to r_h$ corresponds to $\hat{r}\to\hat{r}_h$ for the IR coordinate defined in \eqref{IRmetric}. 
$\\$

 To begin, we note that we can use (\ref{eq:conservation}) to rewrite the $\omega\to 0$ limit of (\ref{conductivityequations}) in the simple form 
\begin{align}
\label{perturbativeequation1}
\frac{d}{dr}\left[C^{d/2-1}\sqrt{\frac{D}{B}}A_t^2\left(\frac{a_{\hat{x}}}{A_t}\right)'\right] = -(sT) \frac{A_t'}{C}a_{\hat{x}}.
\end{align}
where primes denote derivatives with respect to the radial coordinate $r$.
$\\$

From here, it is clear that at $T=0$, the regular solution is
\begin{align}
a_{\hat{x}} = \frac{a_{\hat{x}}^{(0)}}{\mu}A_{t,0}
\end{align}
where $A_{t,0}$ is the zero temperature solution to (\ref{eq:eqsofmotion}).
$\\$

While this equation is very simple, it slightly obscures the role of the complex scalar $\eta$. To restore this dependence, we introduce a function
\begin{align}
R = -\frac{C^{d/2}A_t'}{\sqrt{BD}}\,,\quad R'=-2q^2\sqrt{\frac{B C^d}{D}}\eta^2 A_t
\end{align}
which has the property that at the horizon $R(r_h) = \rho_{in}$ and at the UV boundary $R(r=0) =\rho$. Using \eqref{eq:conservation} and integrating by parts, we may write a slightly messier but more convenient equation
\begin{align}
\label{perturbativeequation2}
\left[C^{d/2-1}\sqrt{\frac{D}{B}}A_t^2\left(1+\frac{sT}{A_t R}\right)\left(\frac{a_{\hat{x}}}{A_t}\right)'-sT\frac{D}{C}\left(\frac{a_{\hat{x}}}{A_t}\right)\right]' =(sT)\frac{2q^2\eta^2 C^{d-1} A_t^2}{R^2}\left(\frac{a_{\hat{x}}}{A_t}\right)'
\end{align}
As an aside, setting the condensate to zero $\eta=0$, $R = \rho$ and using the bulk identity \eqref{eq:conservationsT}, the regular solution for any $T$ is \cite{Davison:2015taa}
\begin{equation}
a^{reg}_{\hat{x}} = a^{reg,(0)}_{\hat{x}} \frac{sT+\rho A_{t}(r)}{sT+\mu\rho}
\end{equation}
from which we immediately get
\begin{align}
\label{eq:noetaexpansion}
Z = \frac{\rho^2}{sT+\mu\rho} \sim \frac{\rho}{\mu} - \frac{sT}{\mu^2} + \frac{(sT)^2}{\mu^3\rho}+...
\end{align}
as expected for a normal fluid.
$\\$

Returning to \eqref{perturbativeequation2}, we treat $sT$ as an expansion parameter. To clean up our expressions, we define $\mathcal{A} = \frac{\mu}{a_{\hat{x}}^{(0)}}\frac{a_{\hat{x}}}{A_t}$, and our expansion reads
\begin{align}
\label{temperatureexpansion}
\mathcal{A} = \mathcal{A}_0 + (sT) \mathcal{A}_1 + (sT)^2\mathcal{A}_2 +...
\end{align}
Notably, 
\begin{align}
\label{ZfrommathcalA}
r^{2-d}\frac{d\mathcal{A}}{dr}\biggr|_{r\to 0^+} =  \frac{\rho}{\mu}-Z.
\end{align}
We can then find $\rho_n$ from (\ref{rhonZdefinition}).
$\\$

In this expansion, we must be careful to distinguish between explicit temperature dependence in $sT$ and implicit dependence in the background metric and matter functions. We start by writing the expansion in terms of the finite temperature background metric and matter field.
$\\$

The functions $\mathcal{A}_i$ can be found in terms of the lower order functions $\mathcal{A}_{(i-1)}$,
\begin{align}
\label{Aexpansionequation}
\left(C^{d/2-1}\sqrt{\frac{D}{B}}A_t^2\mathcal{A}_i'\right)' = \left(\left[\frac{D}{C}\mathcal{A}_{i-1}-C^{d/2-1}\sqrt{\frac{D}{B}}\frac{A_t}{R}\mathcal{A}_{i-1}'\right]' + \frac{2q^2\eta^2 C^{d-1}A_t^2}{R^2}\mathcal{A}_{i-1}'\right).
\end{align}
The first few terms are
\begin{align}
\label{Aexpansion}
\mathcal{A}_0 &= 1\nonumber\\
\mathcal{A}_1 &= \int_{0}^{r} \frac{\sqrt{BD}}{C^{d/2}A_t^2}\;dr'\nonumber\\
\mathcal{A}_2 &= \int_{0}^{r}\frac{\sqrt{BD}}{C^{d/2}A_t^2}\;dr' \int_{0}^{r'} \frac{\sqrt{BD}}{C^{d/2}A_t^2}d\tilde{r}-\int_{0}^r  \frac{\sqrt{BD}}{C^{d/2}A_t^3R} dr' \nonumber\\
&\quad\quad- \int_{0}^r \sqrt{\frac{B}{D}}\frac{1}{C^{d/2-1}A_t^2} dr'\;\int_{r'}^{r_h} \frac{2q^2\eta^2\sqrt{BD} C^{d/2-1}}{R^2}d\tilde{r} \,.
\end{align}
In writing this solution, we have fixed that $a_{\hat{x}}=a^{(0)}_{\hat{x}}$ at the UV boundary always. Notably, because $A_t \simeq A_h(r_h-r)+O\left((r_h-r)^2\right)$, the integrals diverge as $r\to r_h$. Nevertheless, because $\lim_{r\to r_h}a_{\hat{x}}$ is finite, $\lim_{r\to r_h} A_t \mathcal{A}$ cannot diverge. It is easily seen that
\begin{align}
\lim_{r\to r_h} A_t \mathcal{A}_0 = 0,\quad\lim_{r\to r_h} A_t\mathcal{A}_1 = \frac{1}{\rho_{in}}\,.
\end{align} 
For $\mathcal{A}_2$, the first and second terms in \eqref{Aexpansion} seem to diverge as $A_t^{-2}$. However, the first term is
\begin{align}
\label{convergenceintegral2}
\lim_{r\to r_h}\int_{0}^{r}\frac{\sqrt{BD}}{C^{d/2}A_t^2}dr' \int_{0}^{r'} \frac{\sqrt{BD}}{C^{d/2}A_t^2}d\tilde{r} \approx -\int^{r_h} \frac{\sqrt{BD}}{C^{d/2}A_t^3\rho_{in}}dr'+...
\end{align}
so that the potential divergences cancel against each other. The remaining terms diverge no faster than $A_t^{-1}$ so that $\lim_{r\to r_h}A_t\mathcal{A}_2$ is finite.
$\\$

As we stated earlier, our expansion contains explicit temperature dependence in the form of $sT$ as well as implicit temperature dependence from the metric and matter fields. If our expansion is well-behaved, then the implicit temperature dependence should be subleading to the explicit dependence. In particular, to leading order in the temperature, the explicit dependence is dominant when the metric and matter fields appearing in $\mathcal{A}_i$ can be approximated by their zero temperature limit. Here we run into an issue. Noting that $T\to 0$ is $r_h\to \infty$, we find that the term arising from the condensate behaves as
\begin{align}
\label{divergentintegral}
\lim_{r_h\to \infty}\int^{r_h}_0\frac{2q^2\eta_0^2\sqrt{B_0D_0}C_0^{d/2-1}}{R_0^2}dr' \sim \frac{2q^2\eta_h^2 }{\rho_{in}^2(r_h)} r_h^{-z-d+2}\,.
\end{align}
Recalling the behavior of $\rho_{in}$ in (\ref{rhoinscaleinvariant}) and (\ref{rhoinscalecovariant}), we find that for $z<d+2$, the integral diverges as $r_h\to \infty$. This indicates that our expansion breaks down as $T\to 0$. 
$\\$

We thus need to distinguish between two cases, $z<d+2$ and $z>d+2$.
\subsubsection{Vanishing $\rho_n^{(0)}$}
For $z<d+2$, the integral \eqref{divergentintegral} diverges and contributes to the low temperature expansion at subleading order to $(sT)^2$. We now extract the precise temperature dependence and the associated prefactor. 
$\\$

In the integral (\ref{divergentintegral}), we unfortunately cannot simply replace the metric and matter fields with their zero temperature limits. This is because the finite temperature versions of the zero temperature AdS \eqref{AdSIRmetric} and Lifshitz \eqref{IRmetric} metrics  cannot be found in closed form. While simply substituting the zero temperature forms of the background fields in the integrand and introducing temperature through the upper bound $r_h$ would get the correct scaling with $T$, the prefactor would not be reliable. Nevertheless, we can still evaluate the integral in the following way. Let us first note that
\begin{align}
\label{rewrittenintegral}
\int_{0}^{r_h}d\tilde{r}\frac{2q^2\eta^2\sqrt{BD}C^{d/2-1}}{R^2} = \int_{0}^{r_h}d\tilde{r}\left\{-\left(\frac{D}{CR}\right)'\frac{1}{A_t}+ \frac{1}{RA_t}\left(\frac{D}{C}\right)'\right\}
\end{align}
The last term in the integral is
\begin{align}
\int_{0}^{r_h} d\tilde{r}\frac{1}{RA_t}\left(\frac{D}{C}\right)'  = \int_{0}^{r_h}d\tilde{r}\frac{\sqrt{BD}}{C^{d/2+1}}+(sT)\int_{0}^{r_h} d\tilde{r} \frac{BD}{C^{d+1}A_t'A_t}
\end{align}
In the limit $r_h\to\infty$, we show in Appendix \ref{app:importantintegral} that the first term gives
\begin{align}
\label{leadingintegral}
\int_{0}^{r_h}d\tilde{r}\frac{\sqrt{BD}}{C^{d/2+1}}= \frac{c_{ir}^2}{sT}+...
\end{align}
where the $...$ indicate subleading terms and where we have defined
\begin{align}
\label{cirdef}
c_{ir}^2 = \frac{L_t^2}{L_x^2}\left(\frac{r_h}{L}\right)^{2-2z}\sim T^{2-2/z}.
\end{align}
To leading order, the second piece gives
\begin{align}
\label{leadingordersecondpiece}
(sT)\int_{0}^{r_h} d\tilde{r}\frac{BD}{C^{d+1}A_t'A_t} \sim \frac{sT}{C_h\rho_{in}^2}+...
\end{align}
Returning to the first term in \eqref{rewrittenintegral}, the leading order temperature dependence is
\begin{align}
\int_{0}^{r_h} d\tilde{r} \left(\frac{D}{CR}\right)'\frac{1}{A_t} \sim -\frac{sT}{C_h\rho_{in}^2}+...
\end{align}
which cancels \eqref{leadingordersecondpiece}. Thus, the temperature dependence left after this cancelation is subleading to \eqref{leadingintegral} so that
\begin{align}
\lim_{r_h\gg L}\int_{0}^{r_h}d\tilde{r}\frac{2q^2\eta^2\sqrt{BD}C^{d/2-1}}{R^2} \approx \frac{c_{ir}^2}{sT}+...
\end{align}  
for $z<d+2$.
$\\$

From \eqref{cirdef}, $c_{ir}^2\sim T^{2-2/z}$. This is more relevant than $sT\sim T^{1+\frac{d}{z}}$ for this range of $z$ and hence must be considered before subleading $sT$ terms. Then, \eqref{ZfrommathcalA} leads to
\begin{align}
Z-\frac{\rho}{\mu} = -\frac{sT}{\mu^2}(1-c_{ir}^2)+ ...
\end{align}
Using (\ref{rhonZdefinition}), we find that the leading temperature dependence is
\begin{align}
\rho_n = \frac{sT}{\mu}\frac{1-c_{ir}^2}{c_{ir}^2} +...
\end{align}
where $c_{ir}$ can be temperature dependent as indicated in (\ref{cirdef}).
$\\$

This result is exactly the same as found using the EFT for relativistic superfluids at low temperatures \cite{Delacretaz:2019brr,Delacretaz:2020nit}. Here, we see that it also holds when $1<z<d+2$, away from the relativistic limit $z=1$.
$\\$

If we attempt to use this expression for $z>d+2$, something bizarre seems to occur. Including the explicit temperature dependence of $c_{ir}$, $\rho_n \sim T^{\frac{d+2}{z}-1}$ naively diverges for $z>d+2$. However, for this range of exponents $c_{ir}^2$ is now less relevant than $sT$. Instead of $\rho_n$ vanishing as $T\to 0$, we find a finite limit, as we shall now explain.

\subsubsection{Non-vanishing $\rho_n^{(0)}$}
For $z>d+2$, all the integrals in \eqref{Aexpansion} converge so that the expansion \eqref{temperatureexpansion} is well-defined.
Then, combining with \eqref{ZfrommathcalA}, we find
\begin{align}
Z-\frac{\rho}{\mu} = -\frac{sT}{\mu^2} + \frac{(sT)^2}{\mu^3\rho_n^{(0)}} +...
\end{align}
where
\begin{align}
\label{nonvanishingrhon0}
\rho_{n}^{(0)} = \rho^{(0)}\times\left[1+\mu\rho^{(0)}\int^{\infty}_0 \frac{2q^2\eta^2\sqrt{B_0D_0}C_0^{d/2-1}}{R_0^2}dr' \right]^{-1}.
\end{align}
The functions appearing in the integral are the zero-temperature functions of (\ref{IRmetric}).  
$\\$

From \eqref{nonvanishingrhon0}, we conclude that for $z>d+2$, $\rho_n$ does not vanish at $T=0$, in stark constrast to $z<d+2$.
The subleading temperature dependence to the leading non-vanishing constant arises from the same integral with a finite upper bound of $r_h$ or from explicit $sT$ dependence. We write this schematically as
\begin{align}
\rho_{n} \sim \rho_{n}^{(0)} + \# T^{\frac{z-d-2}{z}}+ \# T^{\frac{d+z}{z}} +...
\end{align}
For the examples in the main text, the first subleading temperature is always dominant. On the other hand, in Appendix \ref{convergenceappendix} we give an example where this integral has a temperature scaling subleading to the $sT$.

\subsubsection{Competing broken symmetries}
The criteria $z>d+2$ for $\rho_n^{(0)}\neq 0$ can be considered to arise from a competition between $sT$ and $c_{ir}^2$. These quantities naturally arise in other transport observables, specifically the low-energy sound and diffusion modes, which we discuss in \cite{Gouteraux:2019kuy}. We have seen that the convergence of $\lim_{T\to 0} sT/c_{ir}^2$ gives $\rho_n^{(0)}=0$ and its divergence signals $\rho_n^{(0)}\neq 0$. Unfortunately, this is not a sufficient condition. Instead, the necessary and sufficient condition for the existence of a non-zero $\rho^{(0)}_n$ is the convergence of the integral
\begin{align}
\label{convergenceintegral}
\lim_{r_h\to\infty}\int^{r_h}_0  \frac{2q^2\eta_0^2\sqrt{B_0D_0}C_0^{d/2-1}}{R_0^2}dr'\,,
\end{align}
which appeared in our low temperature expansion, see \eqref{Aexpansion} and \eqref{convergenceintegral2}.
When this diverges, $\rho_n^{(0)}=0$, and when it converges, $\rho_n^{(0)}\neq 0$. 
$\\$

We can unpack the integral a little by looking at the fluctuation equation
\begin{align}
\frac{1}{C^{d/2-1}}\frac{d}{dr}\left[C^{d/2-1} \sqrt{\frac{D}{B}}a_{\hat{x}}'\right]
- \sqrt{\frac{B}{D}}\left(2q^2D\eta^2 + \frac{(A_t')^2}{B}\right)a_{\hat{x}} &=0.
\end{align}
Here, we see that there are two mass-like terms. The first arises from the breaking of the $U(1)$ symmetry and a non-trivial condensate $\eta$. The second arises from the non-zero density $\rho$. The ratio of the two terms is
\begin{align}
\frac{2q^2 BD \eta^2}{(A_t')^2} = \frac{C^{1+d/2}}{\sqrt{BD}}\times\left[\frac{2q^2\eta^2\sqrt{BD}C^{d/2-1}}{R^2}\right]
\end{align}
In terms of the zero temperature fields, we write
\begin{align}
\label{alphadef}
\frac{2q^2B_0D_0\eta_0^2}{(A_{t,0}')^2}|_{r=r_h} \sim \left(\frac{\hat{r}_h}{L}\right)^\alpha
\end{align}
we must have
\begin{align}
\alpha - z + d+ 2 < 0
\end{align}
for the integral to converge and $\rho_n^{(0)}\neq 0$. If this criteria is not met, then $\rho_n^{(0)} = 0$.  We note that $\alpha = d+1-2\tilde{\Delta}_{A_0}$ for conformal phases and $\alpha = 0$ for Lifshitz phases so that this reproduces our earlier results. In Appendix \ref{convergenceappendix}, we will show that modified superfluid actions can give non-trivial $\alpha$ which in turn give rise to $\rho_n^{(0)} \neq 0$ for spacetimes with any $z$ by guaranteeing the convergence of (\ref{convergenceintegral}) and vice-versa.

\section{Numerical examples \label{section:numerics}}
\begin{figure}[t]
\begin{center}
\includegraphics[scale=.19]{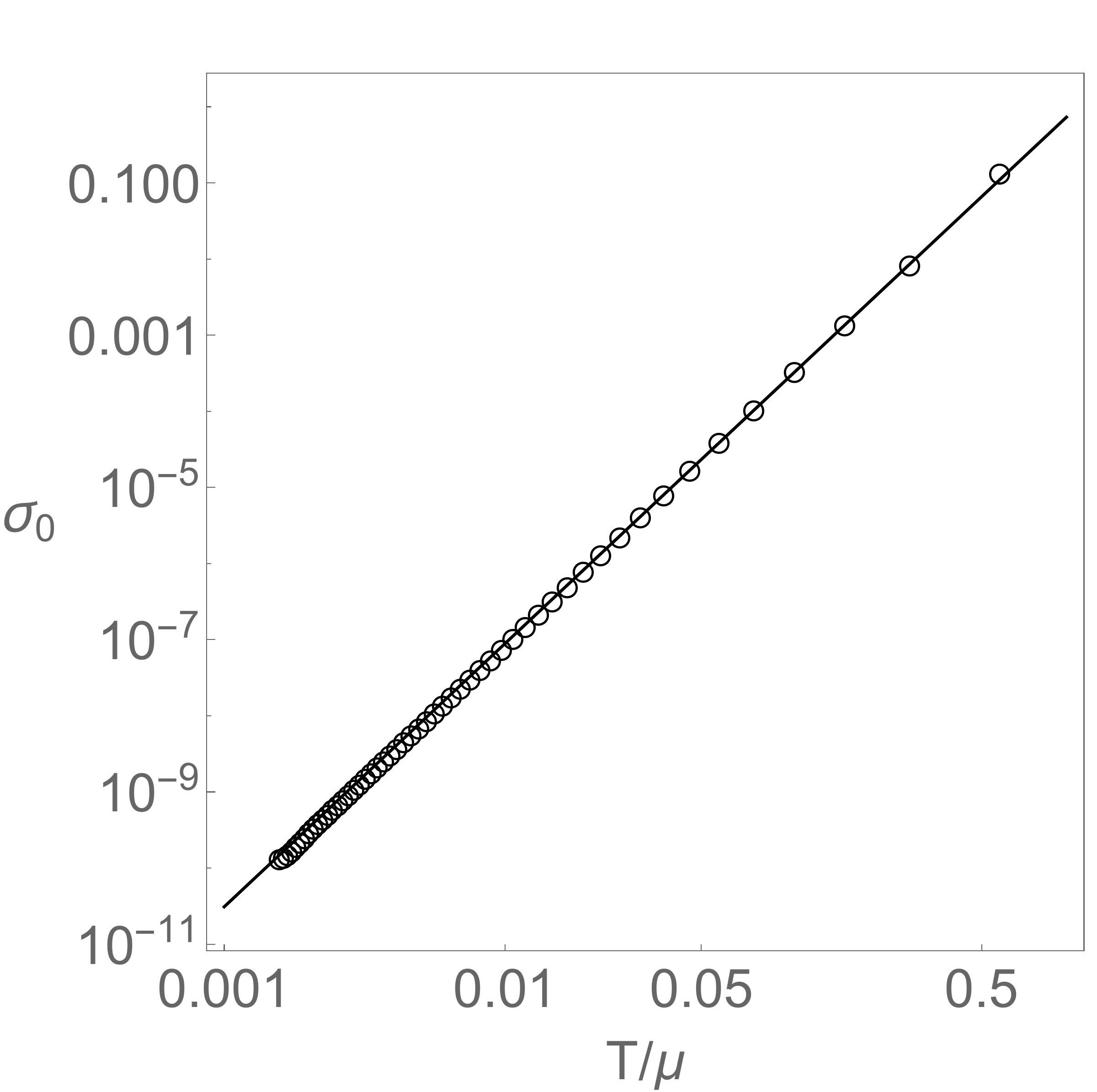}
\includegraphics[scale=.23]{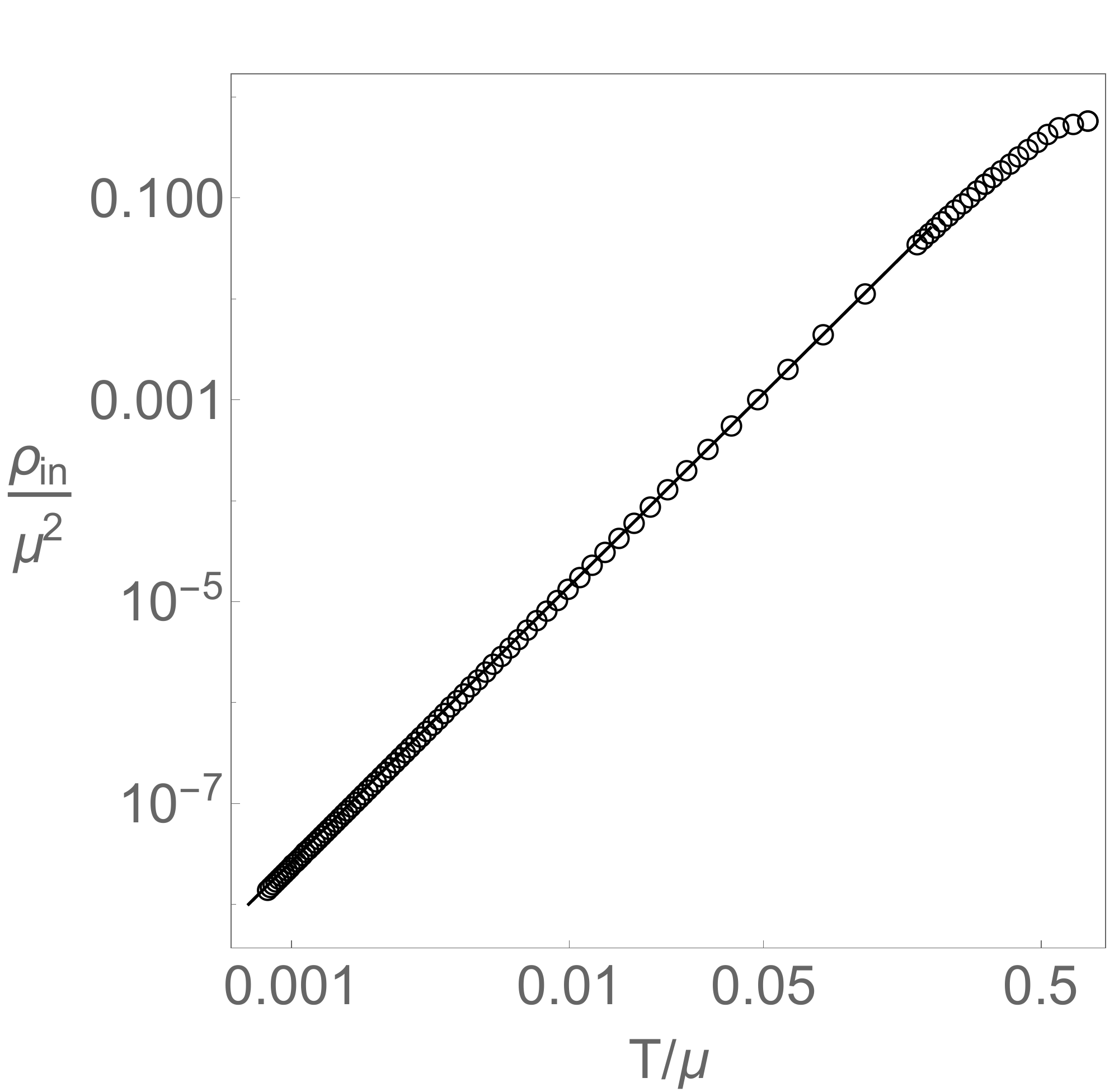}
\includegraphics[scale=.23]{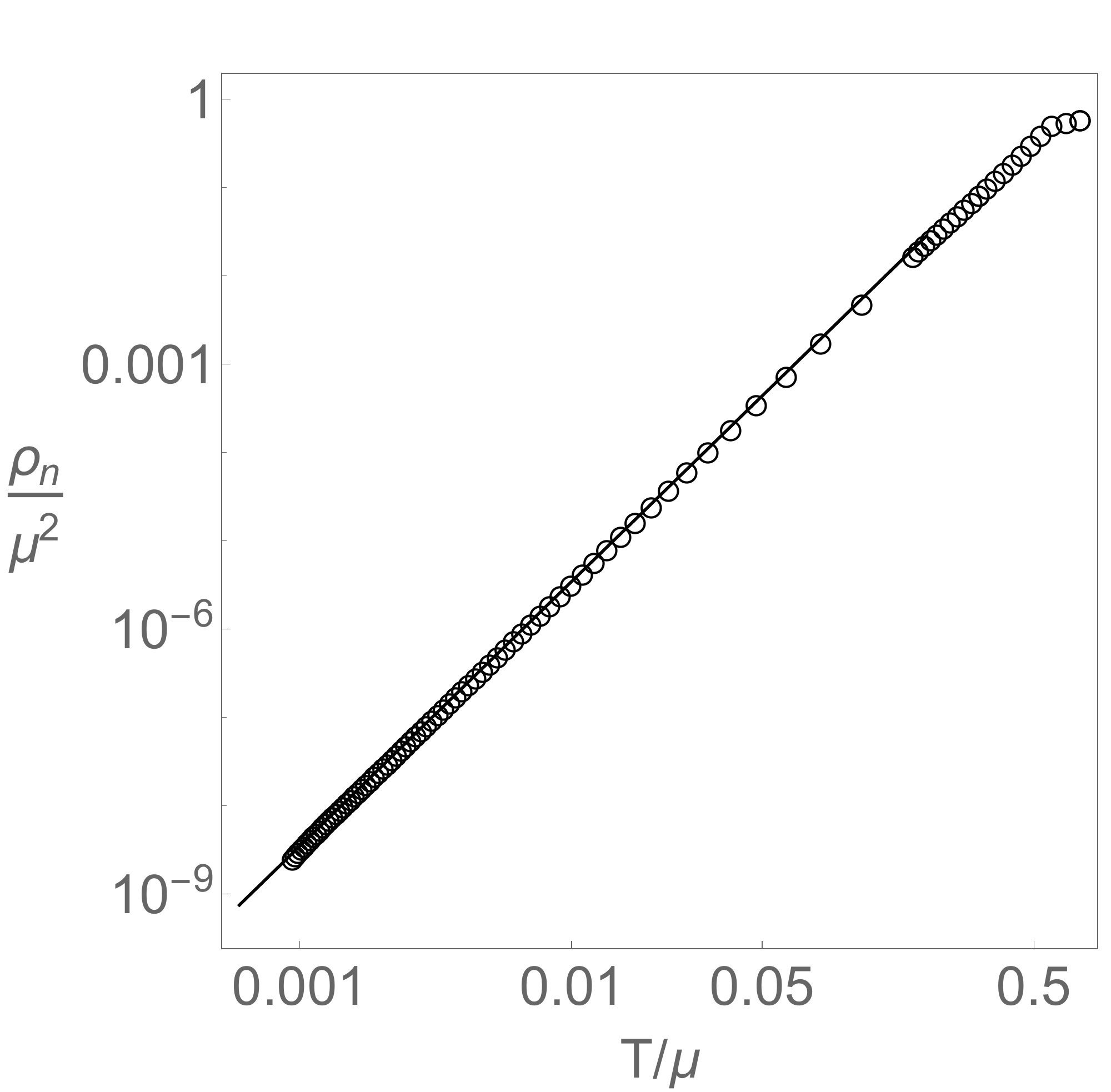}
\caption{$z=1$: $\sigma_0 \sim \# T^{3.46}$, $\rho_{in} \sim \# T^{2.73}$, $\rho_n \sim \# T^{3}$ \label{z1figure}}
\end{center}
\end{figure}

\begin{figure}[h]
\begin{center}
\includegraphics[scale=.19]{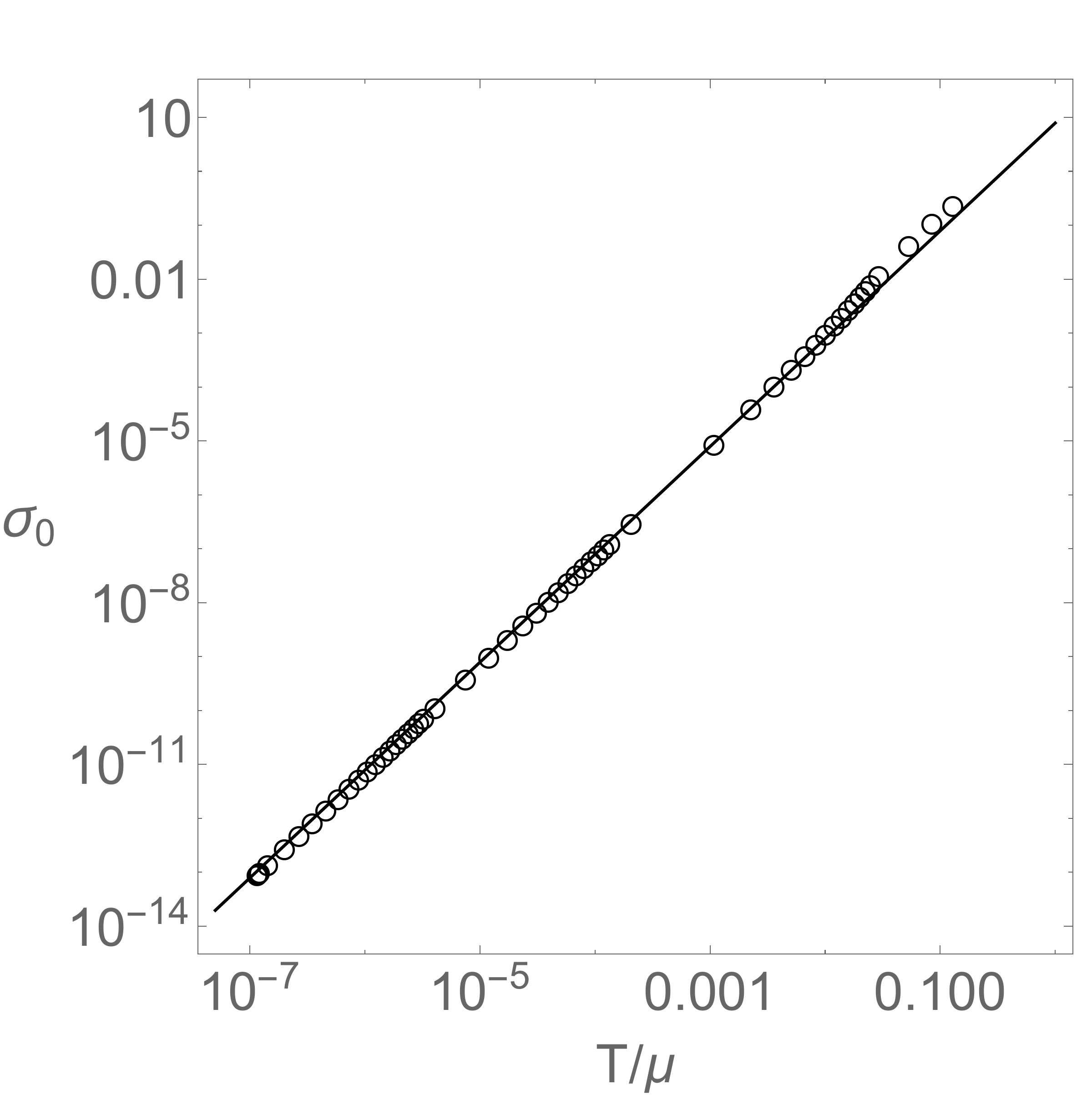}
\includegraphics[scale=.23]{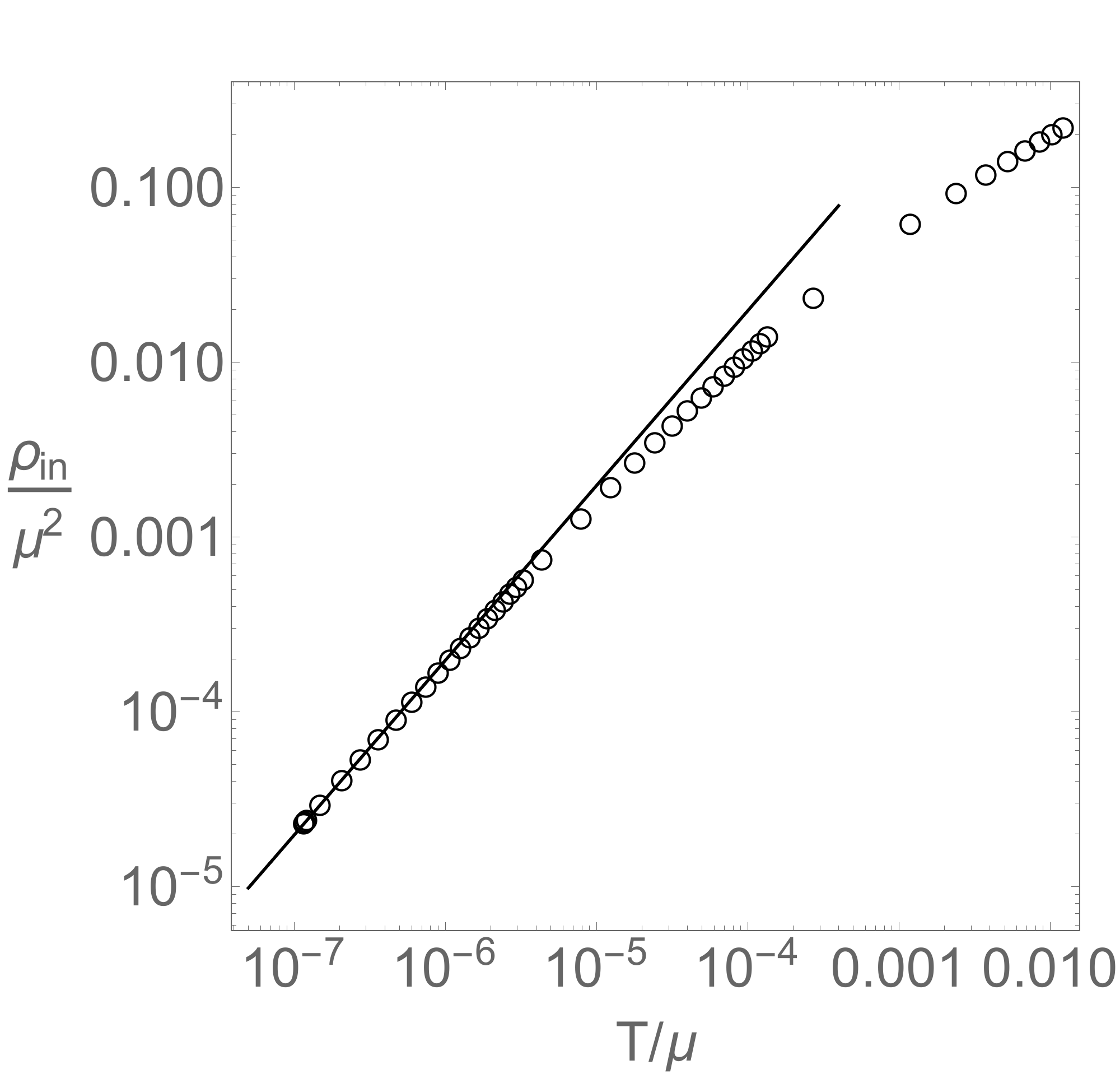}
\includegraphics[scale=.23]{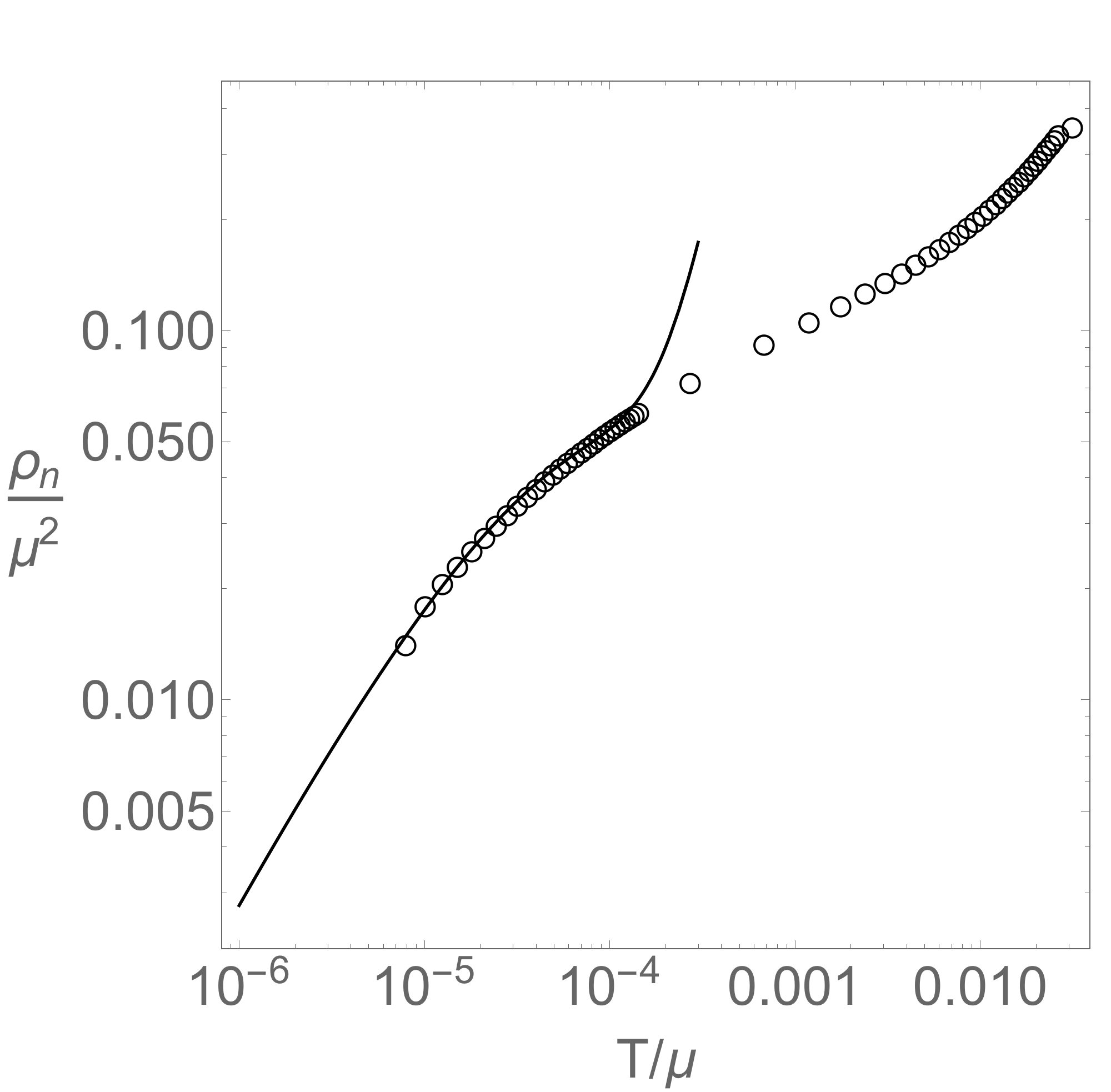}
\caption{$z=2$: $\sigma_0 \sim \# T^{2}$, $\rho_{in}\sim \# T $, $\rho_n \sim b_1 T(1+b_2 T^{.36} + b_3 T^{.62})$ \label{z2figure}}
\end{center}
\end{figure}

\begin{figure}[h]
\begin{center}
\includegraphics[scale=.21]{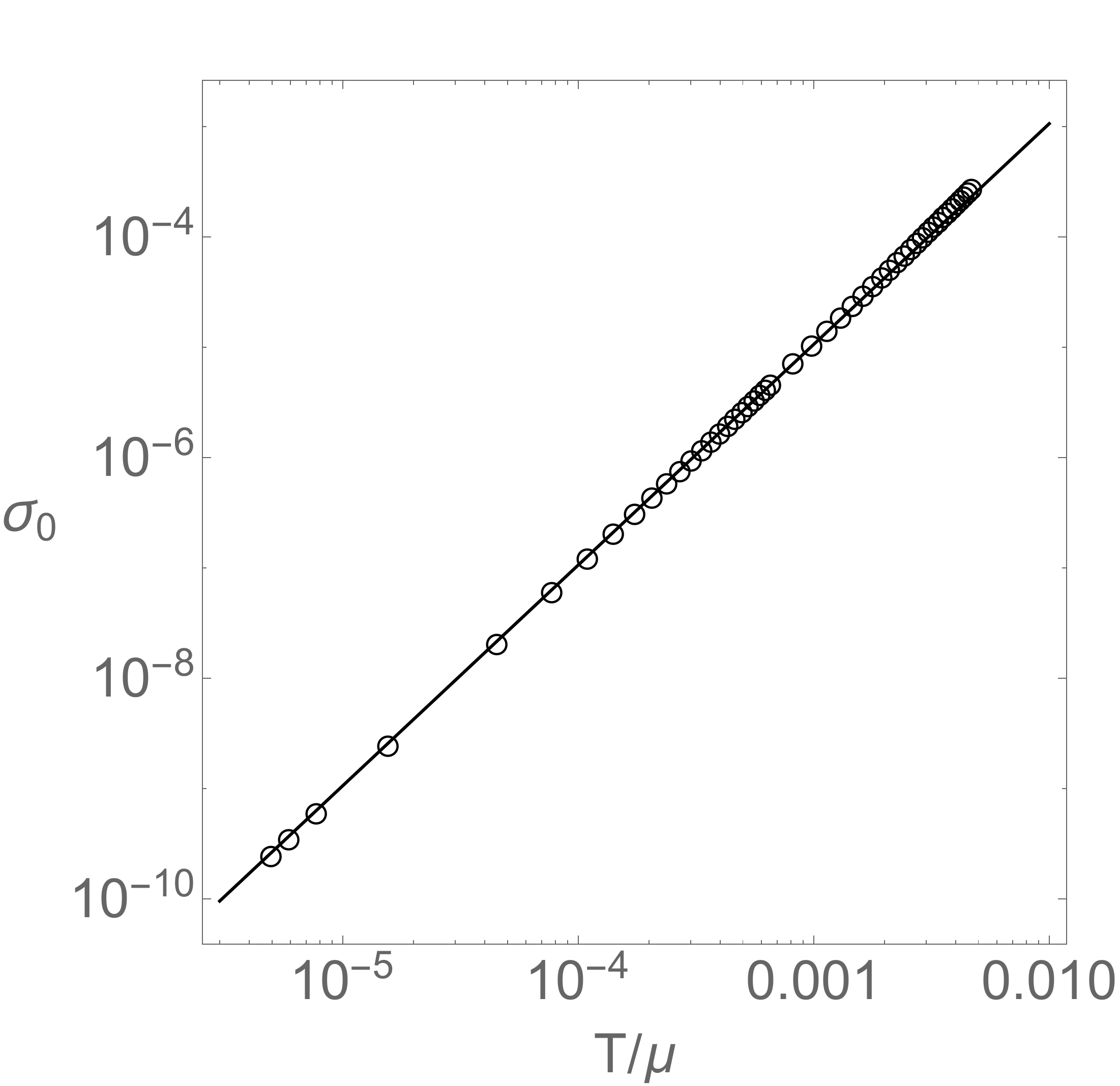}
\includegraphics[scale=.23]{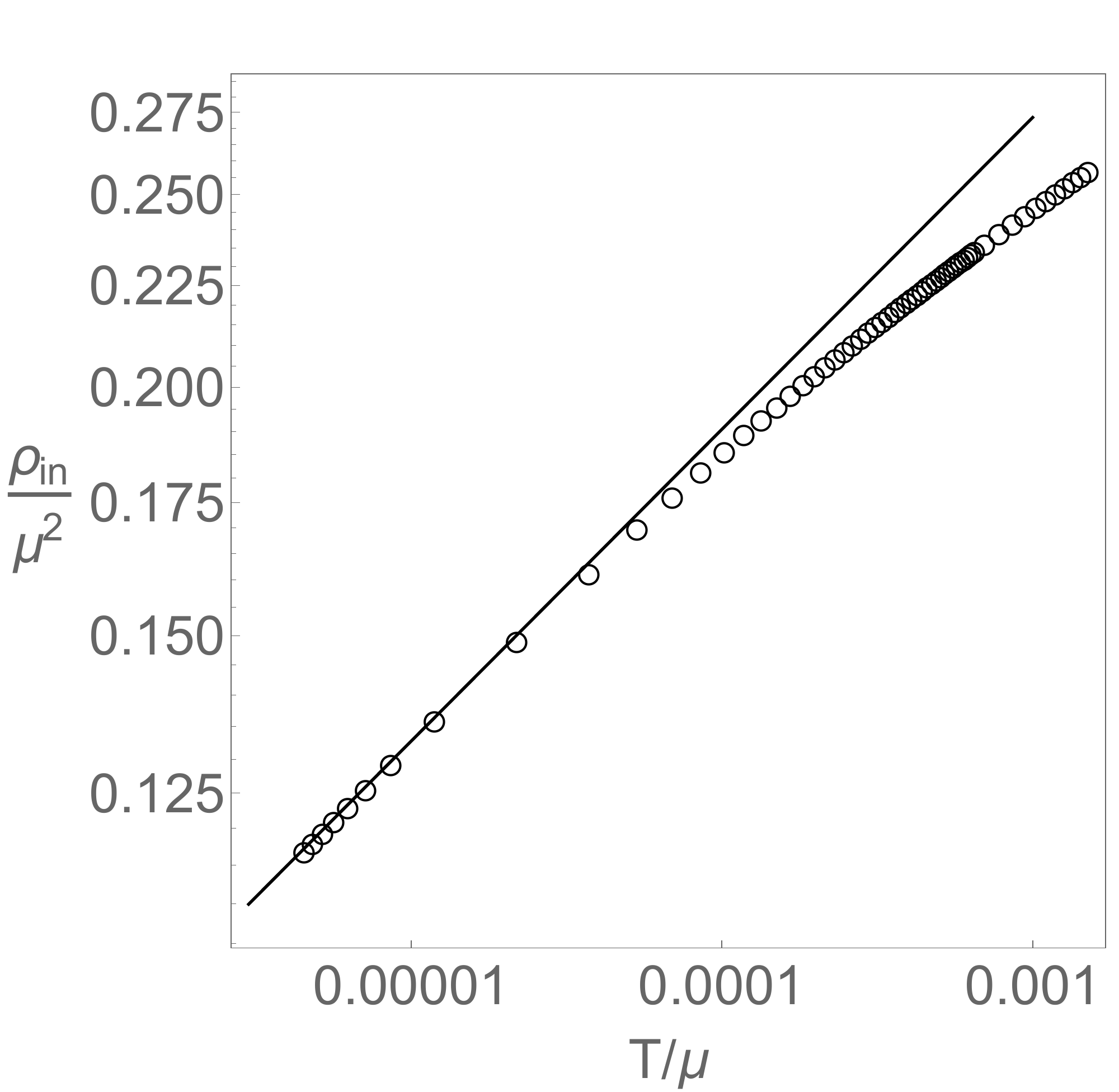}
\includegraphics[scale=.24]{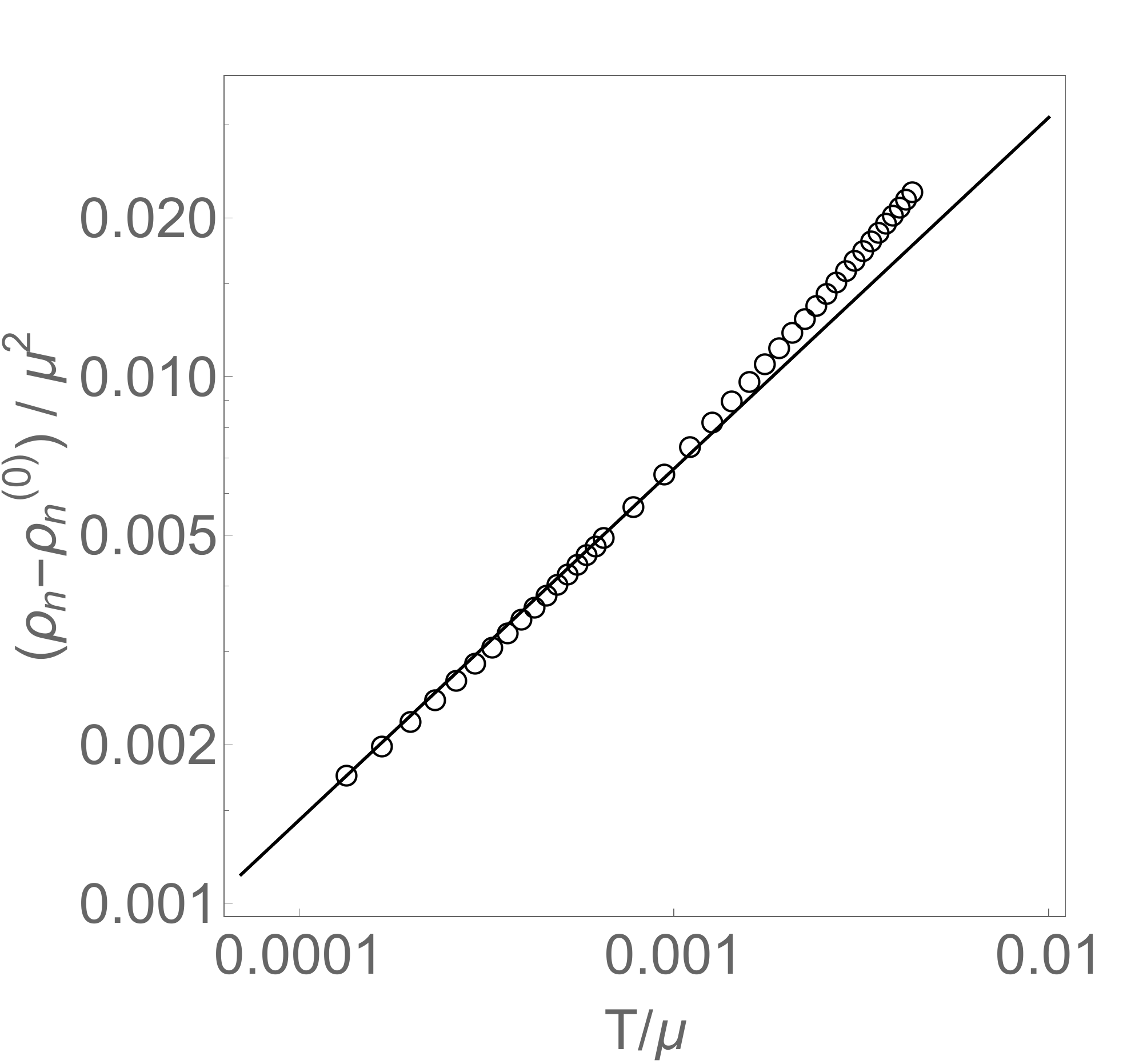}
\newline
\includegraphics[scale=.2]{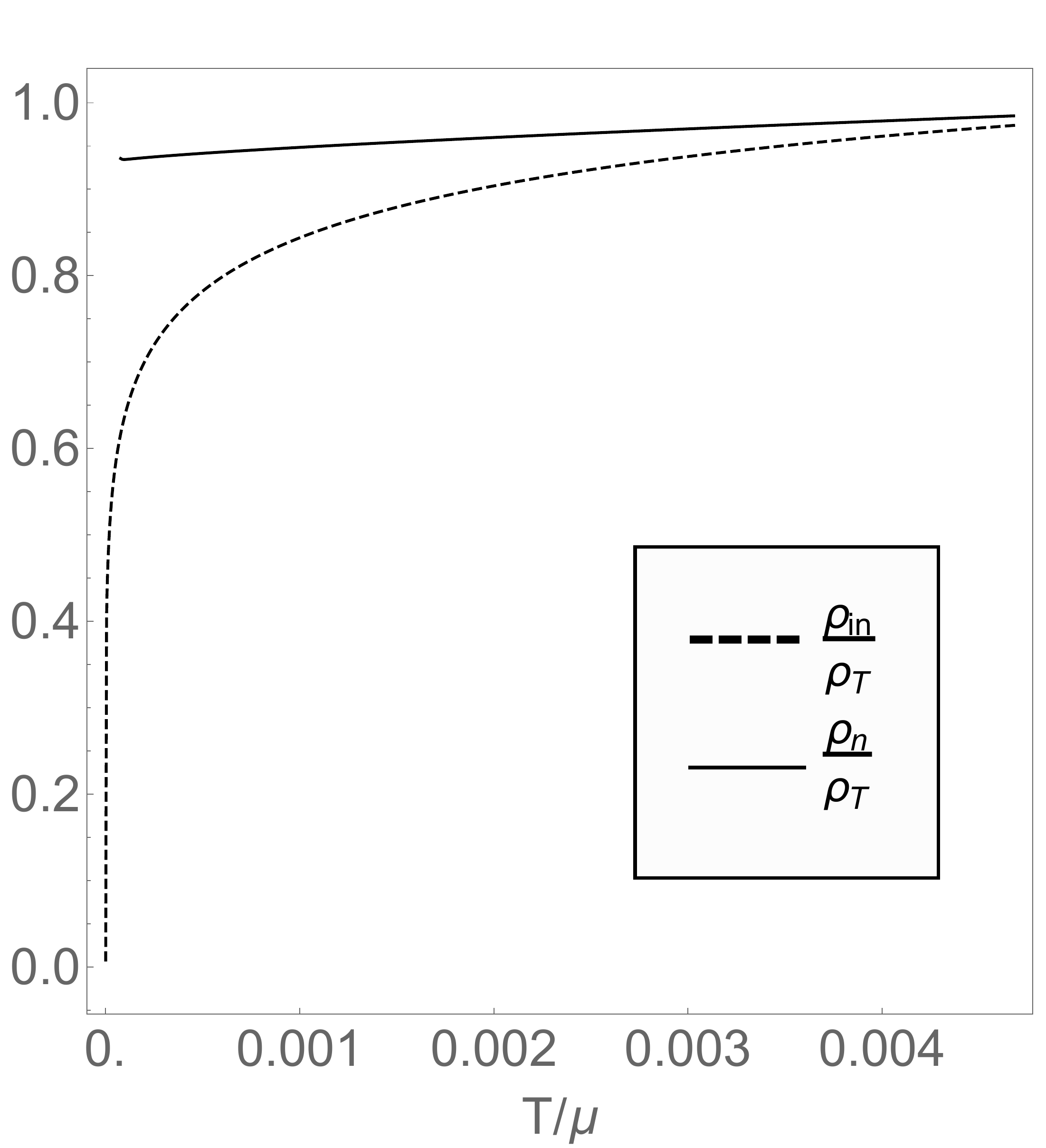}
\caption{$z=12$: $\sigma_0\sim \#T^2$, $\rho_{in}\sim \#T^{1/6}$, $\rho_n \sim \rho_n^{(0)} + \# T^{2/3}$  \label{z12figure}}
\end{center}
\end{figure}

\begin{figure}[h]
\begin{center}
\includegraphics[scale=.2]{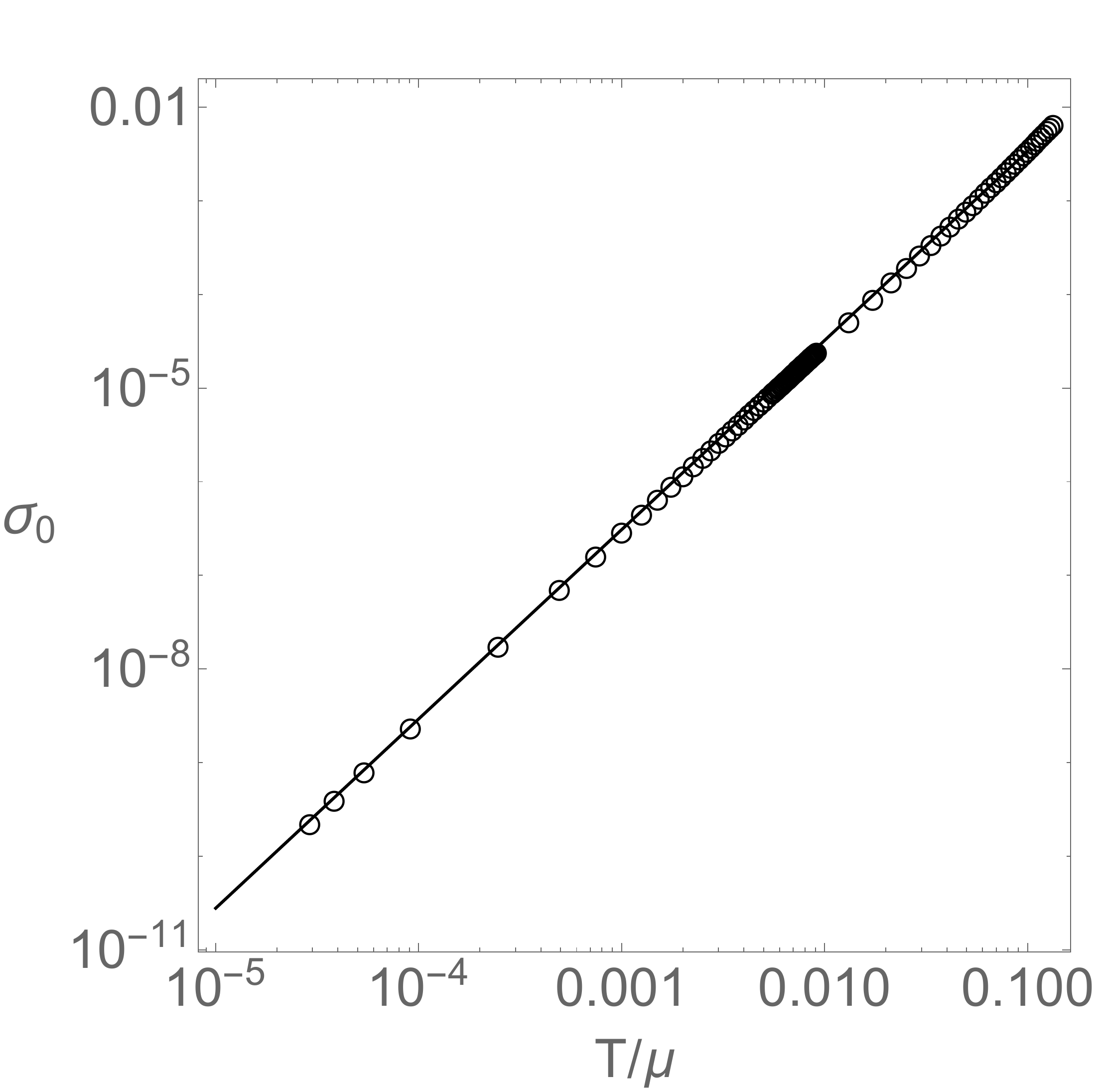}
\includegraphics[scale=.24]{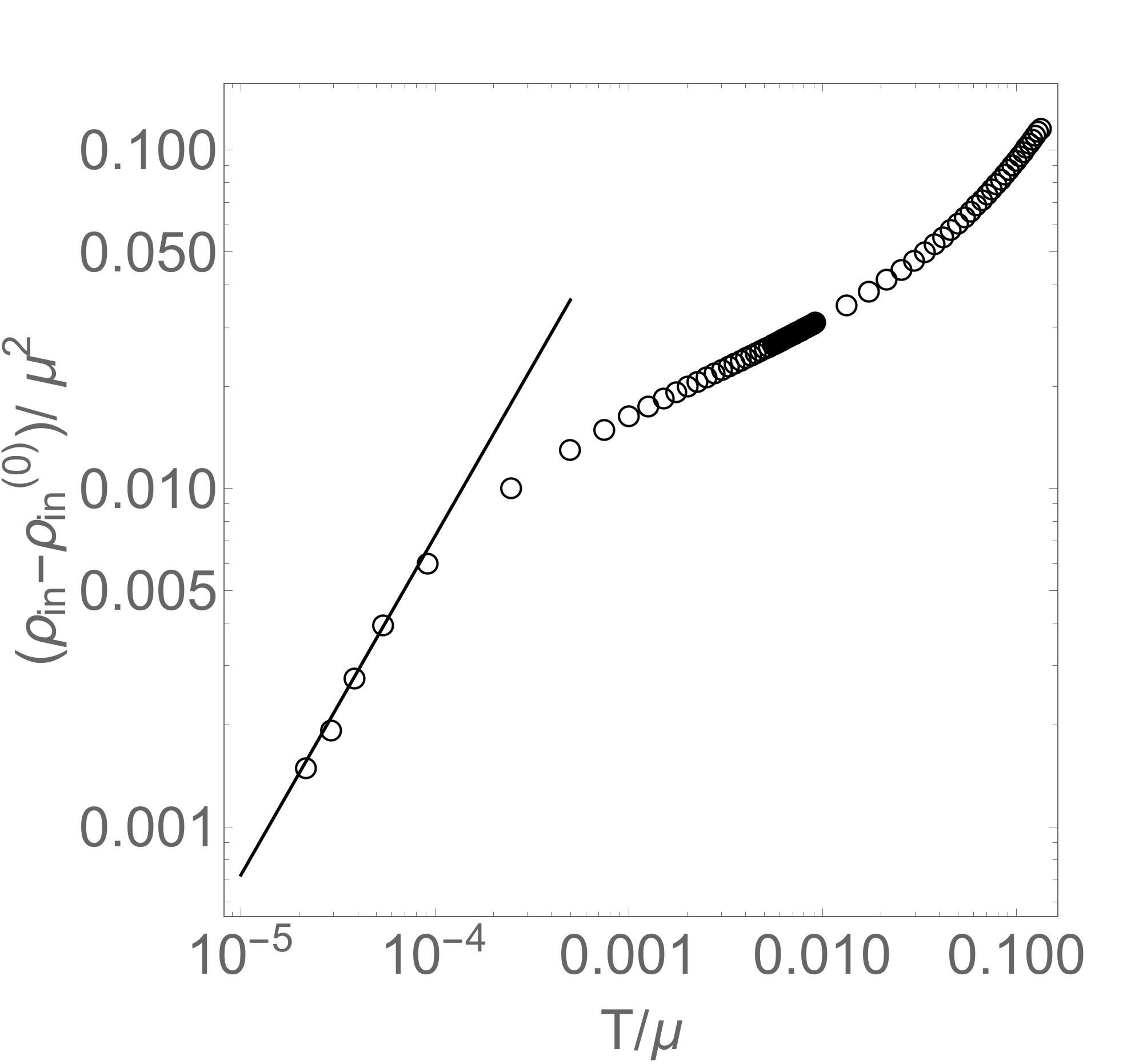}
\includegraphics[scale=.26]{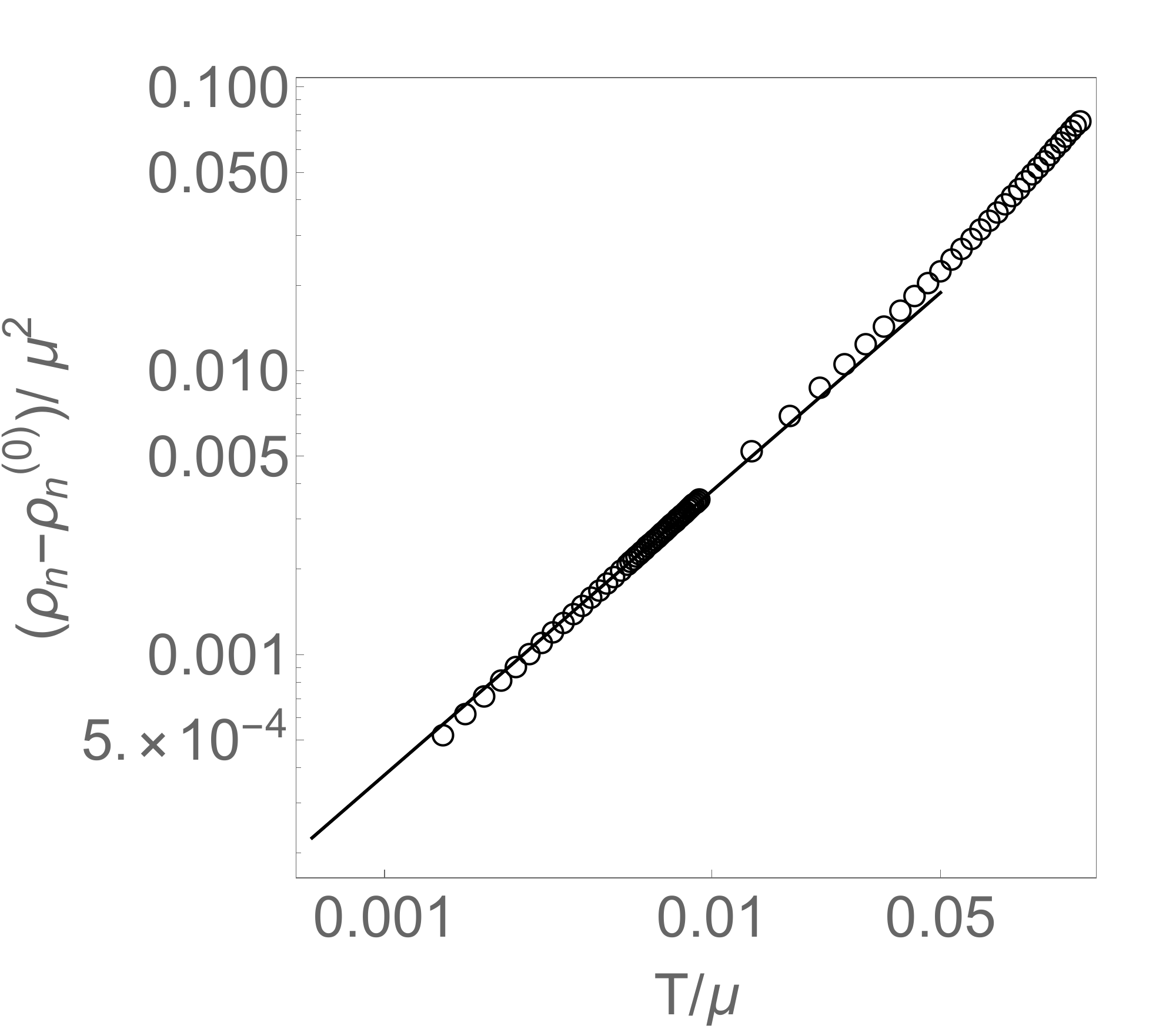}
\newline
\includegraphics[scale=.2]{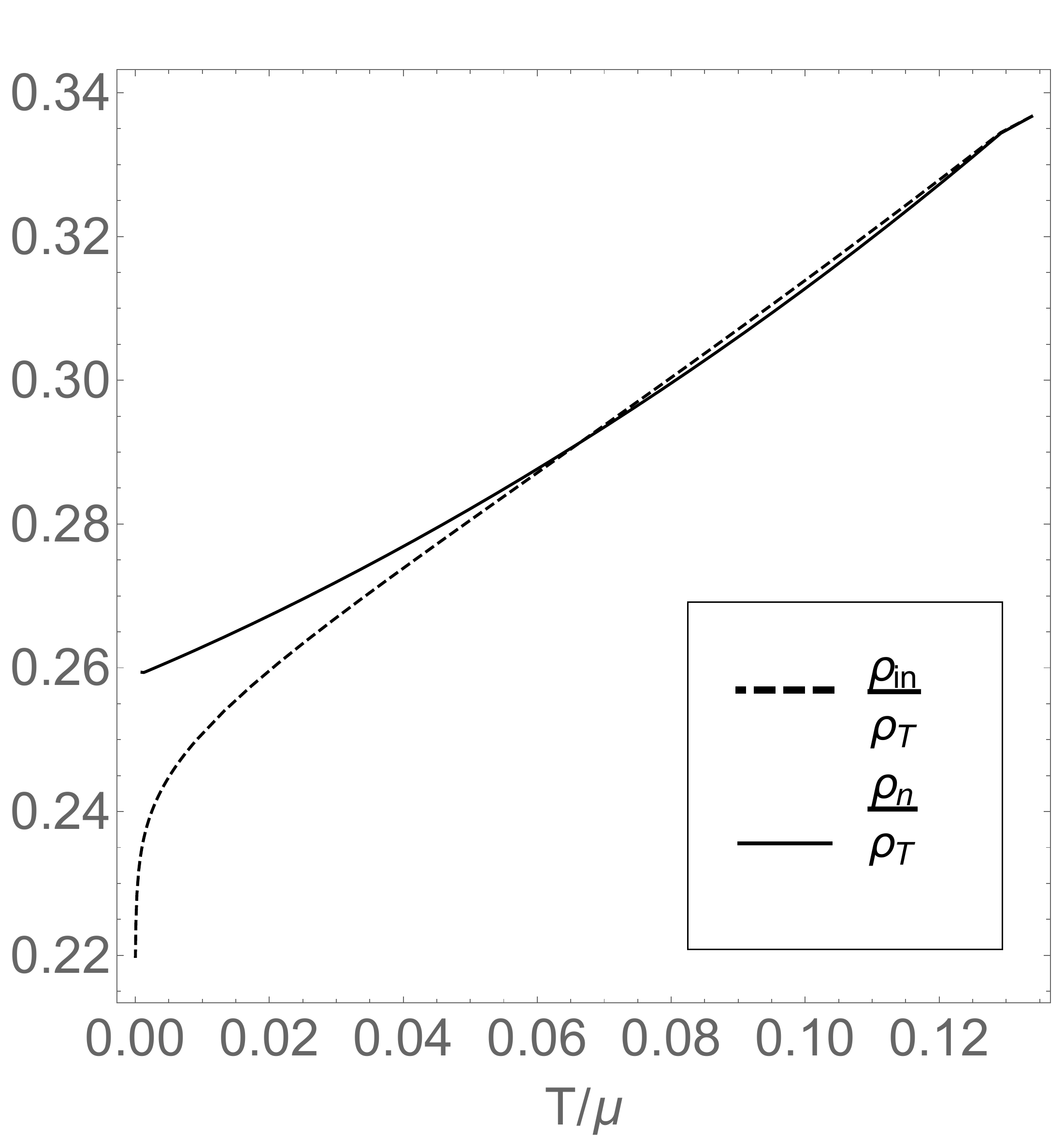}
\caption{$z\to\infty$: $\sigma_0 \sim \# T^2$, $\rho_{in}\sim \rho_{in}^{(0)} + \#T$, $\rho_{n}\sim \rho_{n}^{(0)} + \# T$ \label{semilocalfigure}}
\end{center}
\end{figure}

To demonstrate our results, we choose a specific scalar potential following \cite{Gubser:2009cg} in $d=2$ and set $L=1$,
\begin{align}
V(|\eta|) = -6 -2\eta^*\eta + g_\eta^2(\eta^*\eta)^2
\end{align}
where $g_\eta = 3/2$. We must solve for $q, \eta_0, \tilde{L}$ for a given value of $z$. For $z=1$, we chose $q=2$. We show results for $z=1$ in figure \ref{z1figure}, $z=2$ in figure \ref{z2figure}, $z=12$ in figure \ref{z12figure}, and $z\to \infty$ in figure \ref{semilocalfigure}. In these plots, numerical results are plotted in open circles wheras solid lines denote fits to the appropriate temperature scaling.

$\\$
 To accurately compute $\rho_n$ for $z>1$, we needed to use very high precision numerics. We used a Newton-Raphson method with double floating point precision and up to $N=1000$ points. Even with this level of precision, it was difficult to observe strong $\rho_n \sim T$ behavior for $z=2$ since the a large region of the IR is well-approximated by (\ref{IRmetric}) only below $(T/\mu)\sim 10^{-6}$. In figure \ref{z2figure}, we fit to $\rho_n \sim b_1 T \left(1+ b_2 T^{\nu_\eta^{+}/2}+b_3 T^{\nu_\eta^{-}/2}\right)$ for constants $b_1$, $b_2$, $b_3$ and $\nu_\eta^{\pm} = \nu_\eta(-1,\pm 1)$. In order to clearly observe a single temperature scaling, we can also break translation invariance by adding an extra term to the action as was done in \cite{Gouteraux:2019kuy}. Then extracting $\rho_n$ requires lower precision and we can reach lower temperatures, allowing us to further confirm that to leading order $\rho_n \sim\# T$. We discuss this in detail in Appendix \ref{translationbreaking}. 
 $\\$
 
As we mentioned earlier, the $z\to\infty$ case requires a slightly modified action, the details of which we relegate to Appendix \ref{semilocal}. In the notation of this Appendix, we choose $a=2$ for the plot in figure \ref{semilocalfigure} that shows an example with $z\to+\infty$. For this choice of the parameter $a$, we find an irrelevant deformation with $\tilde{\nu}_\eta = -1$ and that other choices of $a$ lead to $\tilde{\nu}_\eta \neq -1$. Nevertheless, for any choice of $a$, we find that $\rho_n-\rho_n^{(0)}$ is always linear for phases with $z\to\infty$.
 
 $\\$
 Other cases will be discussed further in Appendix \ref{semilocal}.

\section{Connection to previous literature \label{section:connections}}
In this work, we discuss transport in clean, quantum critical, holographic superfluids. Our main object of interest is the normal charge density which controls dissipative transport. In these systems, we establish the criteria for $\lim_{T\to0}\rho_n \neq 0$. Our conclusion is that this is possible for quantum critical systems described by Lifshitz symmetries with dynamical critical exponent $z>d+2$. In Appendix \ref{scalecovariant}, we generalize this to systems that exhibit hyperscaling and find new criteria which depend on the spectrum of irrelevant deformations.
$\\$

With this result, we are able to explain some previously mysterious results in the holographic literature. The first result is from \cite{Herzog:2009md} who looked at various sound modes in a system with action \ref{bulkaction} in $d=3$. The superfluid second sound is given by
 \begin{align}
 c_2^2= \left(\frac{s}{\rho}\right)^2\frac{\rho_s}{(sT+\mu\rho_n)(\partial[s/\rho]/\partial T)_\mu} \simeq \begin{cases}\frac{z}{3}c_{ir}^2 & z<d+2,\\
 \frac{z}{3}\frac{\rho_s^{(0)}}{\rho^{(0)}}\frac{sT}{\mu\rho_n^{(0)}} & z>d+2.
 \end{cases}
 \end{align}
 while the fourth sound is given by
 \begin{align}
 c_4^2 = \frac{\rho_s}{\mu\left(\frac{\partial\rho}{\partial\mu}\right)_s} \simeq \frac{1}{3}\left[1-\frac{\rho_n}{\rho}\right].
 \end{align}
In both equations above, we obtain the last relation by substituting the scalings following from the critical scaling of the holographic groundstate, allowing for $z\geq1$. In figure 2 of their paper, for large enough $q$, $c_2^2 \to \frac{1}{3}$.  On the other hand, for sufficiently small $q$, $c_2\to 0$. As we have seen, for large $q$, $\tilde{\Delta}_{A_0}$ is sufficiently negative to allow for an emergent $z=1$ IR geometry. Hence, $\lim_{T\to 0}c_{ir}$ is a constant in this limit. On the other hand, for small $q$, the system flows to a geometry with $z>1$ which has $\lim_{T\to 0}c_{ir}= 0$. In figure 3, the behavior of $c_4$ mirrors this analysis. For sufficiently small $q$, $c_4^2 \neq \frac{1}{3}$ indicating $\rho_n^{(0)}\neq 0$. It is interesting to look at the upper right panel of figure 2 and the right panel of figure 3 for $q=1$ and $q=2$. The first of these has $z = 18.3$ while the second has $z= 1.22$. Both are Lifshitz phases so that $\lim_{T\to 0}c_2 = 0$. However, for $q=1$, $z>5$ so we expect $\rho_n^{(0)}\neq 0$ and $c_4^2\neq \frac{1}{3}$. This is exactly what we see.
$\\$

In \cite{Sonner:2010yx}, the authors considered the holographic superconductor of Hartnoll, Herzog, and Horowitz \cite{Hartnoll:2008kx,Hartnoll:2008vx}. Importantly, the IR of this system is not quantum critical. Nevertheless, our analysis can explain their results as well. For $q^2>|m^2|/6$, the authors of \cite{Horowitz:2009ij} found that in the IR,
\begin{align}
\label{IRmetricHHH}
ds^2 = -L_t^2\left(\frac{L}{\hat{r}}\right)^2dt^2 + \tilde{L}^2\frac{d\hat{r}^2}{\hat{r}^2 (\ln [\hat{r}/L])} + L_x^2\left(\frac{L}{\hat{r}}\right)^2d\vec{x}^2
\end{align}
and
\begin{align}
\label{IRetaHHH}
\eta = \eta_0\left[\ln \left(\frac{\hat{r}}{L}\right)\right]^{1/2},\quad A_t = A_t^{(0)} \left(\frac{\hat{r}}{L}\right)^{-\beta}\left[\ln\left(\frac{\hat{r}}{L}\right)\right]^{1/2},\quad \beta = \frac{1}{2}+\frac{1}{2}\sqrt{1+\frac{48q^2}{|m^2|}}.
\end{align}
 From this scaling, we find
\begin{align}
\frac{2q^2BD\eta^2}{(A_t')^2} \sim \# \left(\frac{\hat{r}}{L}\right)^{\alpha}\times \left[\ln \left(\frac{\hat{r}}{L}\right)\right]^{-1}, \quad \alpha = -3+\sqrt{1+\frac{48q^2}{|m^2|}}>0.
\end{align}
and 
\begin{align}
\label{HHHIR}
\lim_{r\to\infty}\frac{\sqrt{BD}}{C^2} \sim \# \left(\frac{\hat{r}}{L}\right)^{2}\times \left[\ln\left(\frac{\hat{r}}{L}\right)\right]^{-1/2}.
\end{align}
Hence, the integral (\ref{convergenceintegral}) will diverge and $\rho_{n}^{(0)} = 0$. This solution has a finite $c_{ir}$ and we expect $\rho_n = \frac{1-c{ir}^2}{c_{ir}^2}sT$. We checked numerically, using (\ref{rhonZdefinition}), that this is true for $q=1$, extending the results of \cite{Sonner:2010yx}. 
$\\$

{On the other hand, \cite{Sonner:2010yx} also considered backgrounds with a finite superfluid current. They observed that $\rho_n^{(0)}=0$  for any value of the superfluid current (below the Landau critical velocity at which superfluid disappears), and $q>1$ in $d=2$. However, for $q=1$, the authors found that $\rho_n^{(0)}$ does not vanish, at least for the superfluid velocities they considered. We have checked that in this case, the ground state is not well-described by (\ref{IRmetricHHH}), but we leave the full analysis for this case to later work. In \cite{Arean:2010wu}, superfluid flows in a top-down Type IIb embedding were considered and similar results obtained: above a certain value of the superfluid velocity, the infrared geometry ceases to be another copy of Anti de Sitter spacetime. It would be interesting to further study these systems and determine the relation, if any, to the formation of Bogoliubov Fermi surfaces in weakly-coupled superfluids at large superfluid velocities, \cite{Autti_2020}.}
\newpage

\section{Discussion \label{section:discussion}}

From a physical perspective, our criteria points to an interesting competition between two competing phenomena at low energies--the spontaneous breaking of the $U(1)$ associated with superfluidity and the breaking of charge-conjugation symmetry that gives rise to a finite charge density. Both effects contribute to a zero frequency pole in the imaginary part of the optical conductivity. When the effects of spontaneous symmetry breaking are weaker than the charge-conjugation symmetry breaking, $\rho_n^{(0)}\neq 0$. Otherwise, $\rho_n^{(0)}$ vanishes. For holographic superfluids, the degree to which spontaneous symmetry breaking must be weaker is established by the convergence of an integral in the deep IR of the spacetime. Nevertheless, we can understand a strong charge-conjugation symmetry breaking as indicating strong charge renormalization effects. Our work suggest that it may not be suprising to see $\rho_n^{(0)} = 0$ in systems like $^4$He and in BCS superconductors which are considered weakly interacting. On the other hand, in systems which exhibit strong electron interactions, our work suggests that it is possible for $\rho_n^{(0)}\neq 0$. The relevance of this property to recent experiments on cuprate high T$_c$ superconductors was discussed in \cite{Gouteraux:2019kuy}.
$\\$

 The normal density is the relevant quantity to discuss dissipative effects at low energies and low temperatures. In this work, we demonstrate that in this limit, its behavior depends only on properties of the underlying quantum critical groundstate and the spectrum of irrelevant deformations. Much of the literature to date has worked explicitly with $z=1$ phases and assumed that the low temperature spectrum is dominated by linearly dispersing superfluid phonons \cite{Carter:1995if, Leggett, Leggett2, Schmitt:2014eka}. With this starting point, the result $\rho_n \simeq sT/c_{ir}^2$ is easily obtained. On the other hand, the results of this work suggest that, at least for Lifshitz theories with $z>d+2$, we must include other low energy contributions. Fortunately, given the universal dependence on the underlying quantum critical phase, it seems possible that quantum critical superfluids are amenable to an effective field theory treatment in the vein of \cite{Son:2002zn,Delacretaz:2019brr}. In particular, for Lifshitz phases, we expect at sufficiently low temperatures the linear dispersion is modified to $\omega\sim k^z$. Hence, higher order derivative contributions must be included in a quantum effective action treatment analogous to \cite{Son:2002zn} as well as a modified equation of state. We are currently at work on including these terms.
$\\$
 
An interesting question to ponder is whether our results could be an artifact of the large $N$ limit which underlies the gauge/gravity duality, especially when taking the zero temperature limit. Unfortunately, a precise answer would require a detailed account of finite $N$ corrections, which is difficult. However, it is straightforward to observe that the normal density in \eqref{rhonlowTeft}, which we have shown holds also for Lifshitz-invariant fixed points, diverges when $z>d+2$. Clearly, we expect that the assumptions underlying the superfluid EFT, i.e. that the zero temperature groundstate can be treated as a gas of superfluid phonons, must break down. A possible resolution, as we argue here, is that new degrees of freedom must be taken into account, leading to a non-vanishing normal density.
$\\$

The careful reader might be concerned that, even in the strict large $N$ limit, since we are extracting the normal and superfluid densities by taking a very low temperature limit instead of working at exactly $T=0$, some dramatic drop of the normal density could occur such that $\rho_n^{(0)}$ would be exactly zero at $T=0$ for all cases. Investigating $T=0$ geometries presents challenges, as it is in general difficult to numerically ascertain that the solution is at zero temperature, rather than simply being probed in the region of spacetime where temperature effects are negligible. Spacetimes where an AdS$_2\times$R$^2$ factor arises in the IR behave more nicely, as the timelike Killing vector has double zero and the extremal horizon remains finite. This double zero can be analytically implemented using pseudospectral methods. We construct such flows in appendix \ref{app:T=0} and demonstrate that the zero temperature values for the normal and superfluid densities agree with the values inferred by taking the low temperature limit.
$\\$

Another interesting aspect of holographic superfluids is that, for Lifshitz-invariant fixed points, the superfluid phase itself is critical, while conventional condensed matter treatments usually find that quantum criticality is hidden by a non-critical superfluid phase. For the IR geometries with emergent conformal invariance, as well as the scale-covariant, hyperscaling-violating geometries studied in \cite{Gouteraux:2019kuy} and in appendix \ref{scalecovariant}, the superfluid condensate acts instead as an irrelevant deformation of the underlying quantum critical, normal phase. Hence, as we have described in some detail, the scaling of a number of thermodynamic or transport observables remains controlled by the normal phase. On the other hand, there also exist critical superfluid phases where the superfluid condensate strongly deforms the normal phase, examples of which can be found in \cite{Gouteraux:2012yr,Gouteraux:2013oca}.
$\\$

It is worth contrasting our results with those obtained in a qualitatively different large $N$ limit, where the effects of a gapless boson (a proxy for critical order parameter fluctuations) on a Fermi surface are investigated, \cite{Raghu_2015,Wang_2017,Wang_2018,damia2020thermal}. There, the fermions transform in the fundamental representation of some internal flavor symmetry group SU($N$) while the bosons transform in the adjoint. $N$ is taken to be very large, together with $\epsilon\ll1$ where $\epsilon=3-d$. What these authors find is that for sufficiently large $N\epsilon$, naked metallic quantum critical points with finite BCS couplings appear in the phase diagram, but $T_c=0$ and there is no finite condensate. Instead, for smaller values of $N\epsilon$, the BCS couplings diverge in the IR, triggering a BCS instability which hides the quantum critical point under a superconducting dome, as in more conventional scenarios.
$\\$

By contrast, in the scale-covariant geometries studied in \cite{Gouteraux:2019kuy} and in appendix \ref{scalecovariant}, there is a nonzero $T_c$ with a finite condensate. This condensate only sources an irrelevant deformation of the IR geometry, which retains the same scaling properties as the underlying normal phase -- and hence is not a naked metallic quantum critical point in the sense of \cite{Raghu_2015,Wang_2017,Wang_2018,damia2020thermal}. As we have described in some detail, the scaling of a number of thermodynamic or transport observables in the superfluid phase remain controlled by the normal phase. On the other hand, there also exist critical superfluid IR geometries where the superfluid condensate strongly deforms the normal phase, examples of which can be found in \cite{Gouteraux:2012yr,Gouteraux:2013oca}. These geometries all have in common that they are supported by a logarithmically running scalar in the IR. In the very deep IR, quantum or stringy corrections are expected to become important and can lead to a different fixed point, see eg \cite{Harrison:2012vy}. However, we should re-emphasize that our result on the non-vanishing of the normal density at zero temperature also applies to scale-invariant Lifshitz fixed points, which do not suffer from such instabilities.\footnote{The Lifshitz metric does sport a null singularity, which was argued in \cite{Horowitz:2011gh} to lead to uncontrolled proliferation of test strings. It was later shown in \cite{Bao:2012yt} that scattering of these strings with the matter supporting the Lifshitz metric slows down their proliferation, this circumventing the divergent tidal forces in \cite{Horowitz:2011gh}.}

\begin{acknowledgments}
We would like to thank Elias Kiritsis for helpful discussions. In addition, we would like to thank Tomas Andrade and Richard Davison for initial collaboration at an early stage of this project. This work was supported by the European Research Council (ERC) under the European Union's Horizon 2020 research and innovation programme (grant agreement No.758759).

\end{acknowledgments}

\newpage
\appendix
\section{Landau-Tisza hydrodynamics \label{hydrosection}}
In \cite{Gouteraux:2019kuy}, we cover relativistic two-fluid hydrodynamics in detail. To keep the discussion self-contained, we reproduce that discussion here. \subsection{Conservation equations and constitutive relations for small superfluid velocities}
$\\$

We follow \cite{Herzog:2011ec} to study linear response in conformal, relativistic superfluid hydrodynamics. Compared to the usual relativistic hydrodynamics, the U(1) symmetry is spontaneously broken, which we model by explicitly introducing the resulting Goldstone boson $\xi_\mu \equiv \partial_\mu\varphi-A_\mu$. This expression is gauge invariant, though for simplicity we will fix a gauge in which $A_\mu=0$.  In equilibrium, ${\xi_\mu}$ acquires a constant value, which can be large and is related to the superfluid velocity. It should be counted at zeroth order in the gradient expansion. The authors of \cite{Herzog:2011ec} work in the simplifying limit where the gradient of the Goldstone boson is small, which is justified by the presence of instabilities in superfluid flows with large velocities.
$\\$

To first order in the gradient expansion, the constitutive relations of the superfluid are 
\begin{equation}
\begin{split}
T_{\mu\nu}&=\left(\epsilon + P\right)u_\mu u_\nu+P\eta_{\mu\nu}+2\rho_s\mu_s n_{(\mu}u_{\nu)}+\zeta\rho_s\mu_s n_\mu n_\nu-\eta\sigma_{\mu\nu}-\eta_s \sigma_{\mu\nu}^s\\
J^\mu&=\rho u^\mu+\zeta \rho_s n^\mu-\sigma_0 T P^{\mu\nu}\partial_\nu\left(\frac\mu T\right)\\
u^\mu\xi_\mu&=-\mu+\zeta_3 \partial_\nu\left(\rho_s v^\nu\right)
\end{split}
\end{equation}
Here $\zeta = \mu_s/\mu$ is the superfluid fraction. The last equation is the Josephson condition for the superfluid. $u^\mu$ is the velocity of the normal fluid, while $v^\mu$ is the superfluid velocity defined by
\begin{equation}
\mu_s n_\mu=P^\nu_\mu\xi_\nu\,,\qquad P^{\mu\nu}=\eta^{\mu\nu}+u^\mu u^\nu
\end{equation}
From this definition, $\zeta n_\mu =\frac{\xi_\mu}{\mu}-u_\mu$.  Writing things in terms of $\xi_\mu$ and $u_\mu$, we find
\begin{align}
T_{\mu\nu}&=\left(\epsilon - \mu\rho_s + P\right)u_\mu u_\nu+P\eta_{\mu\nu}+\frac{\rho_s}{\mu}\xi_\mu\xi_\nu-\eta\sigma_{\mu\nu}-\eta_s \sigma_{\mu\nu}^s\\
J^\mu&=\rho_n u^\mu+\frac{\rho_s}{\mu} \xi^\mu-\sigma_0 T P^{\mu\nu}\partial_\nu\left(\frac\mu T\right)
\end{align}
so that the superfluid fraction drops out of the equations. The thermodynamic quantities obey the Smarr relation
\begin{equation}
\epsilon+P=s T+\mu\rho\,,\qquad \rho=\rho_n+\rho_s\,,
\end{equation}
where $\rho$ is the total charge density, as well as the first law
\begin{equation}
dP=s dT+\rho d\mu-\frac{\rho_s}{2\mu}d\left(\xi_\nu\xi^\nu+\mu^2\right)
\end{equation}
This means that contrarily to the usual hydrodynamics of normal fluids, the thermodynamic quantities like the pressure are functions of say $\mu$ and $T$, but also of $\xi_i\xi^i$, the spatial superfluid velocity squared.
However, working in the limit of small superfluid velocities, we can rewrite
\begin{equation}
d\mu=-\frac{s}{\rho} dT+\frac{d P}{\rho}-\frac{\rho_s}{\mu\rho}\xi_i d\xi^i
\end{equation}
which shows that
\begin{equation}
\mu(P,T,|\xi_i|)=\mu(P,T)-\frac{\rho_s}{2\mu\rho}\xi_i\xi^i+\dots
\end{equation}
We see that in this limit $\mu(P,T,|\xi_i|)$ is the usual chemical potential up to terms quadratic in the superfluid velocity. By differentiating with respect to temperature and pressure, the entropy and the density are obtained as:
\begin{equation}
s(P,T,|\xi_i|)=s(P,T)+\rho\xi_i\xi^i\partial_T\left(\frac{\rho_s}{2\mu\rho}\right)+\dots
\end{equation}
\begin{equation}
\rho(P,T,|\xi_i|)=\rho(P,T)+\rho^2\xi_i\xi^i\partial_P\left(\frac{\rho_s}{2\mu\rho}\right)+\dots
\end{equation}
When substituting in the constitutive relations, we neglect terms cubic in $|\xi_i|$. As we are ultimately interested in linear response, our task is even more simple: terms quadratic in $|\xi_i|$ make no contributions to the linearized equations, so we can drop all the quadratic corrections to the thermodynamic quantities. There will be non-zero contributions to $J^i$ and $T^{0i}$, though.
$\\$

Imposing positivity of the divergence of the entropy current  $J^\mu_s=s u^\mu+\mu\sigma_0\partial_\mu(\mu/T) $ together with conformal invariance implies $\eta_s=0$.

\subsection{Linear response a la Kadanoff-Martin}

Starting in the rest frame of the fluid with no background superfluid velocity, we linearize around equilibrium
\begin{equation}
T(t,x^i)=T+\delta T\,,\quad \mu(t,x^i)=\mu+\delta \mu\,,\quad u^\mu=(1,\delta u^i)\,,\quad n^\mu=\zeta^{-1}(0,\xi^i/\mu-\delta u^i)
\end{equation}
and write the corresponding linearized expressions for the stress-tensor and current
\begin{equation}
\begin{split}
\delta T^{00}&=\delta\epsilon\\
\delta T^{0i}&=(\mu\rho_n+s T)\delta u^i+\rho_s\xi^i\\
\delta T^{ij}&=\left(\delta P+\eta\partial_k \delta u^k\right)\delta^{ij}-2\eta \partial^{(i} \delta u^{j)}\\
\delta J^0&=\delta\rho\\
\delta J^i&=\rho_n \delta u^i+\frac{\rho_s}\mu\xi^i-\sigma_0\left(\partial^i\mu-\frac\mu T\partial^i T\right)\\
\xi_0&=-\mu+\zeta_3\rho_s\left(\frac{\partial_i\xi^i}{\mu}-\partial_i u^i\right)
\end{split}
\end{equation}

We can now write linearized equations for the longitudinal fluctuation of conserved charges $(\delta \epsilon, \pi^x=\delta T^{0x},\delta\rho,\xi^x)$ choosing a wavector in the $\hat x$-direction ${\bf k}=(k,0)$:
\begin{equation}
\label{eq:lineareoms}
\begin{split}
&-i\omega\delta\epsilon+i k\pi^x=0\\
&-i\omega\pi^x+i k\left(\beta_1\delta\epsilon+\beta_2\delta \rho\right)+\frac{\eta}{h}k^2\pi^x-k^2\frac{\rho_s}{h}\eta\;\xi^x=0\\
&-i\omega\delta\rho +i k\frac{\rho_n}{h}\pi^x+\sigma_0k^2\left(\alpha_1\delta\epsilon+\alpha_2\delta\rho\right)+ik\frac{\rho_s s T}{\mu h}\xi^x=0\\
&\delta\xi_0+\left.\frac{\partial\mu}{\partial\epsilon}\right|_{\rho}\delta\epsilon+\left.\frac{\partial\mu}{\partial\rho}\right|_{\epsilon}\delta\rho-ik\frac{\rho_s}{h}\zeta_3\delta \pi^x+ik\frac{h+\mu\rho_s}{\mu h}\rho_s\zeta_3\xi^x=0
\end{split}
\end{equation}
where $h=\mu\rho_n+s T$ and
\begin{equation}
\begin{split}
&\alpha_1=\left(\frac{\partial\mu}{\partial\epsilon}\right)_{\rho}-\frac{\mu}{T}\left(\frac{\partial T}{\partial\epsilon}\right)_{\rho}\,,\quad \alpha_2=\left(\frac{\partial\mu}{\partial\rho}\right)_{\epsilon}-\frac{\mu}{T}\left(\frac{\partial T}{\partial\rho}\right)_{\epsilon}\\
&\beta_1=\left(\frac{\partial p}{\partial\epsilon}\right)_{\rho}\,,\quad \beta_2=\left(\frac{\partial p}{\partial\rho}\right)_{\epsilon}
\end{split}
\end{equation}
We note that with no background field strength, $\partial_\mu \xi_\nu = -\partial _\nu \xi_\mu$ using commutivity of the partial trace. Using 
\begin{equation}
\xi^x=\mu(n^x+\delta u^x)\,,\qquad \pi^x=(h+\mu\rho_s)\delta u^x+\mu\rho_s n^x
\end{equation}
The matrix of static susceptibilities is \cite{Valle:2007xx} is
\begin{equation}
\chi=\left(\begin{array}{cccc}
T\left(\frac{\partial\epsilon}{\partial T}\right)_{\mu/T}&0&\left(\frac{\partial\epsilon}{\partial \mu}\right)_{T}&0 \\
0&w&0&\mu\\
T\left(\frac{\partial\rho}{\partial T}\right)_{\mu/T}&0&\left(\frac{\partial\rho}{\partial \mu}\right)_{T}&0 \\
0&\mu&0&\frac{\mu}{\rho_s}
\end{array} \right)
\end{equation}
with $w=h+\mu\rho_s=\epsilon+P$. The sources for $(\delta\epsilon, \pi^x, \delta \rho)$ are $(\delta T/T, \delta u^x, T\delta[\frac{\mu}{T}])$ as per usual. The susceptibility matrix is consistent with the following source for the Goldstone:
\begin{equation}
s_{\xi}=\rho_s{\bf n}
\end{equation}
and comes from the following Hamiltonian deformation
\begin{equation}
\delta H_{A_t}=-\int d^3 {\bf x}\, \rho_s {\bf n}\cdot{\bf\xi}({\bf x},t)=-\int d^3 {\bf x}\, {\bf n}\cdot  {\bf\pi}_{\xi}({\bf x},t)
\end{equation}
with
\begin{equation}
{\bf\pi}=\tilde{\bf\pi}+{\bf\pi}_{\xi}\,,\qquad {\bf\pi}_{\xi}=\rho_s{\bf \xi}
\end{equation}
so that $\pi_{\xi}$ is the momentum along the superfluid.
$\\$

As expected, the susceptibility matrix is symmetric (Onsager relations). This can be shown using the first law of thermodynamics in the grand-canonical ensemble. We can also derive the following relations
\begin{equation}
\begin{split}
&\beta_1\chi_{11}+\beta_2\chi_{31}=w\,,\quad \beta_1\chi_{13}+\beta_2\chi_{33}=\rho\,\\
&\alpha_1\chi_{11}+\alpha_2\chi_{31}=0\,,\quad \alpha_1\chi_{13}+\alpha_2\chi_{33}=1\,\\
&\left.\frac{\partial\mu}{\partial\epsilon}\right|_{\rho}\chi_{13}+ \left.\frac{\partial\mu}{\partial\rho}\right|_{\epsilon}\chi_{33}=1
\end{split}
\end{equation}

If we make use of conformal symmetry (which means that for instance $p=T^3 f(\mu/T)$), we can further obtain
\begin{equation}
\begin{split}
&\chi_{11}=2w\,,\quad\chi_{31}=\chi_{13}=2\rho\,,\quad \beta_1=\frac12\,,\quad \beta_2=0\\
&\alpha_1=\frac{\rho}{2\rho^2-w\chi_{33}}\,,\quad \alpha_2=\frac{-w}{2\rho^2-w\chi_{33}}\\
&\left.\frac{\partial\mu}{\partial\epsilon}\right|_{\rho}=\frac12\frac{2\rho-\mu\chi_{33}}{2\rho^2-w\chi_{33}}\,,\quad \left.\frac{\partial\mu}{\partial\rho}\right|_{\epsilon}=\frac{-s T}{2\rho^2-w\chi_{33}}
\end{split}
\end{equation}
This can be obtained by manipulations involving the Jacobian. For instance, if we define
\begin{equation}
\frac{\partial(X,Y)}{\partial(U,V)}=\left|\begin{array}{cc}
\left(\frac{\partial X}{\partial U}\right)_{V}&\left(\frac{\partial X}{\partial V}\right)_{U} \\
\left(\frac{\partial Y}{\partial U}\right)_{V}&\left(\frac{\partial Y}{\partial V}\right)_{U}
\end{array} \right|
\end{equation}
then we have
\begin{equation}
\frac{\partial(X,Y)}{\partial(S,T)}=\frac{\partial(X,Y)}{\partial(U,V)}\frac{\partial(U,V)}{\partial(S,T)}\,,\qquad \frac{\partial(X,Y)}{\partial(S,T)}=-\frac{\partial(Y,X)}{\partial(S,T)}
\end{equation}
and also
\begin{equation}
\left(\frac{\partial(X,Y)}{\partial(S,T)}\right)^{-1}=\frac{\partial(S,T)}{\partial(X,Y)}\,,\qquad \frac{\partial(X,Y)}{\partial(U,Y)}=\left.\frac{\partial X}{\partial U}\right|_Y
\end{equation}
We can now obtain the retarded Green's functions following the method of Kadanoff and Martin, \cite{Valle:2007xx, Kovtun:2012rj}. We will denote by
\begin{equation}
\langle A B(\omega,k)\rangle =G^R_{AB}(\omega,k)-G^R_{AB}(\omega=0,k)
\end{equation}
the correlator from which the static susceptibility has been subtracted ($G^R_{AB}(\omega=0,k)=-\chi_{AB}$).
$\\$

Then, the thermoelectric conductivities read
\begin{align}
\sigma&=\frac{i}{\omega}\langle J^x J^x(\omega,0)\rangle=\sigma_0+\frac{i\rho_n^2}{h\omega}+\frac{i\rho_s}{\mu\omega}\\
\alpha&=\frac{i}{\omega T}\left(\langle J^x T^{0x}(\omega,0)\rangle-\mu\langle J^x J^x(\omega,0)\rangle\right)=-\frac\mu T\sigma_0+\frac{i\rho_n s}{h\omega}\\
\bar\kappa&=\frac{i}{\omega T}\left(\langle T^{0x} T^{0x}(\omega,0)\rangle-\mu\langle T^{0x} J^x(\omega,0)\rangle-\mu\langle J^x T^{0x}(\omega,0)\rangle+\mu^2\langle J^x J^x(\omega,0)\rangle\right)\nonumber\\
&=\frac{\mu^2}T\sigma_0+\frac{is^2T}{h\omega}
\end{align}
Note that the delta function due to the presence of a superfluid only appears in the electric conductivity $\sigma$. Moreover the following Ward identities are obeyed
\begin{equation}
\alpha T+\mu\sigma-\frac{i \rho}{\omega}=0\,,\quad \bar\kappa+\mu\alpha-\frac{i s}{\omega}=0
\end{equation}
which take the same form as without a superfluid.

\section{Proof of \eqref{leadingintegral} \label{app:importantintegral}}

In this Appendix, we prove the equation \eqref{leadingintegral}, which we recall here for convenience
\begin{align}
\label{crucialequation2}
\int_0^{r_h} \frac{\sqrt{BD}}{C^{d/2+1}}d\tilde{r} \approx \frac{c_{ir}^2}{sT}+...
\end{align}
 First, we note that the behavior of \eqref{IRmetric} in the deep IR implies
\begin{align}
\int_0^{r_h} \frac{\sqrt{B_0D_0}}{C_0^{d/2+1}}d\tilde{r} \sim b \left(\frac{r_h}{L}\right)^{2+d-z} +...
\end{align}
which diverges when $z<d+2$ and converges otherwise. Thus, it is only when $z<d+2$ that this integral can be reliably approximated using only the near-extremal part of the spacetime.
$\\$

To prove this equation, we must distinguish between when the IR metric is AdS \eqref{AdSIRmetric} with $z=1$ and when it is Lifshitz \eqref{IRmetric} with $1<z<d+2$. This is because of the presence of the irrelevant deformation sourced by the gauge field in the IR AdS case. 
$\\$

We first consider $z=1$, and follow closely the logic in section III.B.1 of \cite{Davison:2018nxm}.
We rewrite \eqref{crucialequation2} using \eqref{eq:conservationsT}:
\begin{align}
\int_0^{r_h} \frac{\sqrt{BD}}{C^{d/2+1}}d\tilde{r}=-\int_0^{r_h} \left(\frac{D}C\right)'\frac1{sT+RA}=-\left[\frac{D}C\frac1{sT+RA}\right]_0^{r_h}+\int_0^{r_h}\frac{D}C \left(\frac1{sT+RA}\right)'
\end{align}
Using that $D(r_h)=0$ and $A(0)=\mu$, $R(0)=\rho$, the first term simplifies to $1/(sT+\mu\rho)$ and can be neglected at low $T$, as we will shortly see that the second term diverges as $T\to0$. We now restrict the integration domain of the second term to $L\ll r_{uv}<r<r_{ir}\ll r_h$. We will soon fix $r_{ir}$ and $r_{uv}$ such that the form of the metric \eqref{AdSIRmetric} and gauge field \eqref{AtIRAdS} are valid in this region. In this region, we thus have $D/C\simeq L_t^2/L_x^2\equiv c_{ir}^2$ and can pull this constant factor out of the integral:
\begin{align}
\int_0^{r_h} \frac{\sqrt{BD}}{C^{d/2+1}}d\tilde{r}\simeq c_{ir}^2\int_{r_{uv}}^{r_{ir}} \left(\frac1{sT+RA}\right)'\simeq c_{ir}^2 \left(\frac{R(r_{uv})A(r_{uv})-R(r_{ir})A(r_{ir})}{(sT+R(r_{ir})A(r_{ir}))(sT+R(r_{uv})A(r_{uv}))}\right)
\end{align}
We now define $r_{ir}=\varepsilon_{ir}r_h$ and $r_{uv}=L/\varepsilon_{UV}$ such that given our assumptions the UV and IR cutoffs are very small $\varepsilon_{ir},\varepsilon_{uv}\ll1$. In the region of integration $R(r)A(r)\sim c_A^2(r/L)^{2\tilde\Delta_{A_0}-d-1}$. This leads to 
\begin{align}
R_{r_{ir}}A(r_{ir})\ll s T\ll R(r_{uv})A(r_{uv})
\end{align}
provided
\begin{align}
\label{conditionircutoff}
c_A^2 T^{-2\tilde\Delta_{A_0}}\ll\varepsilon_{ir}^{d+1-2\tilde\Delta_{A_0}}\,,\qquad c_A^{-2}T^{d+1}\ll\varepsilon_{uv}^{d+1-2\tilde\Delta_{A_0}}
\end{align}
which can always be achieved for small enough $T$, recalling that $\tilde\Delta_{A_0}<0$. Neglecting subleading terms, we then obtain
\begin{align}
\int_0^{r_h} \frac{\sqrt{BD}}{C^{d/2+1}}d\tilde{r}\simeq \frac{c_{ir}^2}{sT}
\end{align}
 as desired.
 $\\$
 
 Ultimately, this derivation is rooted in the fact that there is a non-trivial competition between corrections to the metric \eqref{AdSIRmetric} due to the irrelevant deformation sourced by the gauge field and due to nonzero temperature through the factor $sT+R(r)A(r)$. 
 $\\$
 
 For Lifshitz solutions with $1<z<d+2$, the derivation is different, as can be quickly seen by noticing that in this case $\tilde\Delta_{A_0}=0$ so the condition \eqref{conditionircutoff} cannot consistently be imposed. On the other hand, contrarily to cases without a condensate, a non-perturbative finite temperature solution cannot be found analytically, so we may not simply evaluate \eqref{crucialequation2} on the IR $T=0$ metric \eqref{IRmetric}, introducing the temperature through the upper bound. More precisely, as can be checked numerically,
 \begin{align}
 \label{discrepancy}
\lim_{r_{h}\gg L}\int_0^{r_h} \frac{\sqrt{BD}}{C^{d/2+1}}d\tilde{r}\neq\lim_{r_{h}\gg L}\int_0^{r_h} \frac{\sqrt{B_0 D_0}}{C_0^{d/2+1}}d\tilde{r}\simeq \frac{d+z}{d+2-z}\frac{c_{ir}^2}{sT}
\end{align}
where $0$ subscripts indicate we are using the $T=0$ IR metric \eqref{IRmetric}.
$\\$

Instead, consider integrating  \eqref{eq:conservationsT} between $r=0$ and $r=r_h$:
\begin{align}
\label{NonzeroTconservation}
\int_0^r \frac{A_{t}A_{t}'}{C}d\tilde{r} = \frac{D(r)}{C(r)}-1 + (sT)\int_0^r \frac{\sqrt{BD}}{C^{d/2+1}}d\tilde{r} 
\end{align}
Evaluating this at $r=r_h$ gives
\begin{align}
\label{finiteTconservation}
\int_0^{r_h} \frac{A_{t}A_{t}'}{C}d\tilde{r} = - 1 + (sT)\int_0^{r_h} \frac{\sqrt{BD}}{C^{d/2+1}}d\tilde{r} .
\end{align}

Now, set $T\ll1$ in \eqref{NonzeroTconservation}. Since we want to compare to the zero temperature solutions, we choose a gauge in which $r\to\hat{r}$ in the IR, so that the horizon is at $r = \hat{r}_h \gg L$. In other words, we identify the zero temperature and finite temperature radial coordinates outside the black hole horizon. When $\eta\neq 0$ and $T\ll 1$, finite temperature effects are relevant only in a region $L\ll\hat{r}_{ir}<\hat{r} < \hat{r}_{h}$ where $A_{t,0}-A_{t}\sim A_{t,0}(\hat{r}_h)$.\footnote{This is as opposed to geometries with $\eta = 0$ which have $A_t = A_{t,0}-A_{t,0}(\hat{r}_h)$.}
$\\$

We can then write
\begin{align}
\label{zeroTatapproximation}
\int_0^{\hat{r}_h} \frac{A_{t}A_{t}'}{C}d\tilde{r}-\int_0^{\hat{r}_{h}} \frac{A_{t,0}A_{t,0}'}{C_0}d\tilde{r}= (sT)\int_0^{\hat{r}_h} \frac{\sqrt{BD}}{C^{d/2+1}}d\tilde{r}-c_{ir}^2\simeq\int_{\hat{r}_{ir}}^{\hat{r}_h} \frac{A_{t}A_{t}'}{C}d\tilde{r}-\int_{\hat{r}_{ir}}^{\hat{r}_h} \frac{A_{t,0}A_{t,0}'}{C_0}d\tilde{r}
\end{align}
which is regular in the limit $\hat{r}_{ir} \to \hat{r}_h$ (the integrands on the right are non-divergent as $r\to \hat{r}_h$). To fully justify the approximation in \eqref{crucialequation2}, we must show that the right hand side of \eqref{zeroTatapproximation} vanishes as $T\to 0$ fast enough compared to $T^{2-\frac{2}{z}}$. 
$\\$

In the region $\hat{r}_{ir}<r<\hat{r}_h$, we can use the near-horizon expansion of the background fields
\begin{align}
\label{nearhorizonexp}
A_t(r)\simeq A_h(r_h-r)+A_2(r-r_h)^2+O((r_h-r)^3)\\
B(r)\simeq \frac{1}{4\pi T(r_h-r)}+B_2 + O((r_h-r)^1)\\
D(r)\simeq 4\pi T(r_h-r)+D_2(r_h-r)^2+O((r_h-r)^3)\\
C(r)\simeq C_h+C_2(r_h-r)+O((r_h-r)^2).
\end{align}
Notably, in this gauge, $r_h \sim \hat{r}_h^{-z} \sim T$. 
$\\$

To justify the use of this expansion in the region $r_{ir}<r<r_h$, we must have $r_{ir}=r_h(1-\varepsilon)$, $0<\varepsilon\ll1$ with $\varepsilon$ chosen such that we may neglect the subleading terms ($A_2, B_2, C_2, D_2$) in the expansion. In particular, this requires that the contributions from subleading terms must vanish at $r_{ir}$ as $T\to 0$ parametrically faster than the leading terms. Importantly, plugging the near -horizon expansion \eqref{nearhorizonexp} in the Maxwell equation tells us that
\begin{align}
\mathcal{F}_2|_{r=r_{ir}} \sim 2q^2\eta_0^2\frac{A_h}{4\pi T}
\end{align}
where $\mathcal{F}_2$ is a linear combination of $A_2, B_2, D_2$ and $C_2$. In the near horizon gauge, $A_h \sim T^0$. Hence, when $\eta \neq 0$, in order for the contribution from these terms to vanish parametrically quickly in \eqref{nearhorizonexp}, we must have $\varepsilon \sim T^\alpha$ for $\alpha>0$. 
$\\$

Next, in this near horizon region, we can write
\begin{align}
A_{t,0}(r) \simeq A_{t,0}(r_h) -A_{t,0}'(r_h)(r_h-r)+O((r_h-r)^2).
\end{align}
and we have $A_{t,0}'(r_h)\approx -A_h$. This approximation follows from the fact that, once we identify the zero temperature radial coordinate with the finite temperature radial coordinate, then at any radial position outside of the horizon $\lim_{T\to 0}A_{t}'(r)= A_{t,0}'(r)$ since this limit is non-vanishing. Importantly, this is not the case for $A_t$, since we are required to have $A_t(r_h) = 0 \neq A_{t,0}(r_h)$.\footnote{In fact, this approximation also implies $A_{t}'(r_{ir})\approx A_{t,0}'(r_{ir})\approx A_{t,0}'(r_h)$. If the near horizon expansion is justifed, then $A_{t}(r_h) \approx A_{t}(r_{ir}) - A_{t,0}'(r_h)\varepsilon = 0$, which implies $A_{t,0}(r_{ir}) - A_{t}(r_{ir})\approx A_{t,0}(r_h)$.} Hence, for $T\ll 1$, at leading order in the temperature $A_h = -A_{t,0}'(r_h)$, which is the same approximation we used for $\rho_{in}$. Importantly, this tells us that
\begin{align}
A_{t}\approx A_{t,0} - A_{t,0}(r_h)
\end{align}
in the regions $r_{ir}<r<r_h$. This justifies our earlier statement that the near horion region can be defined as the location where the finite temperature solution differs from the zero temperature approximately by a constant. 
$\\$

Finally, given this approximation, we can write
\begin{align}
\label{crucialjustification}
\int_{r_{ir}}^{r_h}\frac{A_tA_t'}{C}d\tilde{r}- \int_{r_{ir}}^{r_h}\frac{A_{t,0}A_{t,0}'}{C}d\tilde{r} \approx A_{t,0}(r_h)\int_{r_{ir}}^{r_h}\frac{A_{t,0}'}{C_h} \sim \varepsilon T^{2-2/z}.
\end{align}
where we used $C_h\sim T^{2/z}$ and $r_h \sim T$. Because $\varepsilon \sim T^\alpha$, the right hand side of \eqref{zeroTatapproximation} vanishes parametrically faster than $c_{ir}^2\sim T^{2-2/z}$ as $T\to0$ and we have \eqref{crucialequation2}.\footnote{For $z>d+2$, we know that the left hand side of \eqref{zeroTatapproximation} scales as $sT$. Then, using \eqref{crucialjustification}, we would find $\varepsilon \sim T^{-1+(d+2)/z}$ which diverges as $T\to 0$. This divergence indicates a breakdown in the use of the near horizon approximation to evaluate the integral. This is expected, since the integral on the right hand side of \eqref{finiteTconservation} is only dominated by the IR region for $z<d+2$. }
$\\$


 \section{Scale-covariant geometries \label{scalecovariant}}
A more general class of solutions can be described by the action
\begin{align}
\label{bulkactionwithdilaton}
S = \int d^{d+2}x \sqrt{-g}\left\{R - \frac{Z_{F}(\psi)}{4}F_{MN}F^{MN} - M_F(\psi)|D\eta|^2 - \frac{1}{2}(\nabla\psi)^2 - V(\psi,|\eta|)\right\}.
\end{align}
As before, $\eta$ is a complex scalar charged under a $U(1)$ gauge field $A_M$. In addition, there is a neutral scalar, the dilaton $\psi$. The main text is a sub-class of this action where we set $\psi = 0$ and $Z_F(\psi)=1$.
$\\$

The equations of motion are:
\begin{align}
\label{eq:eqsofmotion}
\left[M_F\sqrt{\frac{DC^d}{B}}\eta'\right]' + q^2 M_F\sqrt{\frac{BC^d}{D}}A_t^2 \eta - \sqrt{BDC^d}\,\frac{\partial V}{\partial \eta^*} &=0\nonumber\\
\left[\sqrt{\frac{DC^d}{B}}\psi'\right]' + \frac{Z_F'(\psi)}{2} \sqrt{\frac{C^d}{BD}}(A_t')^2- \sqrt{BDC^d}\,\frac{\partial V}{\partial \psi} +\sqrt{\frac{BC^d}{D}}\left(q^2A_t^2\eta^2-\frac{D}{B}(\eta')^2\right)M_F'&=0\nonumber\\
\left[\sqrt{\frac{C^d}{BD}}A_t'\right]' - 2q^2M_F \sqrt{\frac{BC^d}{D}}\eta^2 A_t &=0\nonumber\\
d\left[\sqrt{\frac{C^d}{BD}}D'\right]' + 2\sqrt{BDC^d}V - (d-1)Z_F\sqrt{\frac{C^d}{BD}}(A_t')^2 - 2dq^2M_F\sqrt{\frac{BC^d}{D}}\eta^2A_t^2 &=0\nonumber\\
\frac{d}{2}\left[\frac{C'}{\sqrt{BDC}}\right]' + M_F\sqrt{\frac{C}{BD}}(\eta')^2 +\frac{1}{2} \sqrt{\frac{C}{BD}}(\psi')^2 + q^2M_F \sqrt{\frac{BC}{D^3}}\eta^2A_t^2&=0\nonumber\\
2M_F(\eta')^2 -(\psi')^2- \frac{d}{2}\frac{C'}{C}\left(2\frac{D'}{D}+(d-1)\frac{C'}{C}\right) - \frac{Z_F}{D}(A_t')^2 - 2B V + 2q^2M_F \frac{B}{D}\eta^2A_t^2 &=0.
\end{align}

The conservation equation for the background geometry now reads
\begin{align}
\frac{d}{dr}\left\{\sqrt{\frac{C^d}{BD}}\left[C\left(\frac{D}{C}\right)' - Z_F(\psi)A_tA_t'\right]\right\} = 0\,
\end{align}
In the UV, we are interested in solutions that have an asymptotic form ($r\to0^+$),
\begin{align}
ds^2 \to \frac{dr^2-dt^2+d\vec{x}_d^2}{r^2}.
\end{align}
For simplicity, we will chose a potential that satisfies
 \begin{align}
 V(|\eta|) = - d(d+1) - d\; |\eta|^2  - \frac{d}{2}\psi^2+..., \quad Z_{F} = 1+ O(\psi^2),\quad M_F = 1+O(\psi^2).
  \end{align}
 This choice of potential enforces that the matter fields have an asymptotic fall-off
 \begin{align}
 \eta &\sim \eta^{(1)}r+ \eta^{(2)}r^d+...\nonumber\\
 \psi&\sim \psi^{(1)}r + \psi^{(2)}r^d+...\nonumber\\
 A_t&\sim \mu - \frac{\rho}{(d-1)}r^{d-1}+...\nonumber\\
 D &\sim r^{-2}- \frac{\left(\eta^{(1)}\right)^2}{2d} - \frac{\epsilon}{d+1}\;r^{d-1}+...\nonumber\\
 C &\sim r^{-2} -  \frac{\left(\eta^{(1)}\right)^2}{2d} + \frac{P}{d+1}\;r^{d-1} +...\nonumber\\
 B &\sim r^{-2} +...
 \end{align}
 
As is well-known, in the conventional quantization scheme, $\eta^{(1)}$ and $\mu$ act as sources for the expectation values $\eta^{(2)}$ and $\rho$. In particular, the choice $\mu \neq 0$ gives rise to a finite charge density and breaks the background charge-conjugation symmetry. With the choice $\eta^{(1)} = 0$, a non-zero $\eta^{(2)}$ indicates spontaneous breaking of the $U(1)$ symmetry and characterizes a holographic superfluid. 
$\\$

Following the holographic renormalization procedure, we define the counterterm action
\begin{align}
S_{ct} = 2\int_{\partial \Sigma} \sqrt{-\gamma}\left(\mathcal{K}- d -\frac{1}{2}|\eta|^2-\frac{\psi^2}{4}\right)
\end{align}
so that the full renormalized action is $S_{ren} = S+S_{ct}$. Here $\gamma_{ab}$ is the boundary metric along a constant radial hypersurface with unit normal $n^r=B^{-1/2}(r)$. This can be used to find the extrinsic curvature and its trace $\mathcal{K} = \gamma^{ab}\mathcal{K}_{ab}$. The full renormalized action allows us to identify $\epsilon$ and $P$ as the thermodynamic pressure and energy density via the stress tensor
\begin{align}
T_{ab} = 2\left[\mathcal{K}_{ab} - \mathcal{K}\gamma_{ab} - d\gamma_{ab}\right] -\left (|\eta|^2+\frac{\psi^2}{2}\right)\gamma_{ab} 
\end{align}
Furthermore, the use of the conservation equation (\ref{eq:conservation}) and the UV asymptotics of the metric and matter fields give the thermodynamic relations
\begin{align}
\epsilon + P = sT + \mu \rho
\end{align}
where $P$ is the Gibbs potential
\begin{align}
P = \frac{\epsilon}{d} - \eta^{(1)}\eta^{(2)}-\frac{\psi^{(1)}\psi^{(2)}}{2}.
\end{align}
When $\psi^{(1)}\neq 0$, boundary conformal invariance is broken. On the other hand, when $\psi^{(1)}=0$ and when the $U(1)$ symmetry is spontaneously broken, as in the main text, the boundary fluid maintains conformal invariance.
 $\\$
 
The derivation of transport coefficients follow the same logic as in the main text, subject to including the new impact of the dilaton. In particular, 
\begin{align}
\rho_{in} &= -\frac{C^{d/2}}{\sqrt{BD}}Z_F(\psi)A_t'\biggr|_{r=r_h},\nonumber\\
\sigma_0 &= \frac{(a_{\hat{x}}^{(0)})^2}{\mu^2}Z_h A_{t,0}(r_h)^2
\end{align}
where $Z_h = Z_F[\psi(r_h)]$.
$\\$

The derivation of $\rho_n$ is also similar. The relevant equation of motion is
\begin{align}
\frac{d}{dr}\left[C^{d/2-1}\sqrt{\frac{D}{B}}Z_FA_t^2\left(1+\frac{sT}{A_t R}\right)\left(\frac{a_{\hat{x}}}{A_t}\right)' - sT \frac{D}{C}\left(\frac{a_{\hat{x}}}{A_t}\right)\right] = (sT)\frac{2q^2\eta^2C^{d-1}M_FZ_F A_t^2}{R^2}\left(\frac{a_{\hat{x}}}{A_t}\right)'
\end{align}
where
\begin{align}
R = -\frac{C^{d/2}}{\sqrt{BD}}Z_FA_t'\,.
\end{align}
From this, we find
\begin{align}
Z-\frac{\rho}{\mu} = -\frac{sT}{\mu^2} + \frac{(sT)^2}{\mu^3\rho} + \frac{(sT)^2}{\mu^2}\int_{0}^{r_h} \frac{2q^2\eta^2\sqrt{BD}M_FC^{d/2-1}}{R^2}dr'+...
\end{align}
as before.

\subsection{IR geometries}
 When the scalar $\psi$ is non-zero, it will generically have runaway behavior towards the IR. In these cases, we choose the IR behavior of the functions to be
\begin{align}
V(\psi) \sim V_0 e^{-\delta\psi}, \quad Z_F(\psi) \sim Z_0 e^{\gamma\psi}.
\end{align}
For now, we set $M_F=1$. The charged scalar field is chosen to obtain a finite value as the temperature vanishes, $\eta_0$. This can be generalized, as in \cite{Horowitz:2009ij}, though for simplicity we will avoid these cases. This choice of potentials leads to an effective action in the IR,
\begin{align}
S_{IR} = \int d^{d+2}x \sqrt{-g}\left(R-\frac{1}{2}(\partial\psi)^2 - V_0e^{-\delta\psi} - \frac{Z_0}{4}e^{\gamma\psi}F^2\right).
\end{align} 
This RG flow will be between a conformally invariant UV and a quantum critical phase with scaling exponents $z$ and $\theta$ which are determined in terms of $\delta, \gamma$ and the behavior of $\psi$.
$\\$

The behavior of the IR critical phase can be captured by the gravitational metric
\begin{align}
\label{IRmetricNew}
ds^2 = \left(\frac{\hat{r}}{L}\right)^{\frac{2\theta}{d}}\left[-\frac{L^{2z}}{\hat{r}^{2z}}fL_t^2 dt^2 + \tilde{L}^2\frac{d\hat{r}^2}{f\hat{r}^2} + \frac{L^2}{\hat{r}^2}L_x^2d\vec{x}_d^2\right], \quad f= 1-\left(\frac{\hat{r}}{\hat{r}_h}\right)^{d+z-\theta}
\end{align}
As before, the coordinate $\hat{r}$ is in general distinct from the radial coordinate $r$ which governs the UV. For $\theta<d$ ($\theta>d$), this metric will accurately capture the IR geometry for $\hat{r}\gg L$ ($\hat{r}\ll L$) and tell us that the UV is at $\hat{r}\to 0$ ($\hat{r}\to\infty$). The entropy and temperature are
\begin{align}
s = 4\pi L_x^d\left(\frac{\hat{r}_h}{L}\right)^{\theta-d},\quad T = \frac{d+z-\theta}{4\pi \tilde{L}}L_t\left(\frac{\hat{r}_h}{L}\right)^{-z}
\end{align}
so that
\begin{align}
\frac{\hat{r}_h}{L} = \left(\frac{T}{r_0}\right)^{-\frac{1}{z}},\quad s = 4\pi L_x^d \left(\frac{T}{r_0}\right)^{\frac{d-\theta}{z}}.
\end{align}
$\\$

When the scalar $\psi$ runs in the IR, the IR exponents are determined as follows,
\begin{align}
\psi = \kappa \ln\left(\frac{\hat{r}}{L}\right),\quad\kappa^2 = (d-\theta)(2z-2-2\theta/d),\quad \kappa\delta = 2\theta/d.
\end{align}
Instead, when the scalar $\psi$ vanishes, $\theta = 0$, as in the main text. 
$\\$

Our holographic superfluids can only have $z$ and $\theta$ that satisfy the inequalities
\begin{align}
\frac{d-\theta}{z}\geq0,\quad (2-\theta)(2z-2-\theta)\geq0,\quad (z-1)(d+z-\theta)\geq 0.
\end{align}
As we will see below, the first inequality is the requirement that the entropy density increase with temperature, i.e. that the system posesses a positive specific heat. The second two inequalities are imposed by the null energy condition along the radial and transverse directions.
$\\$

\subsubsection{Marginal deformations, $z\neq 1$}
In these solutions, the gauge field $A$ can behave as a marginal or relevant deformation depending on the value of $z$ and $\theta$. If $z\neq 1$ and $\theta\neq 0$, the gauge field is a marginal deformation that renormalizes parameters in the IR metric but does not change the overall scaling with $\hat{r}/L$,
\begin{align}
\label{marginaldeformations}
A = A_0L_t\left(\frac{\hat{r}}{L}\right)^{\theta-d-z}dt,\quad A_0^2 &= \frac{2(z-1)}{Z_0[d+z-\theta]},\nonumber\\
\kappa\gamma = 2d-2\frac{d-1}{d}\theta,\quad \tilde{L}^2&=\frac{(d+z-\theta)(d+z-\theta-1)}{-V_0}.
\end{align}

There are irrelevant deformations around these solutions \cite{Gouteraux:2012yr}, but they only source subleading temperature dependencies in the observables we are interested in.
$\\$

These solutions are partially condensed with
\begin{align}
\lim_{T\to 0}\rho_{in}=(d+z-\theta)\frac{L_x^d}{L\tilde{L}}A_0
\end{align}

The subleading temperature scaling arises from either the integral
\begin{align}
\rho_{in}-\rho_{in}^{(0)} \sim \int_{r_h}^{\infty} dr \sqrt{\frac{B_0}{D_0}}2q^2 C_0^{d/2}\eta_0^2 A_{t,0} \sim \# T^{\frac{2}{z}(d-\theta)}
\end{align}
or from $sT$, whichever is least irrelevant. Similarly, for these solutions, $\rho_{n}^{(0)}\neq 0$. The subleading temperature dependence is from the integral
\begin{align}
\rho_n-\rho_n^{(0)} \sim \int_{r_h}^{\infty} dr \frac{2q^2\eta_0^2\sqrt{B_0D_0}C_0^{d/2-1}}{R_0^2}\sim \#T^{1+\frac{d-\theta-2}{z}}
\end{align}
which is always less irrelevant than $sT$. Hence, the transport observables have temperature dependence
\begin{align}
\rho_{in} &= \rho_{in}^{(0)} + \begin{cases} \# T^{\frac{2}{z}(d-\theta)} & z>d-\theta,\\ \#T^{1+\frac{d-\theta}{z}} & z<d-\theta.\end{cases}\\
\rho_n &= \rho_n^{(0)} + \# T^{1+\frac{d-\theta-2}{z}}\nonumber\\
\sigma_0 &= \# T^{2-\frac{2\theta}{dz}}
\end{align}

\begin{figure}[h]
\begin{center}
\includegraphics[scale=.2]{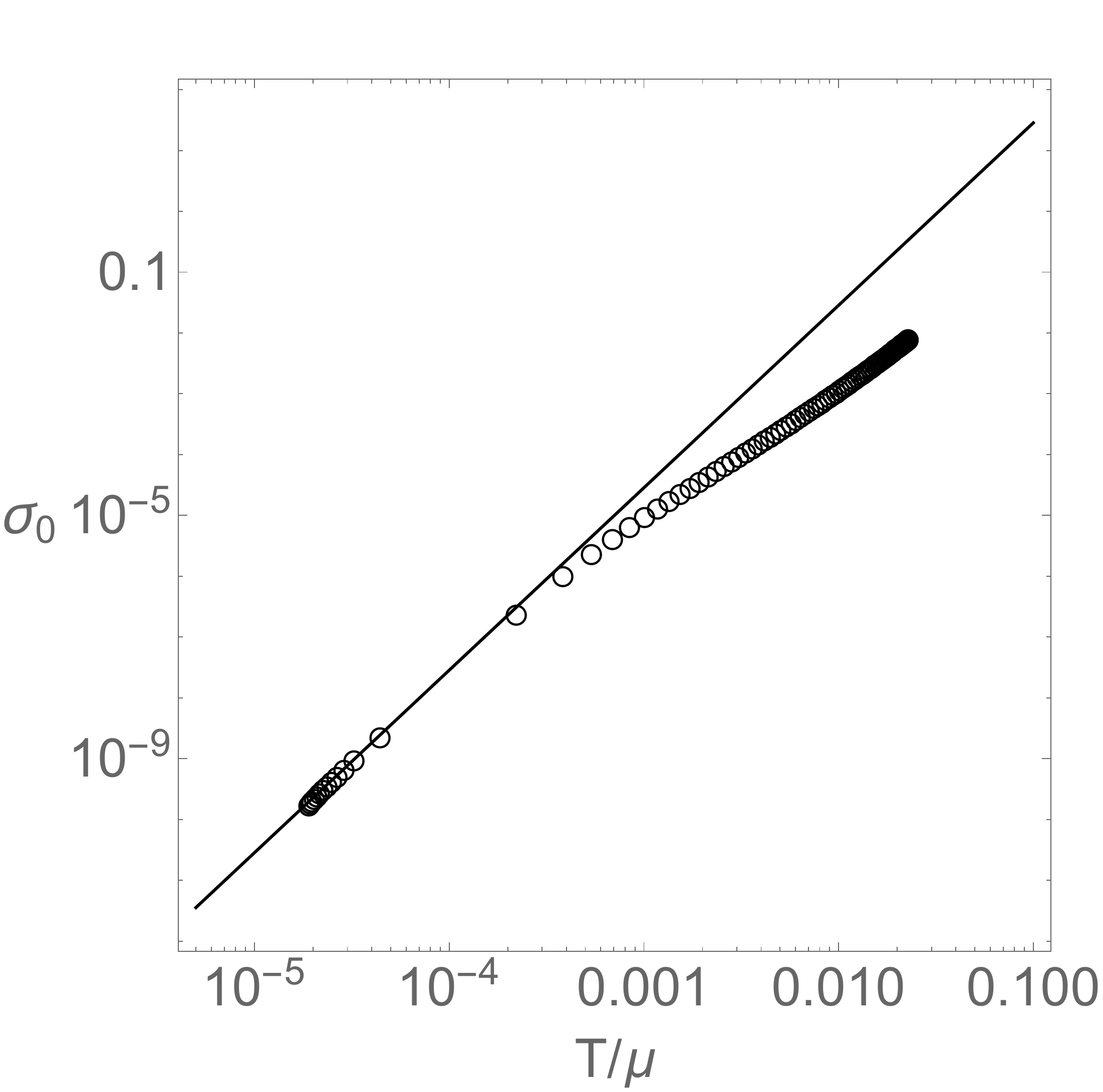}
\includegraphics[scale=.24]{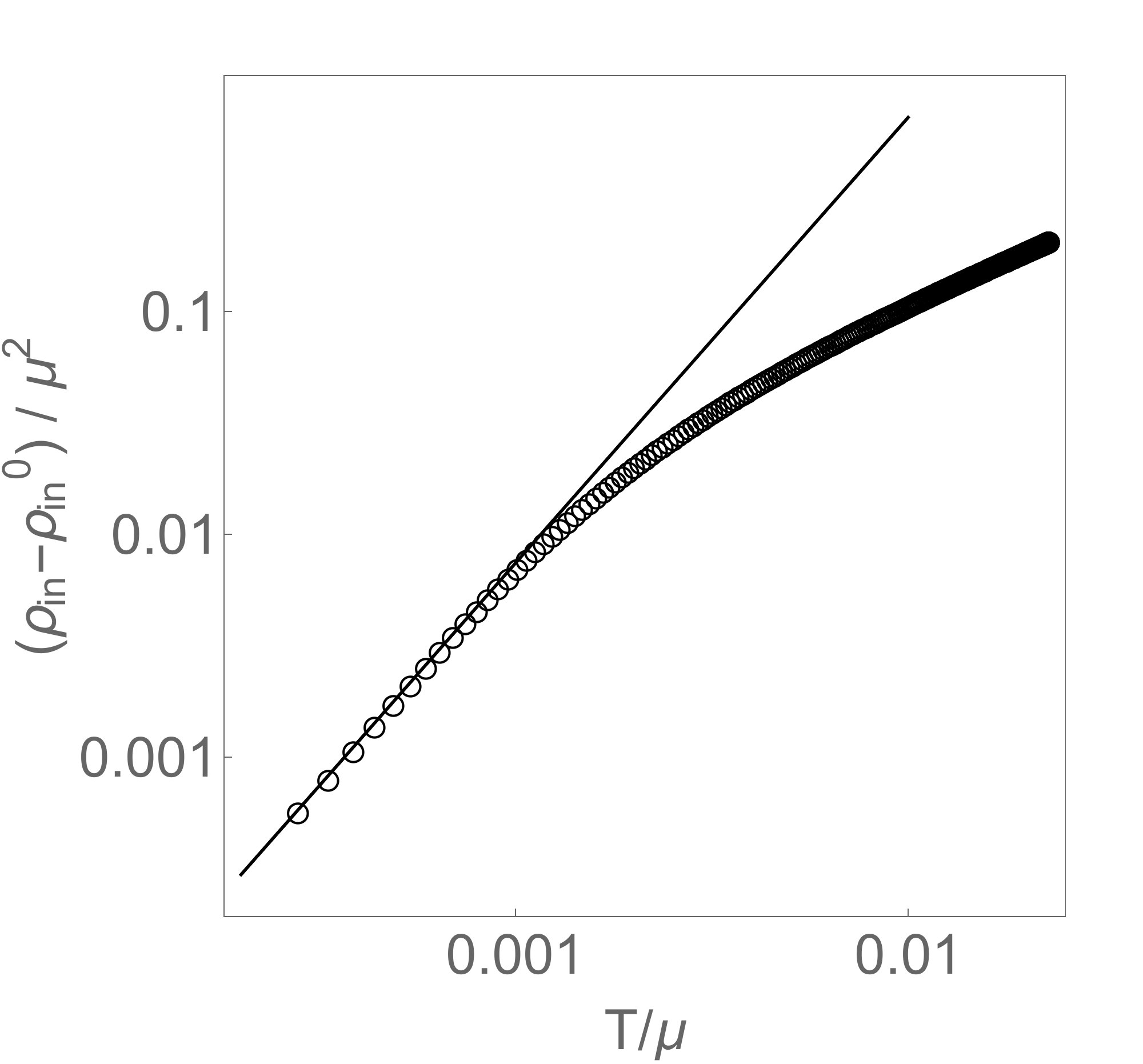}
\includegraphics[scale=.24]{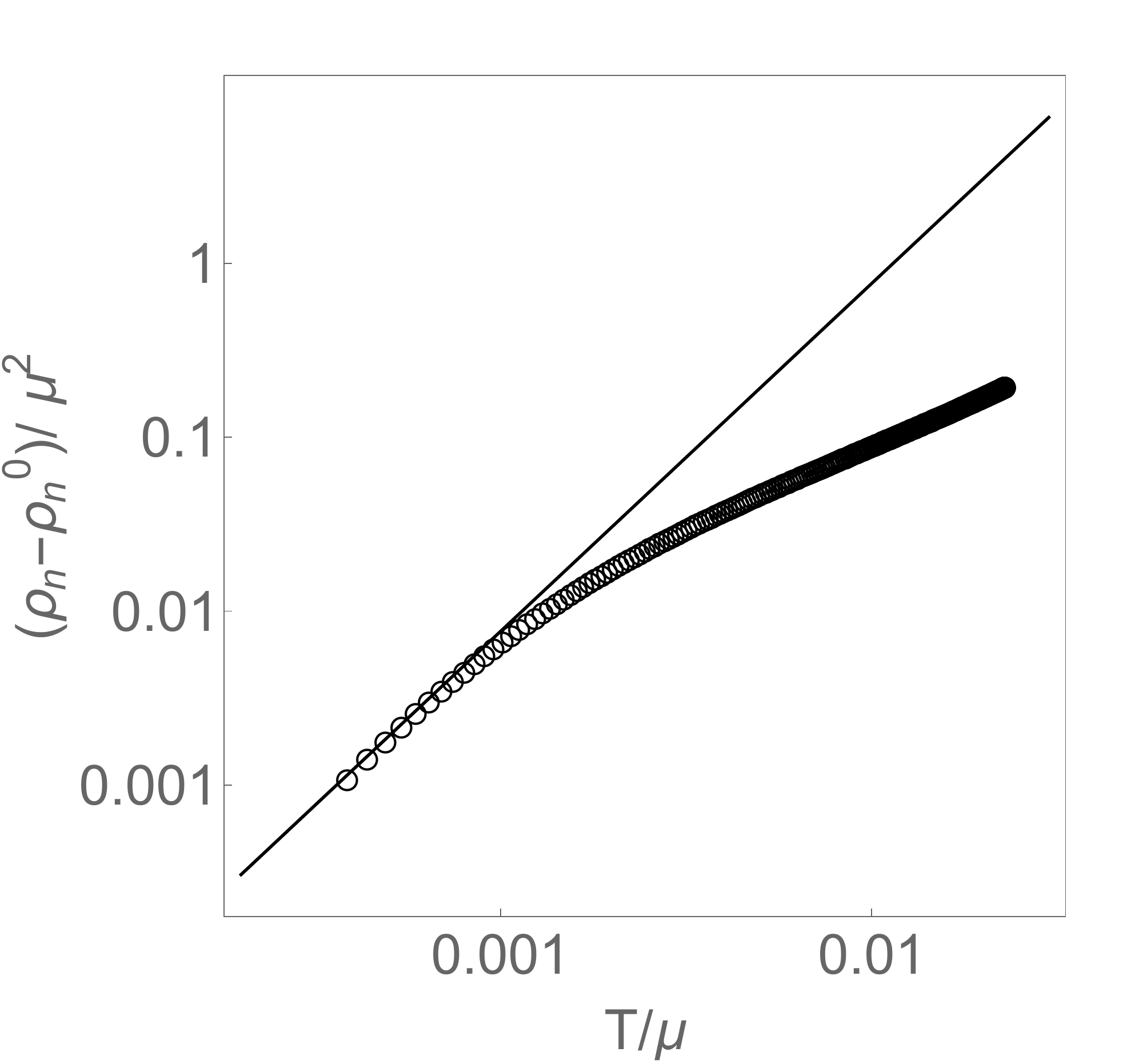}
\newline
\includegraphics[scale=.19]{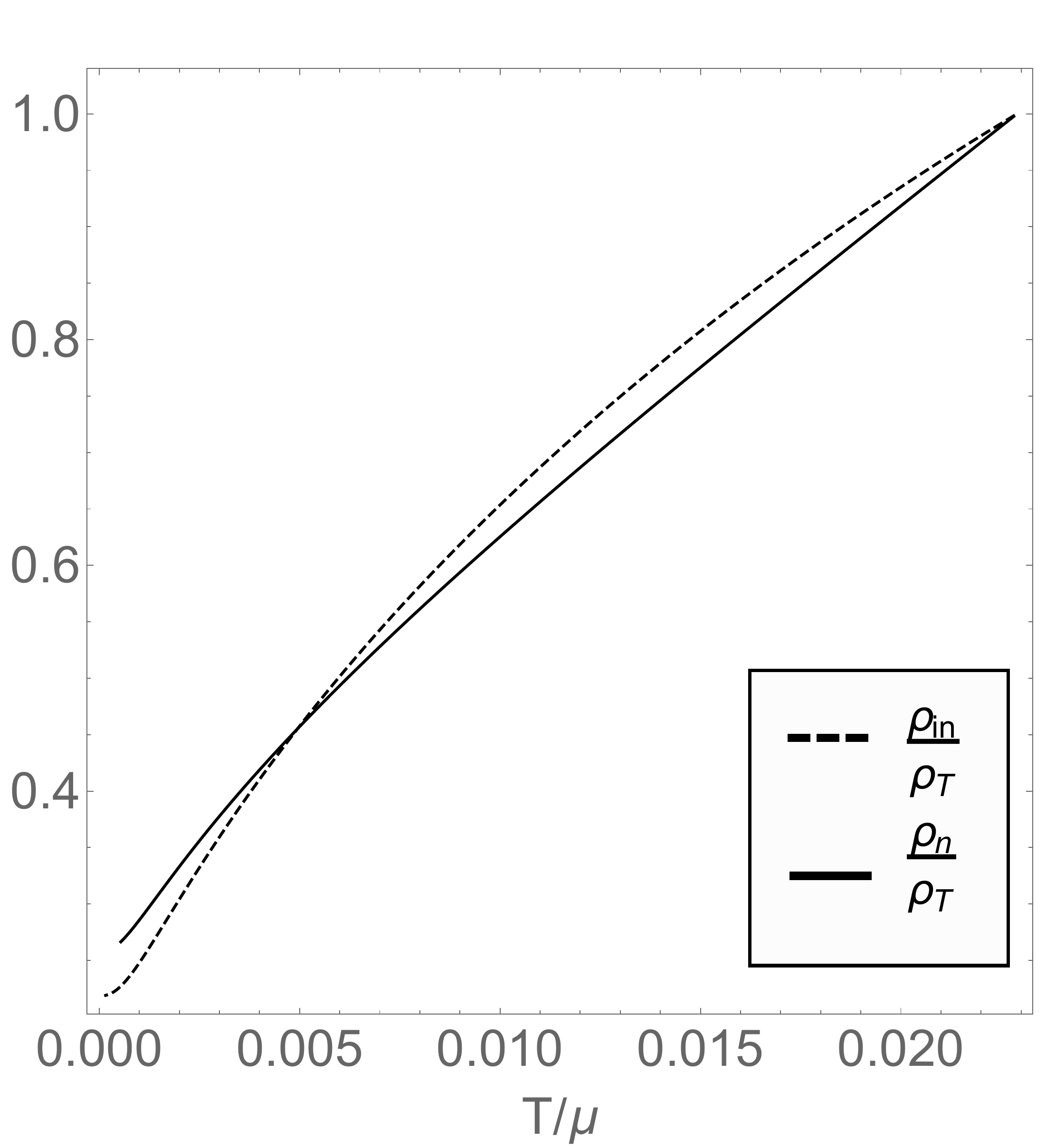}
\caption{$a=1 (z\to \infty, \theta\to -\infty, z/\theta=-1)$: $\sigma_0\sim \#T^{3}$, $\rho_{in}\sim \rho_{in}^{(0)}+\#T^{2}$, $\rho_n \sim \rho_n^{(0)} + \# T^{2}$ \label{a1figurefractionalized}}
\end{center}
\end{figure}

\begin{figure}[h]
\begin{center}
\includegraphics[scale=.2]{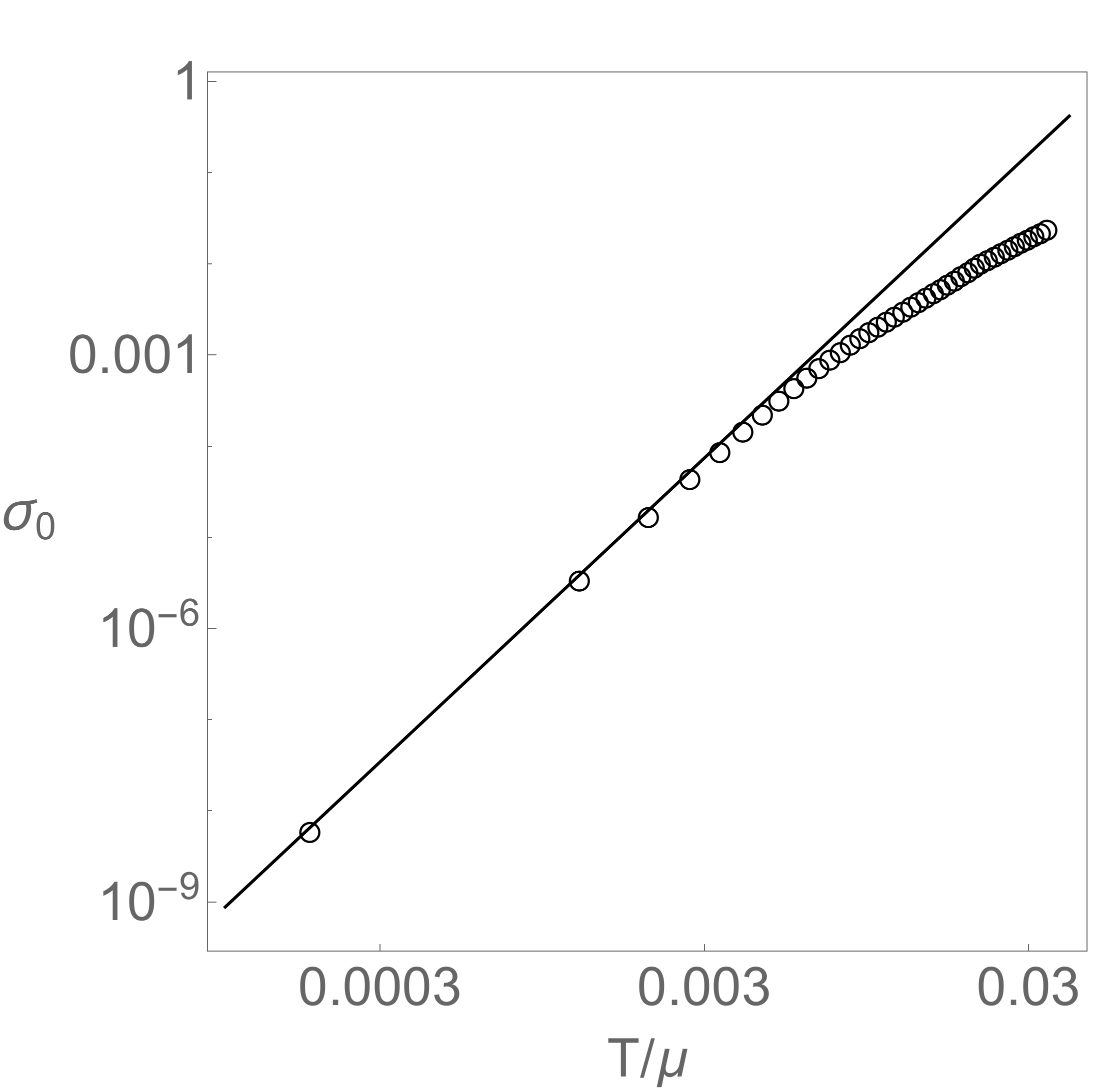}
\includegraphics[scale=.26]{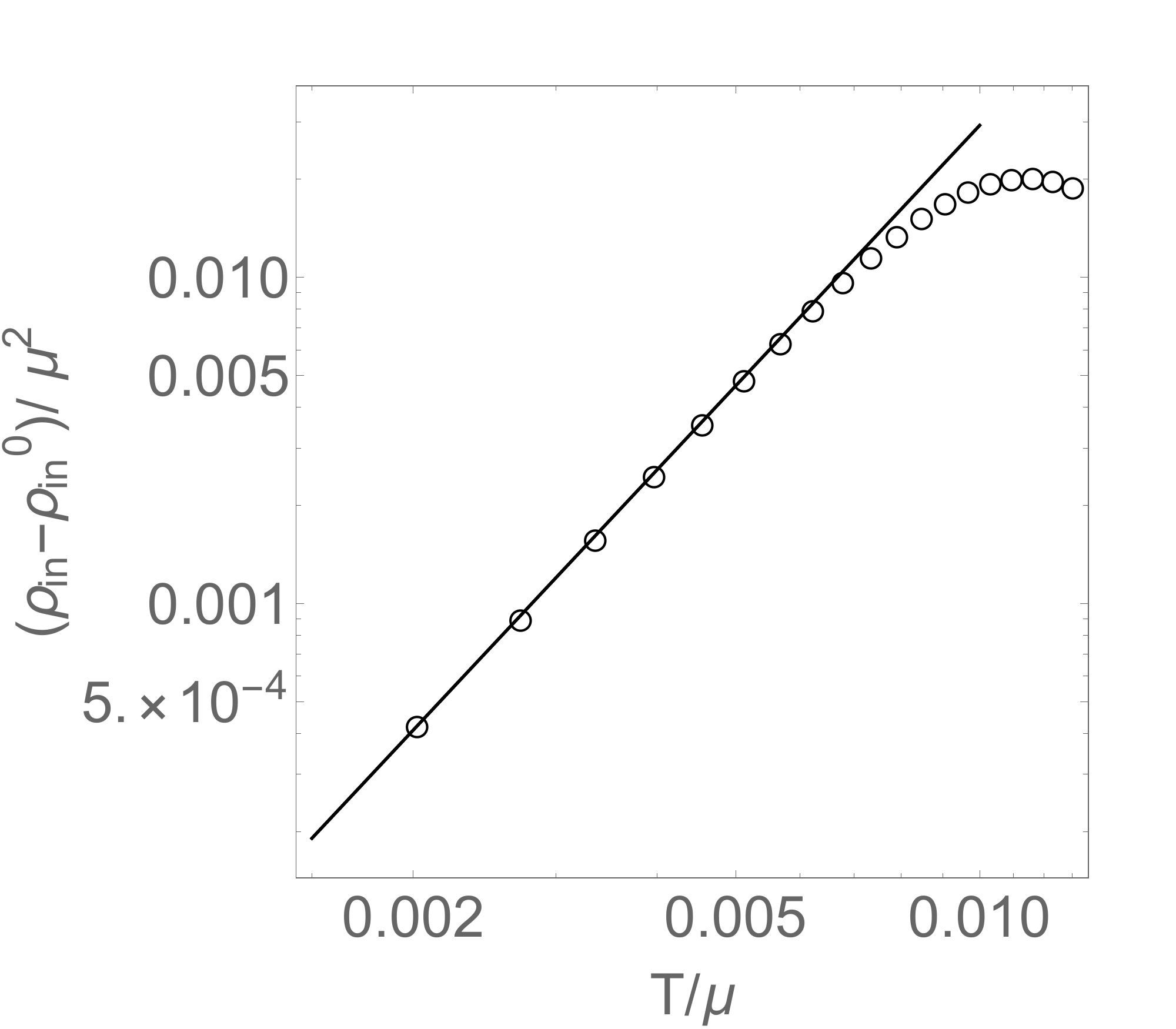}
\includegraphics[scale=.26]{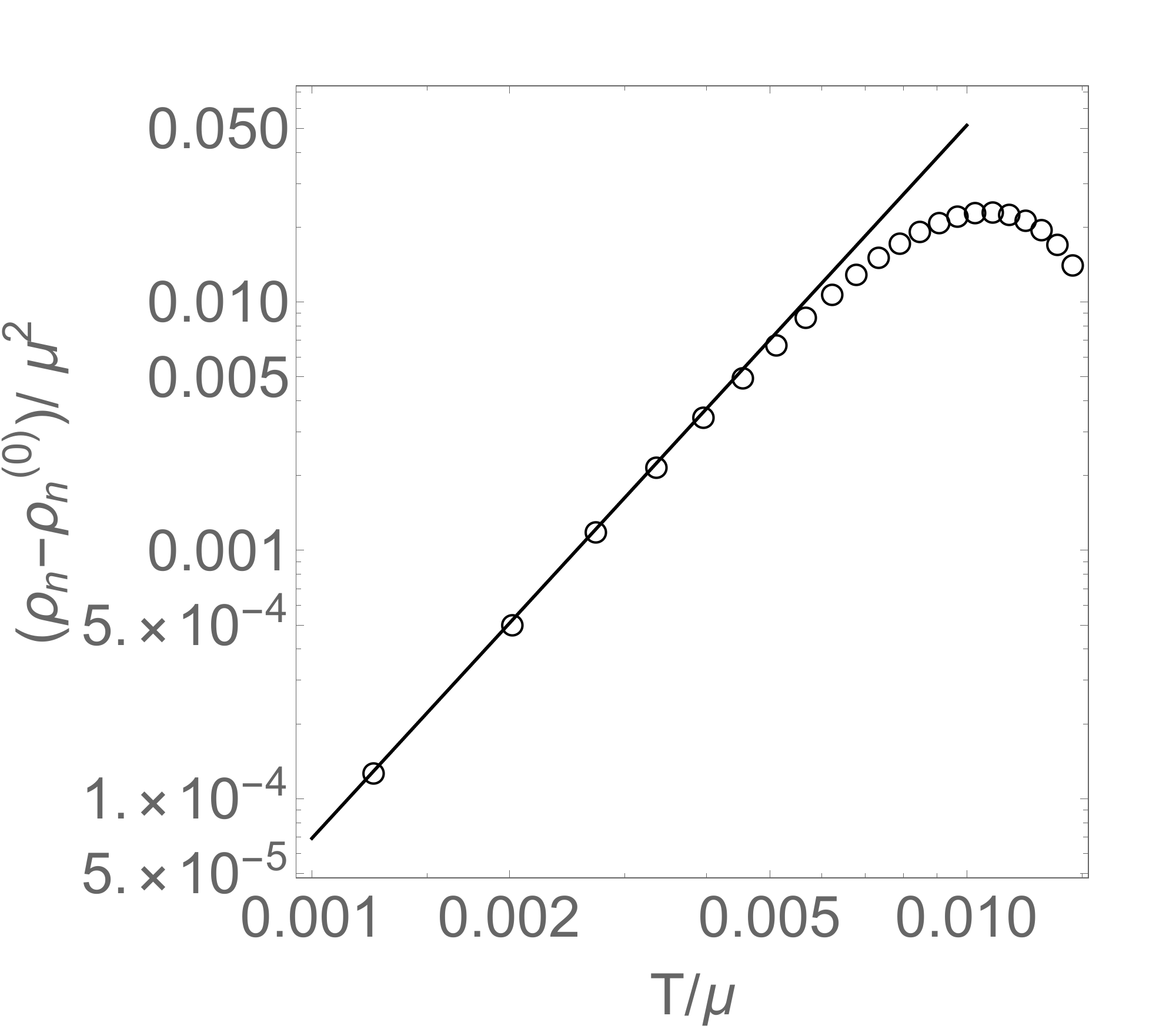}
\newline
\includegraphics[scale=.19]{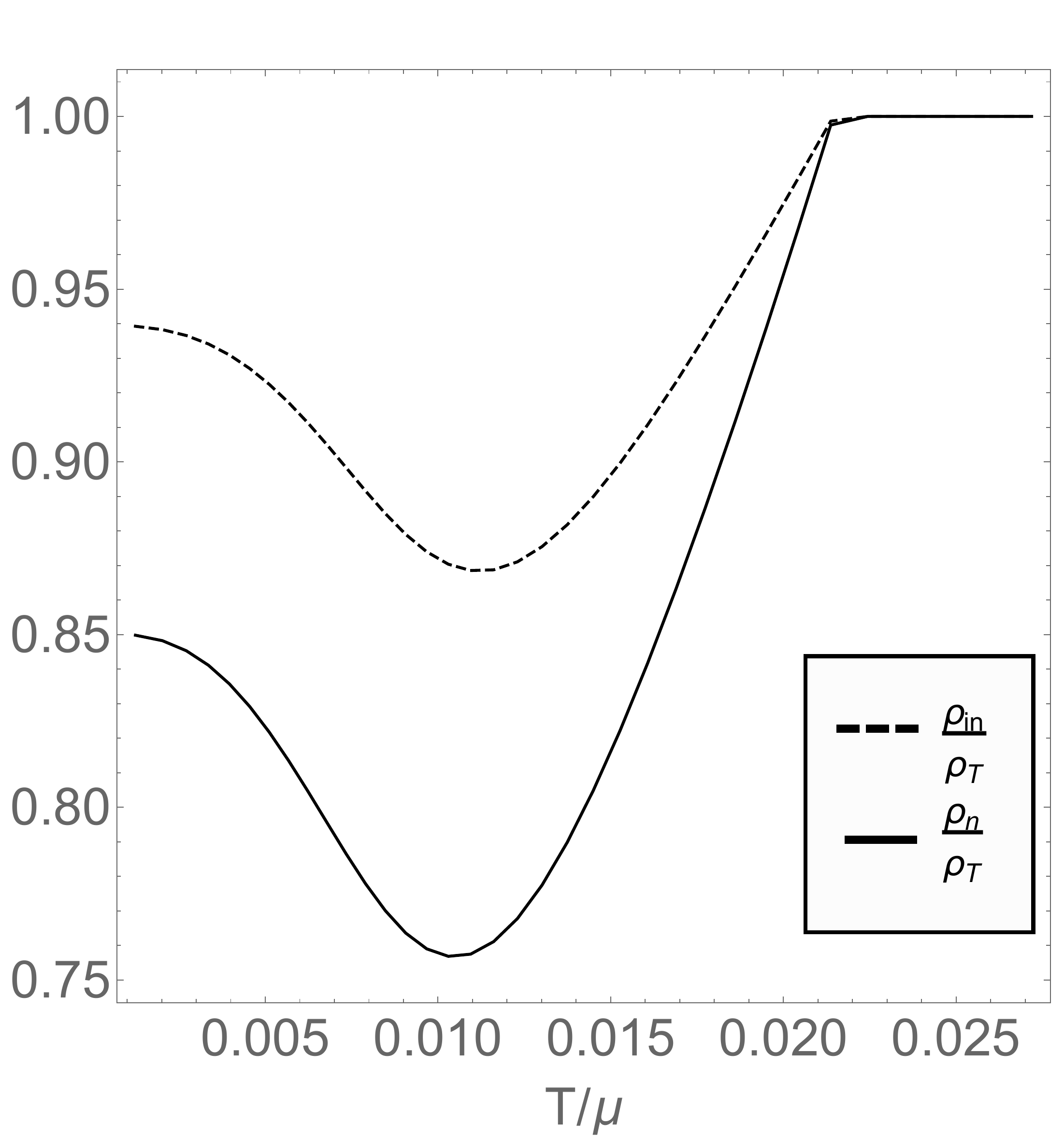}
\caption{$a=2 (z=3, \theta=-4)$: $\sigma_0\sim \#T^{\frac{10}{3}}$, $\rho_{in}\sim \rho_{in}^{(0)}+\#T^{3}$, $\rho_n \sim \rho_n^{(0)} + \# T^{\frac{7}{3}}$ \label{a2figurefractionalized}}
\end{center}
\end{figure}

\subsubsection{Irrelevant deformations, $z=1$}
On the other hand, if $z=1$, the gauge field is irrelevant. Here

\begin{align}
\theta = \frac{d^2\delta^2}{d\delta^2-2},\quad\tilde{L}^2 =\frac{1}{-V_0} \left(d-\theta\right)\left(1+d-\theta\right)
\end{align}

The IR phase can then be thought of as a ``CFT'' in $d-\theta$ dimensions in the presence of an irrelevant deformation with coupling $A_0$. This irrelevant deformation grows toward the IR drives the RG flow to the UV conformal fixed point. The behavior of $A\equiv A_0\phi(\hat{r})dt$ can be determined from solving the Maxwell equation on the background (\ref{IRmetric}) in the presence of a constant condensate $\eta_0$. This is then backreacted on the metric to give an IR solution which can be solved order-by-order in $A_0$. For instance, as recently discussed in \cite{Davison:2018ofp,Davison:2018nxm}, for $\eta = 0$, $A_t(r)$ behaves as a power-law. For instance,
\begin{align}
A_t &= L_t A_0 \left(\frac{\hat{r}}{L}\right)^{\theta-d-1+2\Delta_{A_0}}\left(1+\#A_0^2\left(\frac{\hat{r}}{L}\right)^{2\Delta_{A_0}}+O\left(\hat{r}^{4\Delta_{A_0}}\right)\right),\nonumber\\
2\Delta_{A_0} &=2(d-\theta)-\kappa\gamma+\frac{2}{d}\theta<0.
\end{align}
This gives a metric deformation,
\begin{align}
ds^2 &= ds_{A_0=0}^2\left(1+\#A_0^2\left(\frac{\hat{r}}{L}\right)^{2\Delta_{A_0}}+O\left(\hat{r}^{4\Delta_{A_0}}\right)\right),\nonumber\\
\psi(r) &= \psi_{A_0=0} + \#A_0^2\left(\frac{\hat{r}}{L}\right)^{2\Delta_{A_0}}+O\left(\hat{r}^{4\Delta_{A_0}}\right).
\end{align}
$\\$

When $\eta\neq 0$, the range of $\gamma$ is extended to a regime where $\Delta_{A_0}$ would be positive and we must be more careful. When $\gamma = \delta$, the deformation still behaves as a power law, though the power is different,
\begin{align}
A_t &= L_t A_0 \left(\frac{\hat{r}}{L}\right)^{\tilde{\Delta}_{A_0}-1}\left(1+\#A_0^2\left(\frac{\hat{r}}{L}\right)^{2\tilde{\Delta}_{A_0}}+...\right),\nonumber\\
2 \tilde{\Delta}_{A_0} &= d+1-\theta + (d-1-\theta)\sqrt{1+\frac{8\tilde{L}^2q^2\eta_0^2}{Z_0^2(d-1-\theta)^2}}.
\end{align}
In order for this perturbation to be well-defined, we must have $\tilde{\Delta}_{A_0}<0$ which constrains the value of the condensate,
\begin{align}
\eta_0^2 > \frac{Z_0^2}{8\tilde{L}^2q^2}\left[(d+1-\theta)^2-(d-1-\theta)^2\right]>0
\end{align}
and hence this does not smoothly connect to the $\eta_0\to 0$ case.
$\\$

Interestingly, for $\gamma\neq\delta$, $A_t(r)$ instead decays exponentially. Define $\xi =\gamma/\delta$. For $\xi>1$, there exists a normalizeable solution which to leading order in $(\hat{r}/L)$ has the form
\begin{align}
A_t &= L_t A_0 \left(\frac{\hat{r}}{L}\right)^{\Delta_1}\times \exp\left[-c_\theta \left(\frac{\hat{r}}{L}\right)^{\Delta_2}\right]+...,\nonumber\\
\Delta_1 &= \frac{(d-1)(d-\theta)-\theta\xi}{2d},\quad\Delta_2 = \frac{(1-\xi)\theta}{d},\quad c_\theta^2 = \frac{2d^2q^2\eta_0^2\tilde{L}^2}{Z_0\theta^2(\xi-1)^2}\,.
\end{align}
There is no normalizeable solution for $\xi<1$. Notably, $\Delta_1>0$ and $\Delta_2>0$ for $\theta<0$. The exponential decay is indicative of a gapped system. In particular, the exponential decay in $A_t$ leads to low frequency conductivities that decay exponentially with temperature, reminiscent of s-wave BCS superconductors.
$\\$

We now find the temperature dependence of the various transport coefficients of interest. We will not find the coefficient of the temperature scaling since this depends on the UV parameters through $A_0$, rather we just look at the overall scaling. Recalling that $\xi = \gamma/\delta$,
\begin{align}
\rho_{in} = \begin{cases} \#T^{d-\theta-\tilde{\Delta}_{A_0}}+... &\xi=1\\ \#T^{\Delta_1}\times\exp\left[-\tilde{c}_\theta\left(T\right)^{-\Delta_2}\right]+...&\xi>1\end{cases}
\end{align}
Here, we have rescaled $\tilde{c}_\theta = c_\theta r_0^{\Delta_2}$. Despite the fact that $\Delta_1>0$, we also have $\Delta_2 >0$ so $\rho_{in}^{(0)}$ always vanishes. To derive this temperature scaling, we used the fact that for $\xi>1$,
\begin{align}
\Delta_1 + \Delta_2  - 1 -\frac{d-2}{d}(d-\theta) - \frac{2\xi}{d}\theta = -\Delta_1.
\end{align}
Next, the conductivities are
\begin{align}
\sigma_0=\begin{cases} \#T^{2(1-\frac{\theta}{d}-\tilde{\Delta}_{A_0})}+... & \xi=1,\\
\#T^{\Delta_2-1}\times\exp\left[-2\tilde{c}_\theta\; T^{-\Delta_2}\right]+... &\xi>1.\end{cases}
\end{align}
Finally, these solutions have $z=1$ and have, for any $\xi$,
\begin{align}
\mu\rho_n = \frac{1-c_{ir}^2}{c_{ir}^2}(sT)+...
\end{align}

\begin{figure}[h]
\begin{center}
\includegraphics[scale=.2]{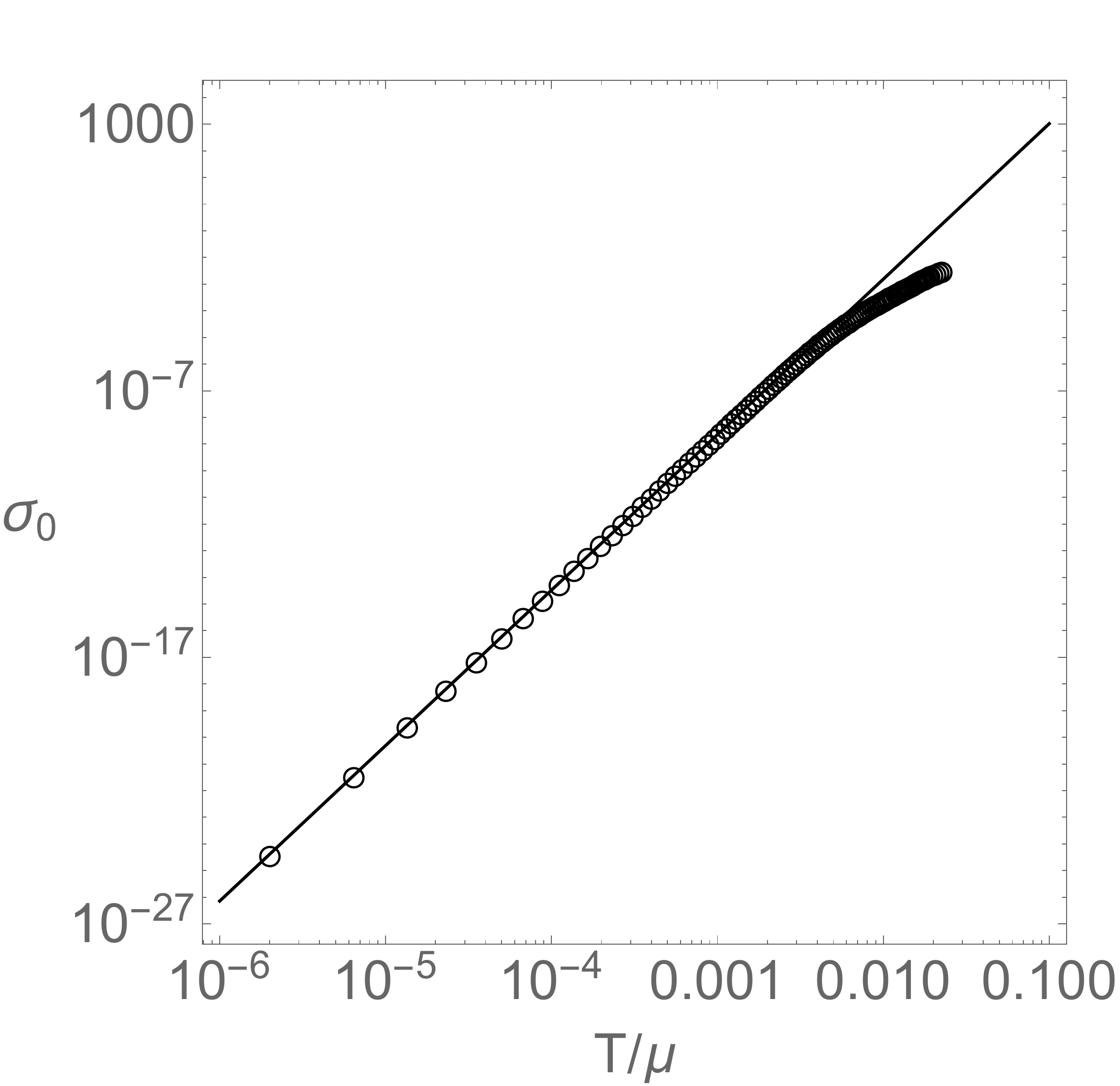}
\includegraphics[scale=.25]{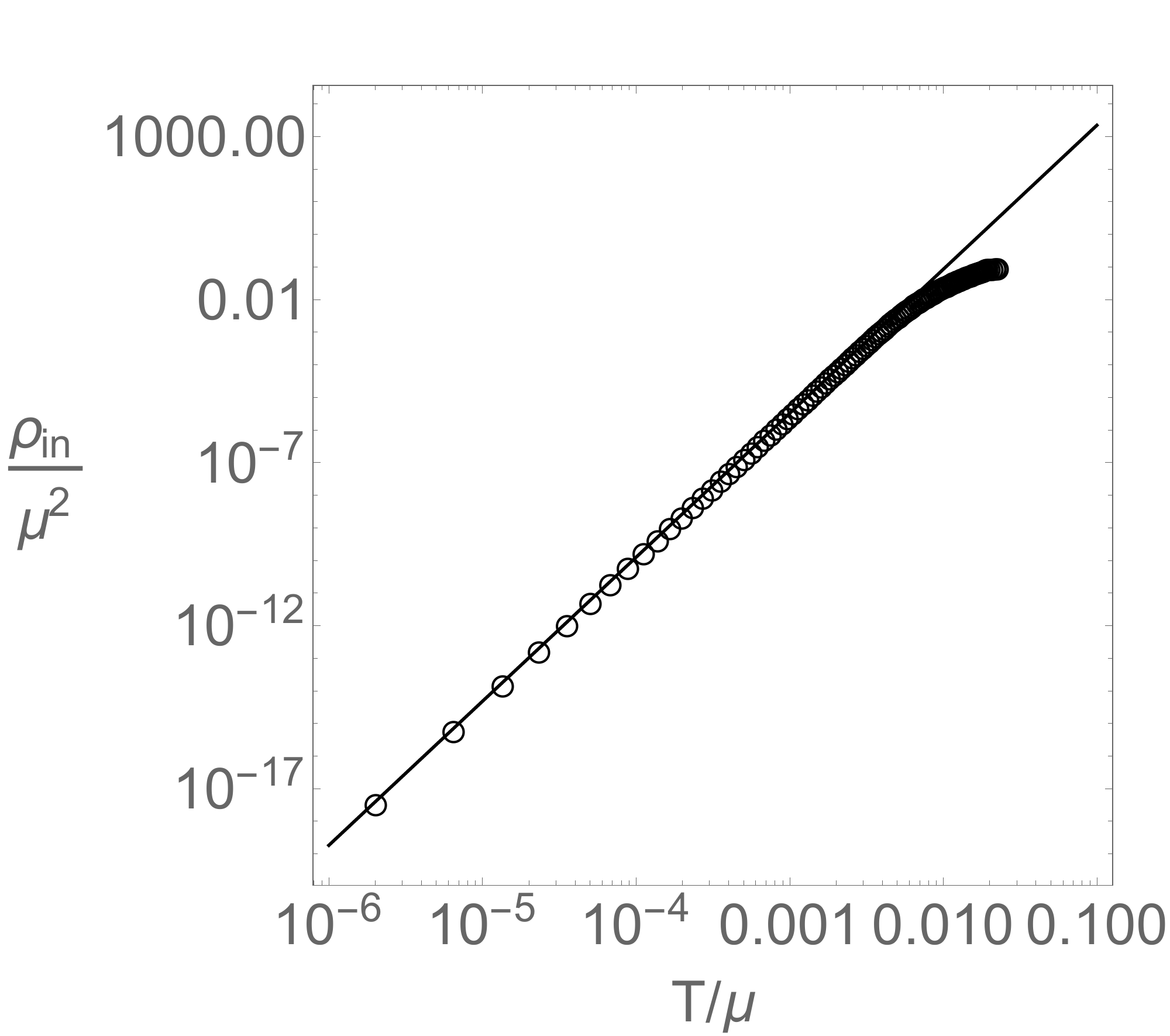}
\includegraphics[scale=.23]{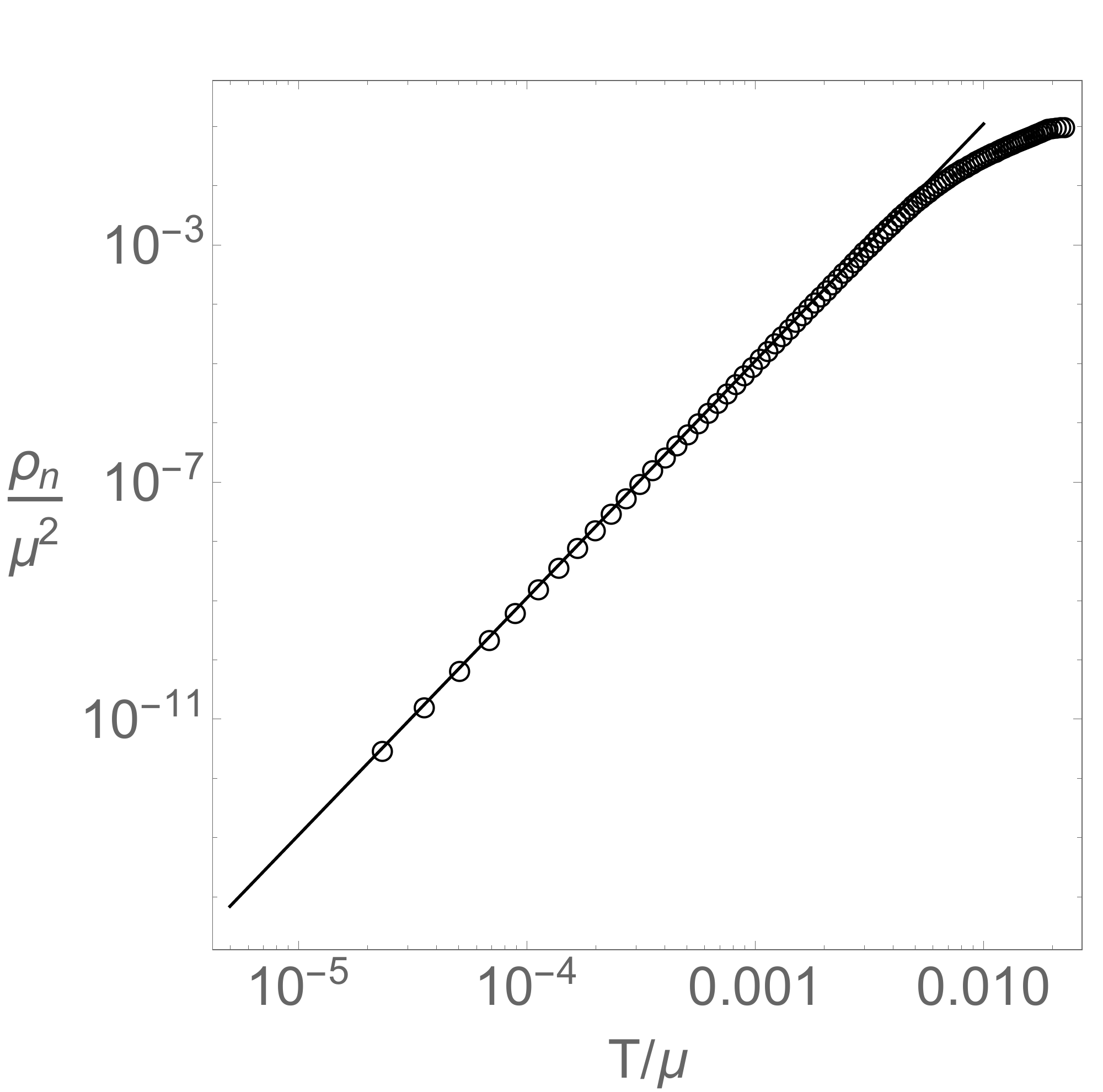}
\caption{$a=1 (z=1,\theta=-1, \xi=1)$: $\sigma_0\sim \#T^{5.83}$, $\rho_{in}\sim \#T^{4.42}$, $\rho_n \sim \# T^{4}$ \label{a1figurecohesive}}
\end{center}
\end{figure}

\begin{figure}[h]
\begin{center}
\includegraphics[scale=.2]{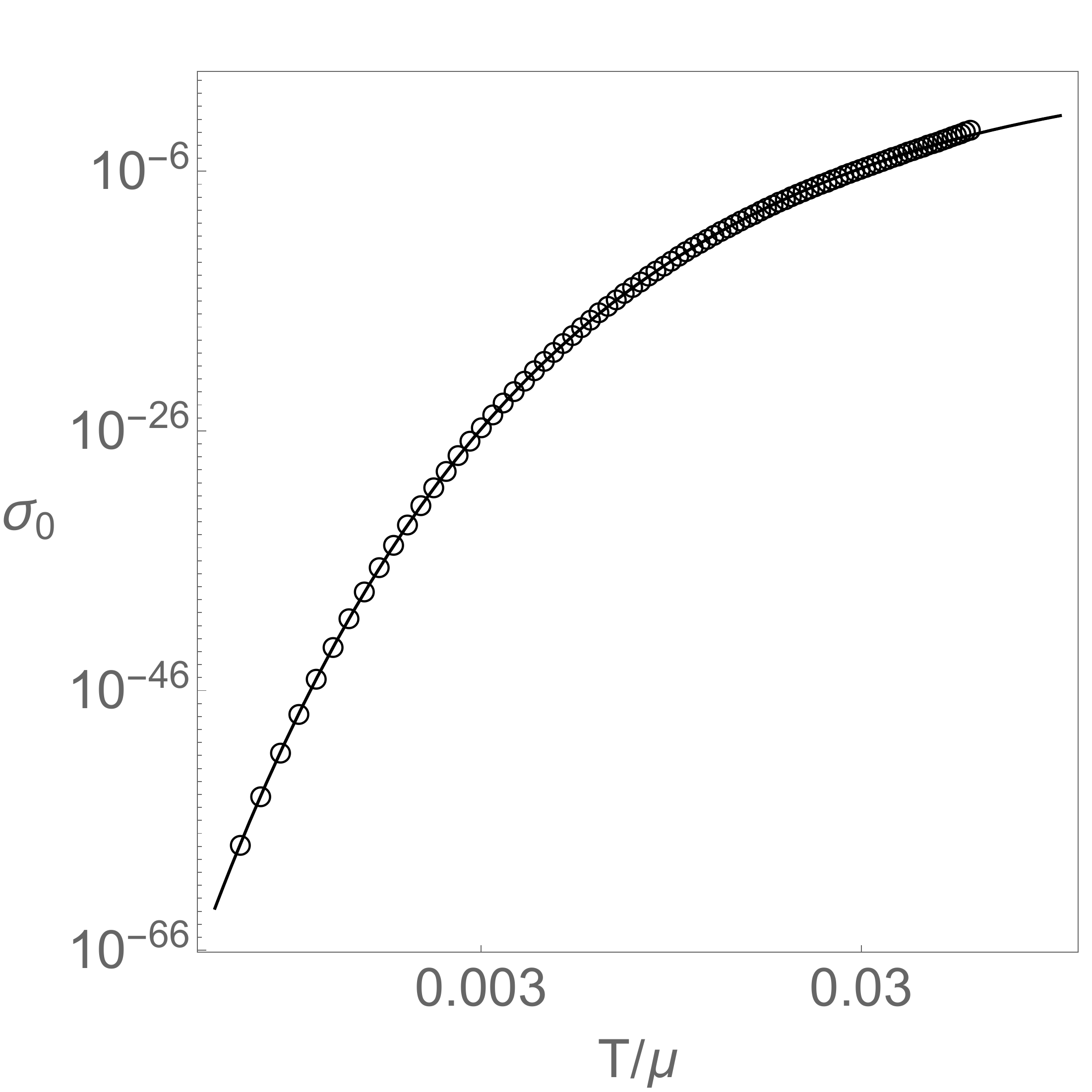}
\includegraphics[scale=.23]{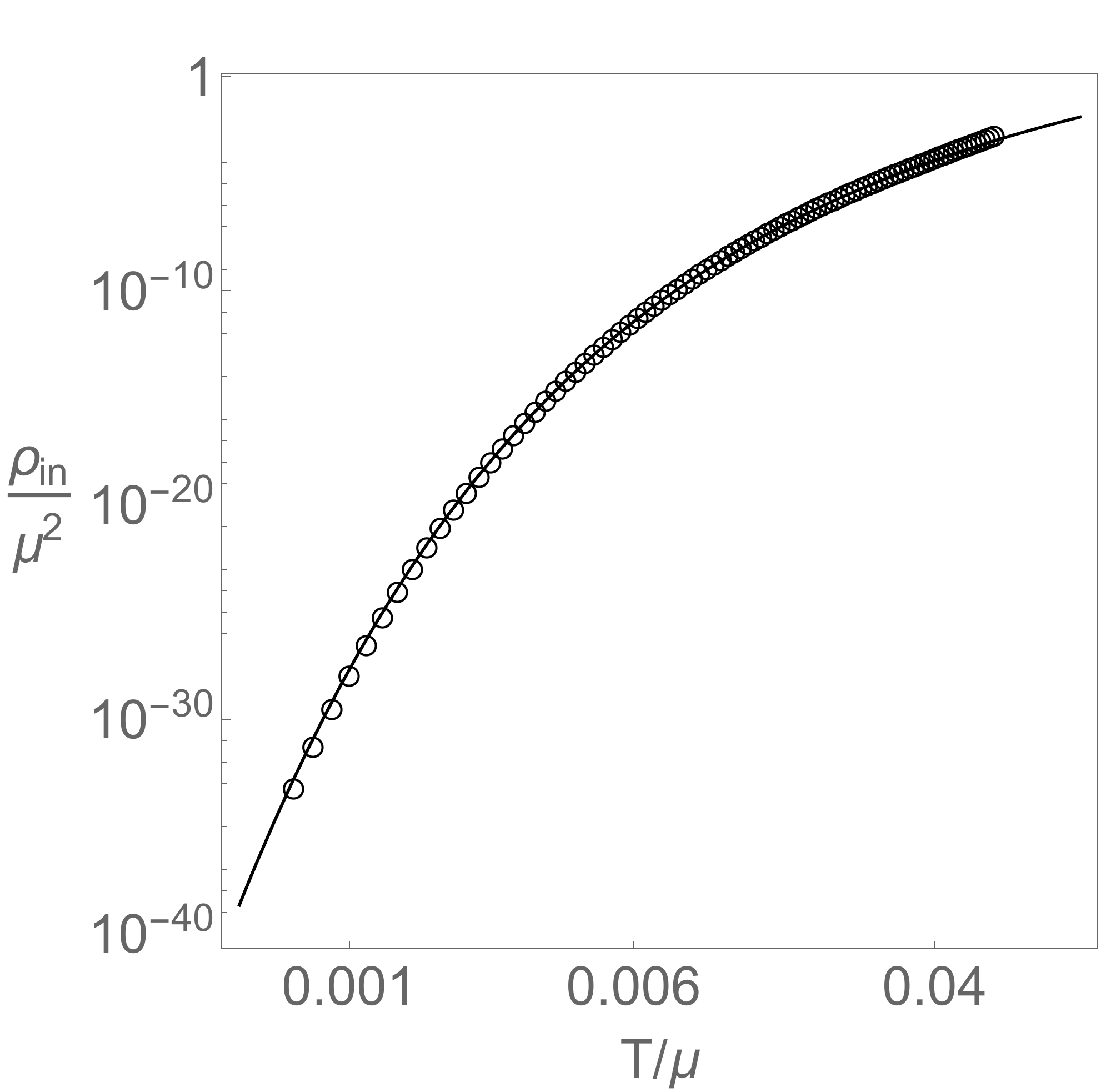}
\includegraphics[scale=.23]{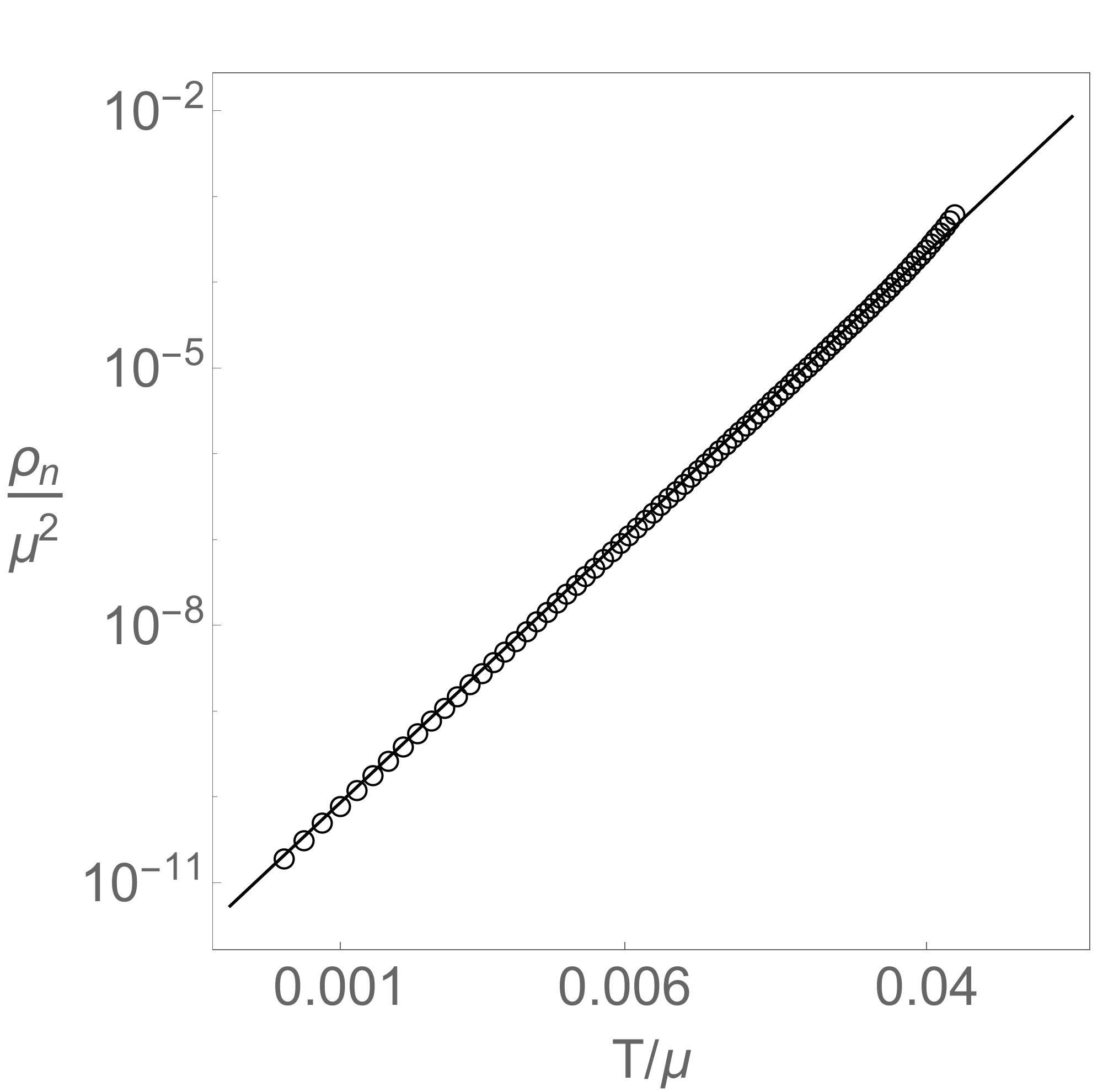}
\newline
\includegraphics[scale=.23]{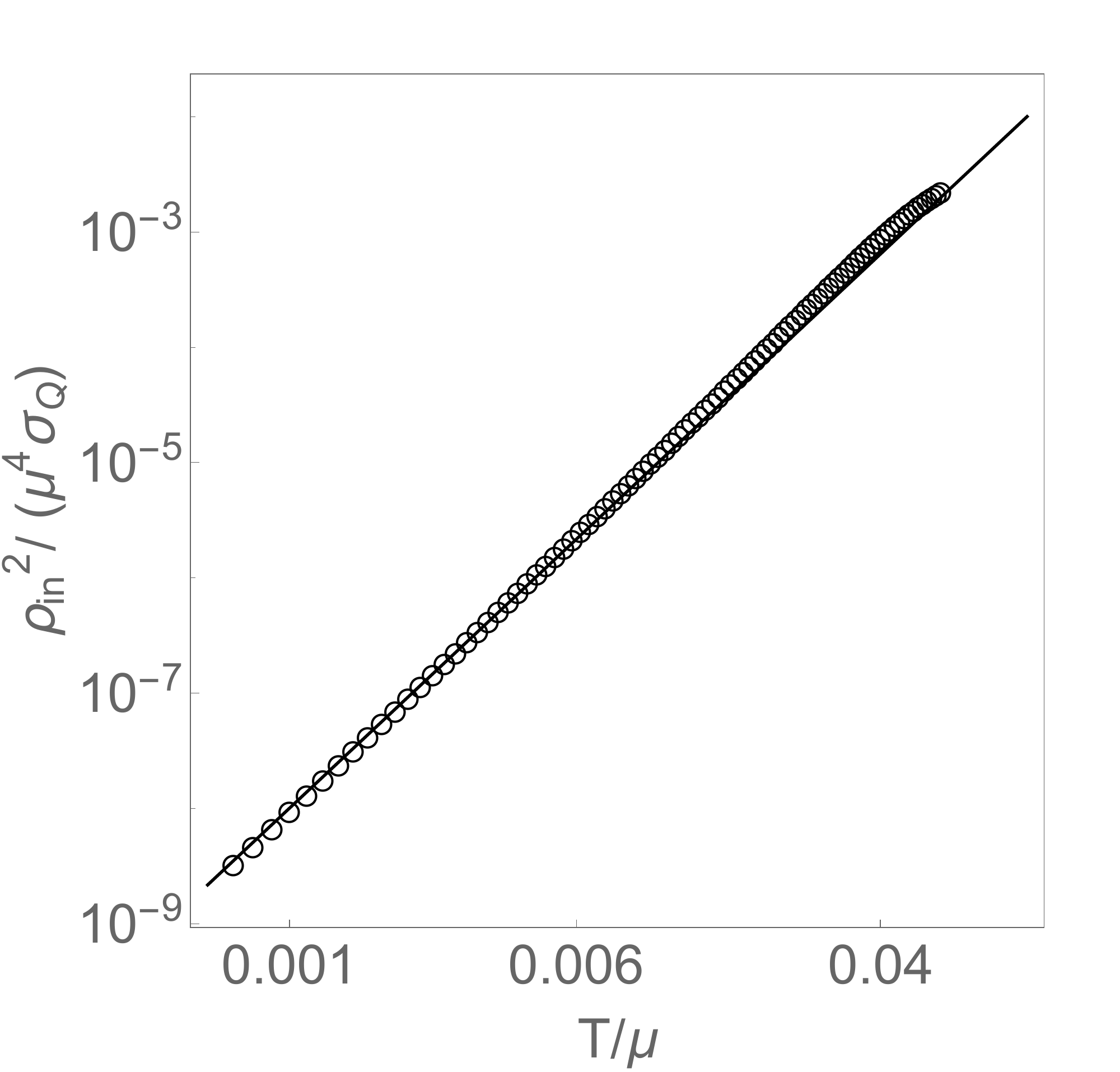}
\caption{$a=2$ $(z=1,\theta=-1,\xi=2)$: $\sigma_0\sim \#T^{-\frac{1}{2}}\times\exp\left[-2\tilde{c}_\theta T^{-\frac{1}{2}}\right]$, $\rho_{in}\sim \#T^{\frac{5}{4}}\times\exp\left[-\tilde{c}_\theta T^{-\frac{1}{2}}\right]$, $\rho_n \sim \# T^{4}, \rho_{in}^2/\sigma_0 \sim \#T^3$ \label{a2figurecohesive}}
\end{center}
\end{figure}

\subsection{Specific example}
As a specific example, we will use the model of \cite{Adam:2012mw} in $d=2$,
\begin{align}
\label{fractionalizedexample}
Z_F(\psi) = Z_0 e^{a\psi/\sqrt{3}},\quad V(\psi,|\eta|) = -6\cosh(\psi/\sqrt{3})-2|\eta|^2 + |\eta|^4
\end{align}
which gives
\begin{align}
z = \frac{12+(a-3)(a+1)}{a^2-1},\quad\theta = \frac{4}{1-a}
\end{align}
for partially condensed phases and 
\begin{align}
z=1,\quad\theta=-1
\end{align}
for fully condensed phases. By turning on a source for the dilaton, $\psi^{(1)}\neq 0$, one can flow to either the partially condensed or full condensed phases in the IR. A qualitative understanding of this behavior is that $Z_F(\psi)$ controls the effective $U(1)$ charge of the complex scalar in the IR. If $Z_F\to 0$, the charge diverges and it becomes easier for the scalar to condense. On the other hand, if $Z_F\to \infty$, the charge vanishes and it is more difficult for the scalar to condense. In fact, there is a special value of $\psi^{(1)}$ above which it is not possible to condense the scalar at $T=0$ and this model will flow to an uncondensed phase. Since $\eta$ is an irrelevant deformation of the IR, the uncondensed and partially condensed phases are characterized by the same $z,\theta$. We do not consider uncondensed phases in this work.
$\\$

In \cite{Gouteraux:2019kuy}, we show how this source controls the magnitude of $\rho_n^{(0)}$ for $a=b=1$. Here, we care only about the temperature scaling, so we choose a value of $\psi^{(1)}$ that leads either to a partially condensed phase in figures \ref{a1figurefractionalized} and \ref{a2figurefractionalized} or to a fully condensed phase in figures \ref{a1figurecohesive} and \ref{a2figurecohesive}.


\section{Translation breaking \label{translationbreaking}}

While we will not go into detail, we note that we can extract the normal and superfluid scaling more straightforwardly by explicitly breaking translations. This is done by modifying the action \ref{bulkactionwithdilaton}
\begin{align}
S' = S + \int d^{d+2}x \sqrt{-g} \frac{Y(\psi)}{2}\sum_{i=1}^{d}\partial_\mu\chi_i\partial^\mu\chi^i
\end{align}
with 
\begin{align}
Y(\psi)=Y_0 \exp(\lambda\psi),\quad\quad \chi_i = m x_i.
\end{align}
This deformation of the action breaks translations homogeneously and allows for the equations of motion to maintain dependence only on the bulk radial coordinate \cite{Andrade:2013gsa}. We choose the parameter $\lambda$ such that this is an irrelevant deformation of the IR critical phase. Hence, our earlier analysis of the IR geometries does not change.

\begin{figure}[h]
\begin{center}
\includegraphics[scale=.3]{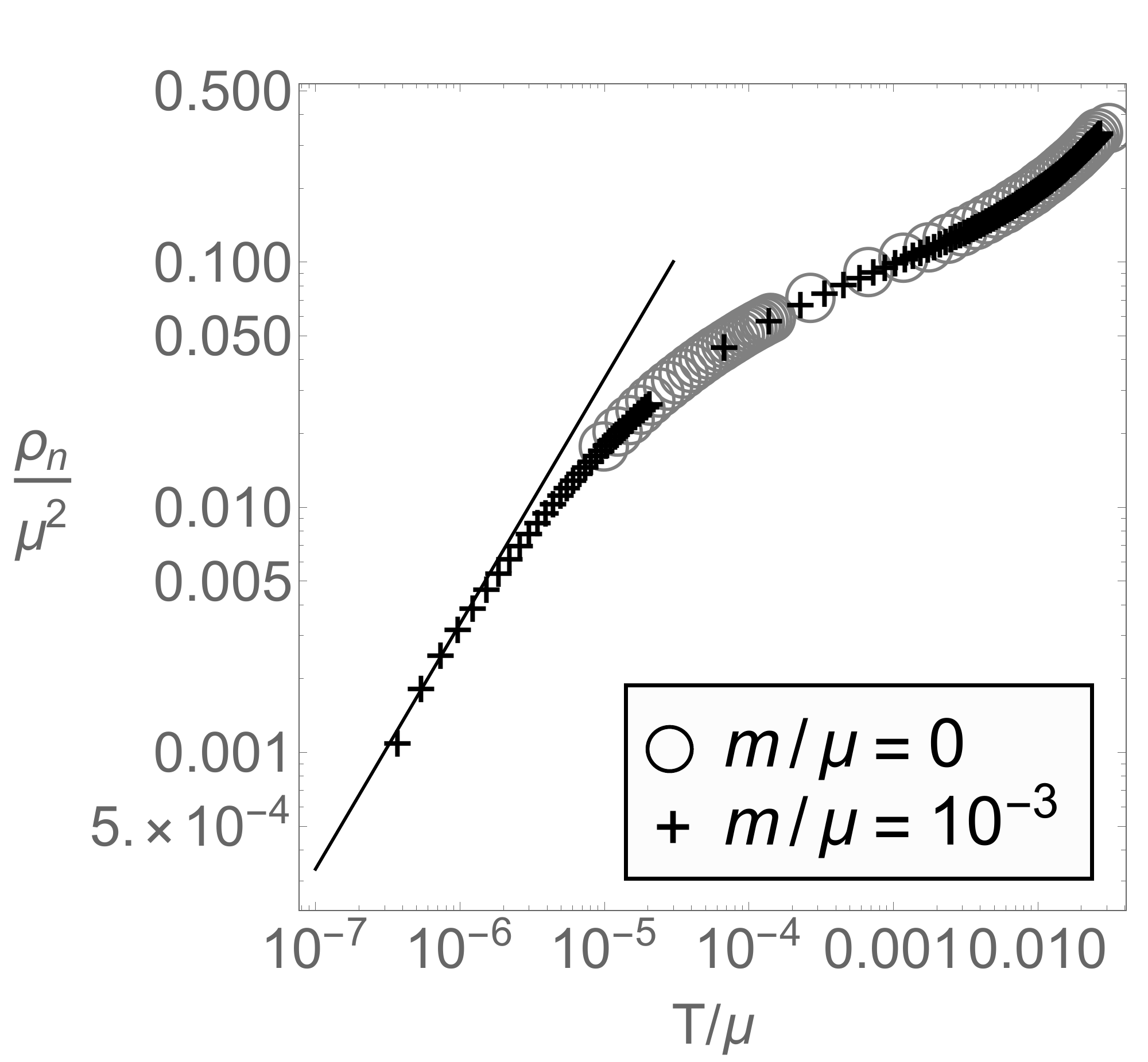}
\caption{Breaking translations allows for lower precision computing and we can find $\rho_n$ to lower temperature. Here, for $z=2$, we plot our result with $m/\mu=0$ in gray and for $m/\mu=10^{-3}$ in black. The points exactly overlap and we confirm $\rho_n \sim \#T$.}
\end{center}
\end{figure}

In the hydrodynamic model, this choice of translation breaking is reflected in the introduction of a dissipation timescale, $\tau$, that only couples to the momentum. The timescale will depend on the function $Y(\psi)$ and the parameter $m$, but we will leave its derivation to later work. For our purposes, the effect of this dissipation is to broaden the normal contribution to the zero frequency pole in the imaginary conductivity, so that the remaining weight only depends on the superfluid. 

\begin{align}
\sigma(\omega) = \frac{i}{\omega}\left[\frac{\rho_s}{\mu}\right] + \frac{\sigma_0}{1-i\omega/\tau}
\end{align}

When breaking translations, care must be taken to work in terms of gauge invariant contributions of the fluctuating fields. Some details are outlined in \cite{Andrade:2013gsa} and in \cite{Gouteraux:2019kuy}. After carefully accounting for these, we can holographically compute the optical conductivity and extract the superfluid density. We confirm that the hydrodynamic expression holds. Furthermore, we find that our earlier analysis of the normal and superfluid densities does not change; that is, the leading temperature dependence of these quantities is fully determined by the IR geometries. For clean holographic superfluids, numerically extracting the normal and superfluid densities requires high numerical precision and accuracy due to subtle cancellations in \eqref{rhonZdefinition}. Translation breaking allows for lower precision computing and nicely confirms our results in clean superfluids.

\section{Semi-local quantum critical geometries \label{semilocal}}
In the main text, we stated that $z\to \infty$ is a boundary case of the Lifshitz spacetimes. It is clear from our general results for $z>1$ that the $z\to\infty$ limit is well behaved for $\rho_{in}$ and $\rho_n$. However, if one looks closely at the conditions for existence of a Lifshitz IR, one finds that $z\to\infty$ implies $q=0$. Hence, no superfluid can exist for this case. 
$\\$

Instead, start with a more general action that has a canonically normalized scalar kinetic term but a modified mass for the gauge field.
\begin{align}
S = \frac{1}{16\pi G}\int d^{d+2}x\sqrt{-g}\left\{R-\frac{1}{2}(\partial\eta)^2 - V(\eta) - \frac{W(\eta)}{2}A^2 - \frac{Z(\eta)}{4}F^2\right\}.
\end{align}
Here $\eta$ is a real scalar field, analogous to the modulus we chose in the main text. For $\eta\to \eta_0$ in the IR, if we have $W(\eta_0) = W'(\eta_*) = 0$, an $AdS_2\times R^d$ solution is possible. For simplicity,
we will choose $V'(\eta_0) = 0$. 
$\\$

A simple class of functions $W(\eta)$ that at lowest order in $\eta$ resembles a superfluid action is
\begin{align}
W(\eta) = q^2\eta^2(1-\frac{\eta^2}{\eta_0^2})^a
\end{align}
with $a>1$. It will turn out that perturbations are best behaved for $a\geq 2$. We would like this 
action to arise from a $U(1)$ invariant effective holographic action for a complex scalar $\zeta = \chi e^{i\theta}$,
\begin{align}
\frac{G(|\zeta|)}{2}|\mathcal{D}\zeta|^2 = \frac{G(\chi)}{2}\left[(\partial\chi)^2+\chi^2(\partial_\mu\theta+qA_\mu)^2\right].
\end{align}
We want $\chi$ and $\eta$ to be related, so
\begin{align}
G(\chi)(\partial\chi)^2 = (\partial \eta)^2\quad\text{and}\quad q^2G(\chi)\chi^2=W(\eta)
\end{align}
For $a=2$, which we use for figure \ref{semilocalfigure},\begin{align}
\chi = a_\eta\frac{\eta}{\sqrt{\eta_0^2-\eta^2}},\quad\eta = \eta_0\frac{\chi}{\sqrt{\chi^2+a_\eta^2}},\quad G(\chi) = \eta_0^2\frac{a_\eta^4}{(\chi^2+a_\eta^2)^3}.
\end{align}
If we want $\eta\approx\chi$ and $G(\chi)\to 1$ as $\chi\to 0$, then we can choose $a_\eta=\eta_0$.
$\\$

We now look at solutions to the Einstein equations with this action. We choose $Z(\eta) = 1$ for simplicity. As we said before, there is a solution which is $AdS_2\times R^d$,
\begin{align}
ds^2 = \frac{L^2}{\tilde{r}^2}\left[-L_t^2dt^2 + \frac{d\tilde{r}^2}{V_0}\right] + L_x^2d\vec{x}_d^2
\end{align}
where $V_0 = L^2V(\eta_0)$ and $V'(\eta_0)=0$. The gauge field is
\begin{align}
A = L_t\sqrt{2}\frac{1}{\tilde{r}}dt.
\end{align}
Here, $\tilde{r}\to \infty$ is the IR boundary.
$\\$

Denoting $\vec{X} = \{D,B,C,A_t\}$, the perturbations all have the form
\begin{align}
\eta=\sim \eta_0 + c_\eta\left(\frac{\tilde{r}}{L}\right)^{\tilde{\nu}_\eta}+...,\quad \vec{X} = \vec{X}_{c_{\eta}=0}\left[1+c_{\vec{X}}\left(\frac{\tilde{r}}{L}\right)^{\tilde{\nu}_\eta}+...\right]
\end{align}
There exist four solutions to these equations with the following exponents
\begin{align}
\tilde{\nu}_{\eta_1} = 0 \quad \tilde{\nu}_{\eta_2} = 2, \quad \tilde{\nu}_{\eta}^{\pm} = \frac{1}{2}\left[1\pm\sqrt{1-4\lambda}\right]
\end{align}
with
\begin{align}
\lambda=\frac{-V''(\eta_0) + W''(\eta_0)}{-V(\eta_0)}
\end{align}
If $\lambda > 0$, there is only one real irrelevant perturbation, $\tilde{\nu}_{\eta_1}$. If $\lambda>\frac{1}{4}$ then there is a set of complex perturbations, indicating a potential instability. Finally, if $\lambda < 0$, then there are two irrelevant perturbations. The least irrelevant of these perturbations is always $\tilde{\nu}_{\eta_1}$ though this will not introduce any new temperature dependence. The perturbation with $\tilde{\nu}_{\eta_2}$ is associated with introducing a finite temperature horizon and requires $c_\eta = 0$. 
$\\$

Following the main text, we can choose a potential in $d=2$,
\begin{align}
V(\eta) = -\frac{6}{L^2} -\eta^2 + \frac{g_\eta^2}{4}\eta^4,\quad \eta_0 = \pm\frac{\sqrt{2}}{g_\eta}.
\end{align}
Then,
\begin{align}
V(\eta_0) = -\frac{6}{L^2} - \frac{1}{g_\eta^2},\quad V''(\eta_0) = 4.
\end{align}
For $a=2$, 
\begin{align}
W_*'' = 8q^2.
\end{align}
Using $L=1$ and $g_\eta = \sqrt{\frac{3}{2}}$ to align with the treatment of Lifshitz phases in the main text,
\begin{align}
\lambda = -\frac{3}{5}(1-2q^2)
\end{align}
and we have a complex perturbation for $q>\frac{\sqrt{3}}{2}$. For smaller $q$, there is an extra irrelevant perturbation which is always less relevant than the universal irrelevant deformation, $\tilde{\nu}_{\eta_1}$. For $a>2$, $W''(\eta_0) = 0$ and $\lambda < 0$ so we have two irrelevant perturbations. Finally, for $a<2$, $W''(\eta_0)$ diverges. 
$\\$

The temperature dependence of the transport coefficients are then
\begin{align}
\sigma_0 &= \# T^2\nonumber\\
\rho_{in} &= \rho_{in}^{(0)} + \# T^{-\tilde{\nu}_\eta^{-}}+...\nonumber\\
\rho_{n} &= \rho_n^{(0)} + \#T +...
\end{align}

\subsection{Zero temperature solution \label{app:T=0}}
A nice feature of the semi-local quantum critical geometries is that they readily allow for constructing numerical solutions at exactly zero temperature since the zeroes of the metric are only second order. Using pseudospectral methods, this double zero can be analytically implemented. To motivate the necessity for developing a zero temperature effective field theory for Lifshitz superfluids, we would like to confirm that the zero temperature limit thermodynamic quantities derived in the main text, for instance $\rho_n^{(0)}$, agree with what we find at exactly zero temperature. But there is a subtlety. From the translation invariant hydrodynamics discussed in Appendix \ref{hydrosection}, the weight of the pole in the imaginary conductivity is always $\rho^{(0)}/\mu$ independent of the critical exponents. A physical rationale for this is that, directly at zero temperature, the translation invariant system has no dissipation and no way to distinguish $\rho_n^{(0)}$ from $\rho_s^{(0)}$. To distinguish the two, we can introduce a finite superfluid velocity as in \cite{Bhattacharya:2011tra} or break translations as in \cite{Gouteraux:2019kuy}. We choose that latter and introduce neutral translation breaking scalars as in Appendix \ref{translationbreaking}. Effectively, this changes the conductivity
\begin{align}
\sigma(\omega) = \frac{i}{\omega}\frac{\rho^{(0)}}{\mu} \to \frac{i}{\omega}\frac{\rho_s^{(0)}}{\mu} + \frac{ \sigma_{DC}^{(0)}}{1 - i\omega/\Gamma}
\end{align}
Furthermore, since $\sigma_0\sim T^2$, the incoherent conductivity does not contribute. Notably, the normal density contributes only to the real part of the conductivity after translation breaking, though there may be extra contributions to the coherent sector \cite{Davison:2015bea}. These will in general be subleading in the strength of translation breaking. In general, at finite temperatures $\Gamma \propto (m^2s)^{-1}$ for the translation breaking parameter $m$ of Appendix \ref{translationbreaking}. For $1\leq z<\infty$, this diverges as $T\to 0$. For semi-local quantum critical states, $s\to s^{(0)}$ and $\Gamma^{(0)}$ remains finite and nonzero, which lets us access a zero temperature regime where $\Gamma^{(0)}\ll\mu$ provided $m\ll\mu$. In this regime, $m$-dependent corrections to the values of $\rho_{n,s}^{(0)}$ should be negligible, and we expect to be able to confirm that their zero temperature values agree with their low temperature limit in the translation-invariant case. In addition, we should see a measurable transfer of spectral weight to the real part of the conductivity at $T=0$ if $\rho_n^{(0)}\neq 0$. This was used in $z\to \infty, \theta\to -\infty$ geometries in \cite{Gouteraux:2019kuy}.
$\\$

\begin{figure}[h]
\begin{center}
\includegraphics[scale=.47]{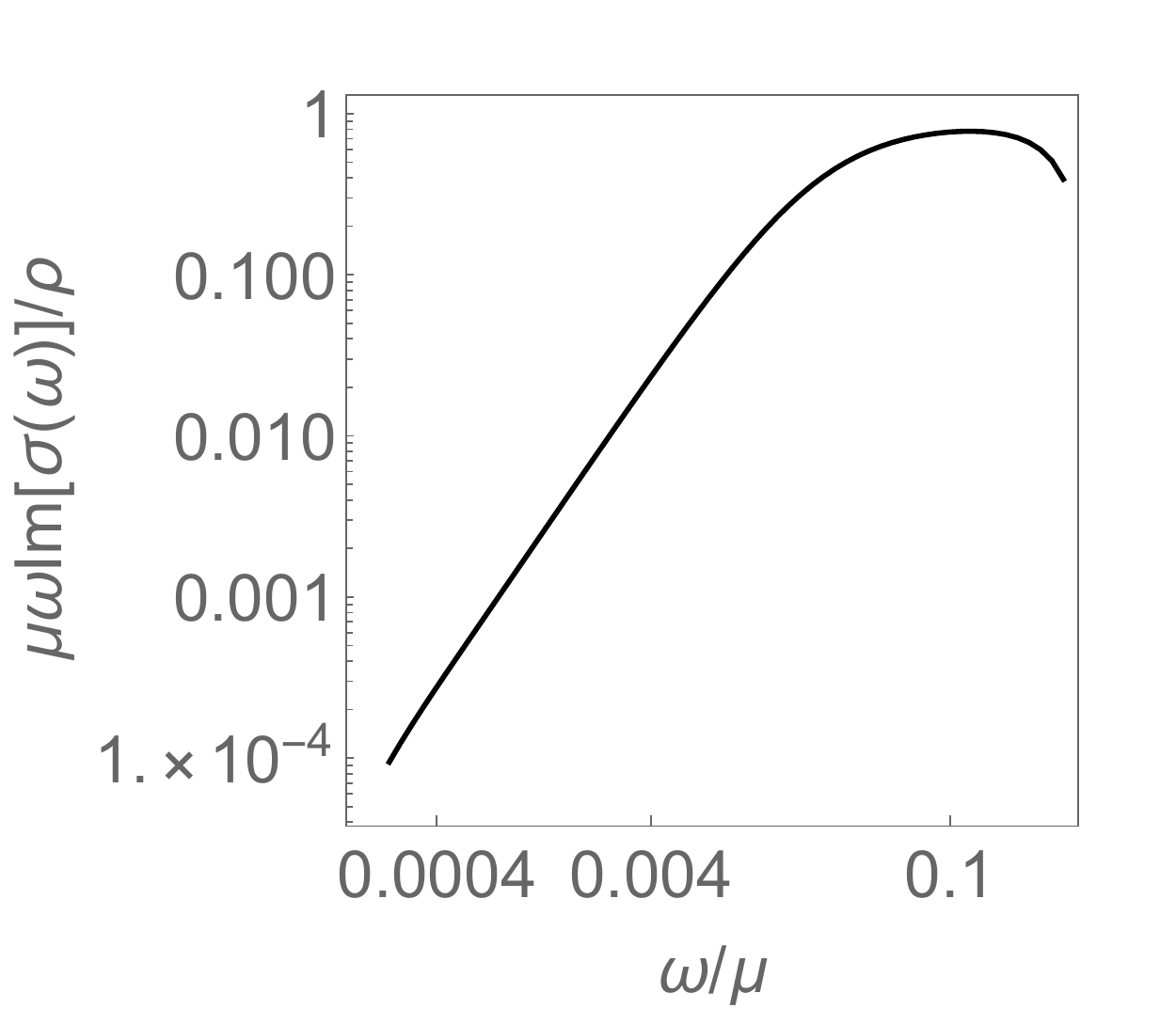}
\includegraphics[scale=.4]{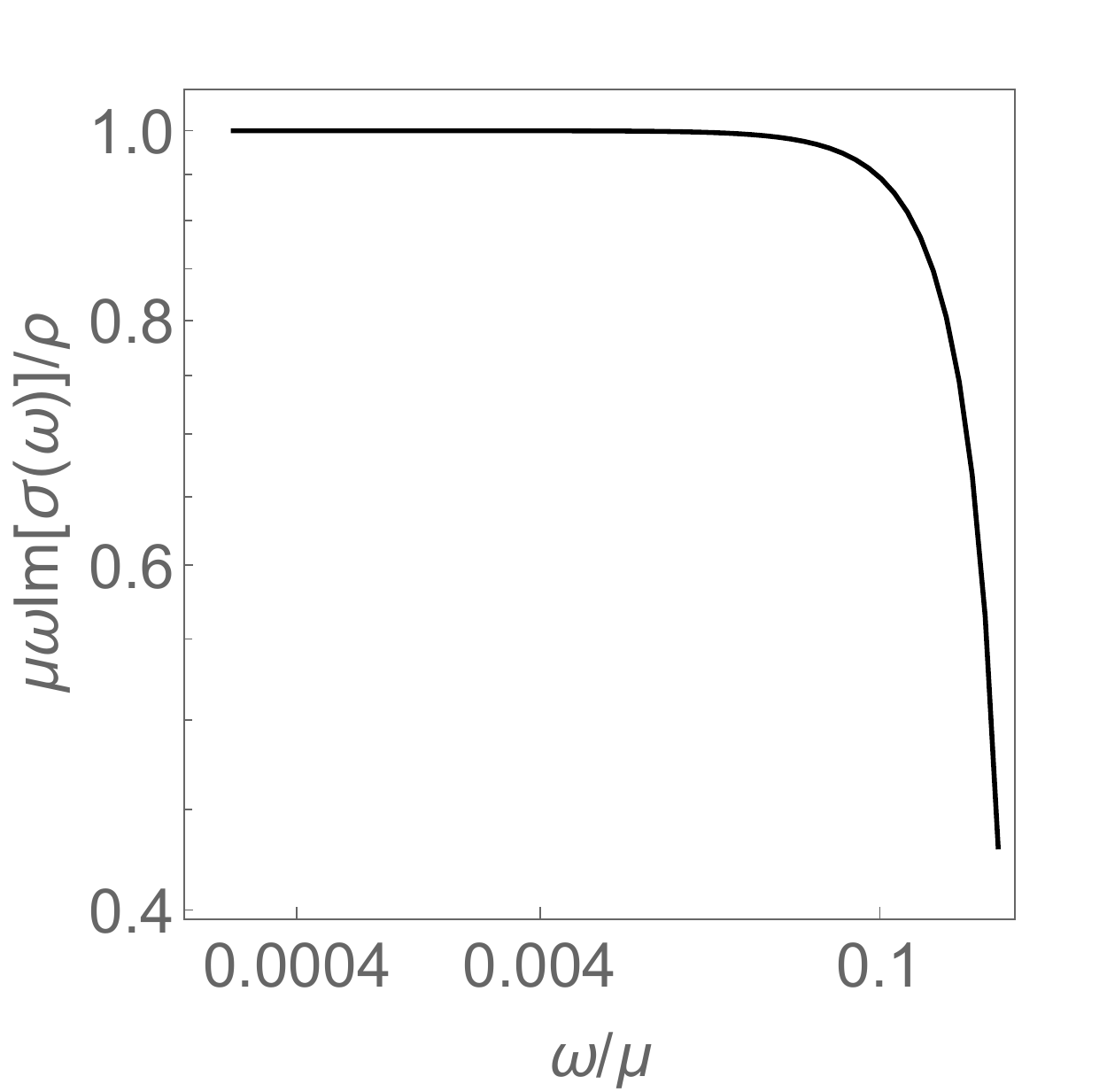}
\caption{\label{AdSRNresults} Complete depletion of spectral weight in the $T=0$ AdS-RN conductivity from translation breaking. (Left) Weight of the $T=0$ imaginary pole with $m/\mu=1/\sqrt{10}$ showing complete depletion of spectral weight. (Right) Weight of the $T=0$ imaginary pole with $m=0$ showing no depletion of spectral weight.}
\end{center}
\end{figure}

\begin{figure}[h]
\begin{center}
\includegraphics[scale=.4]{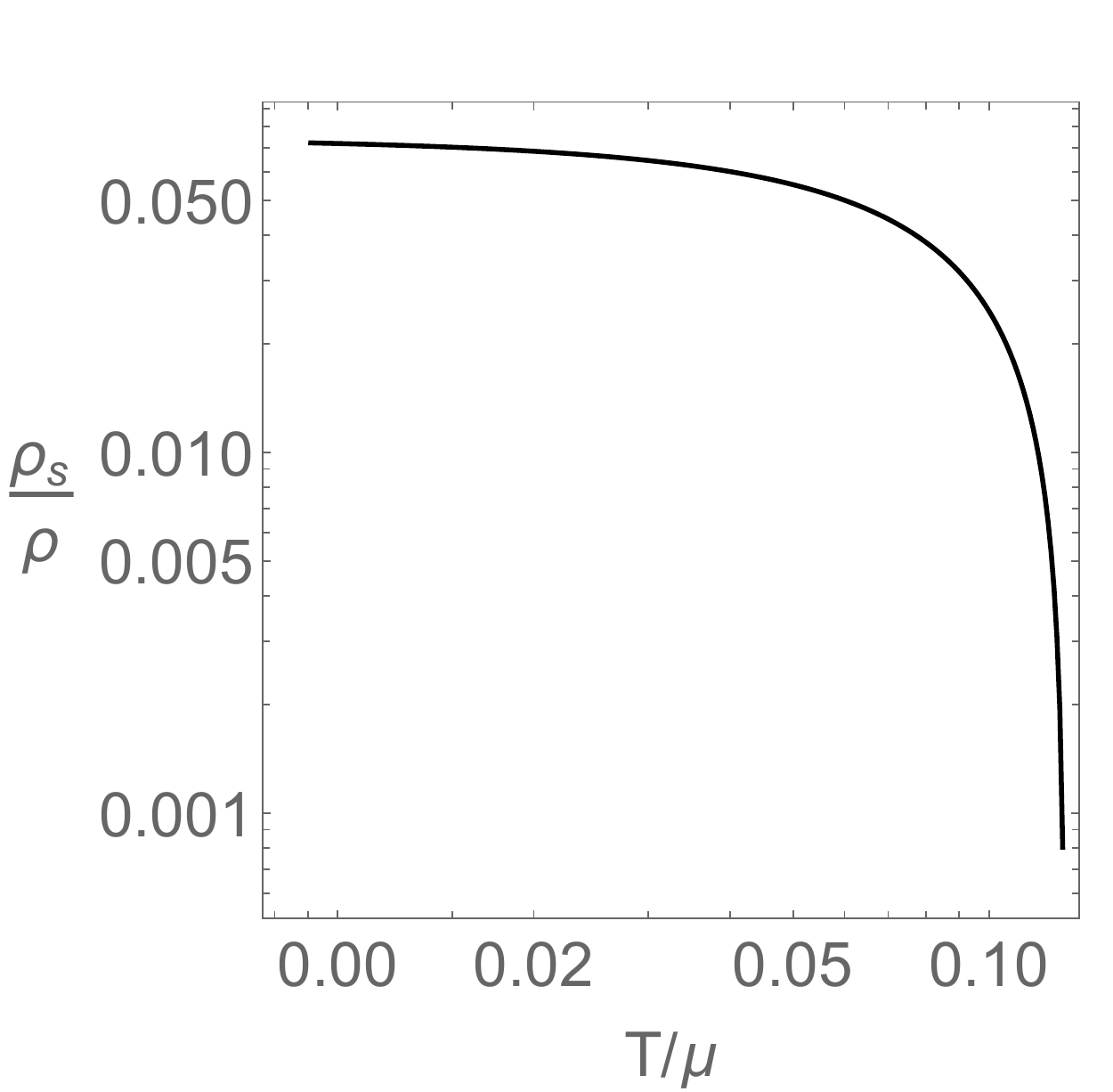}
\includegraphics[scale=.4]{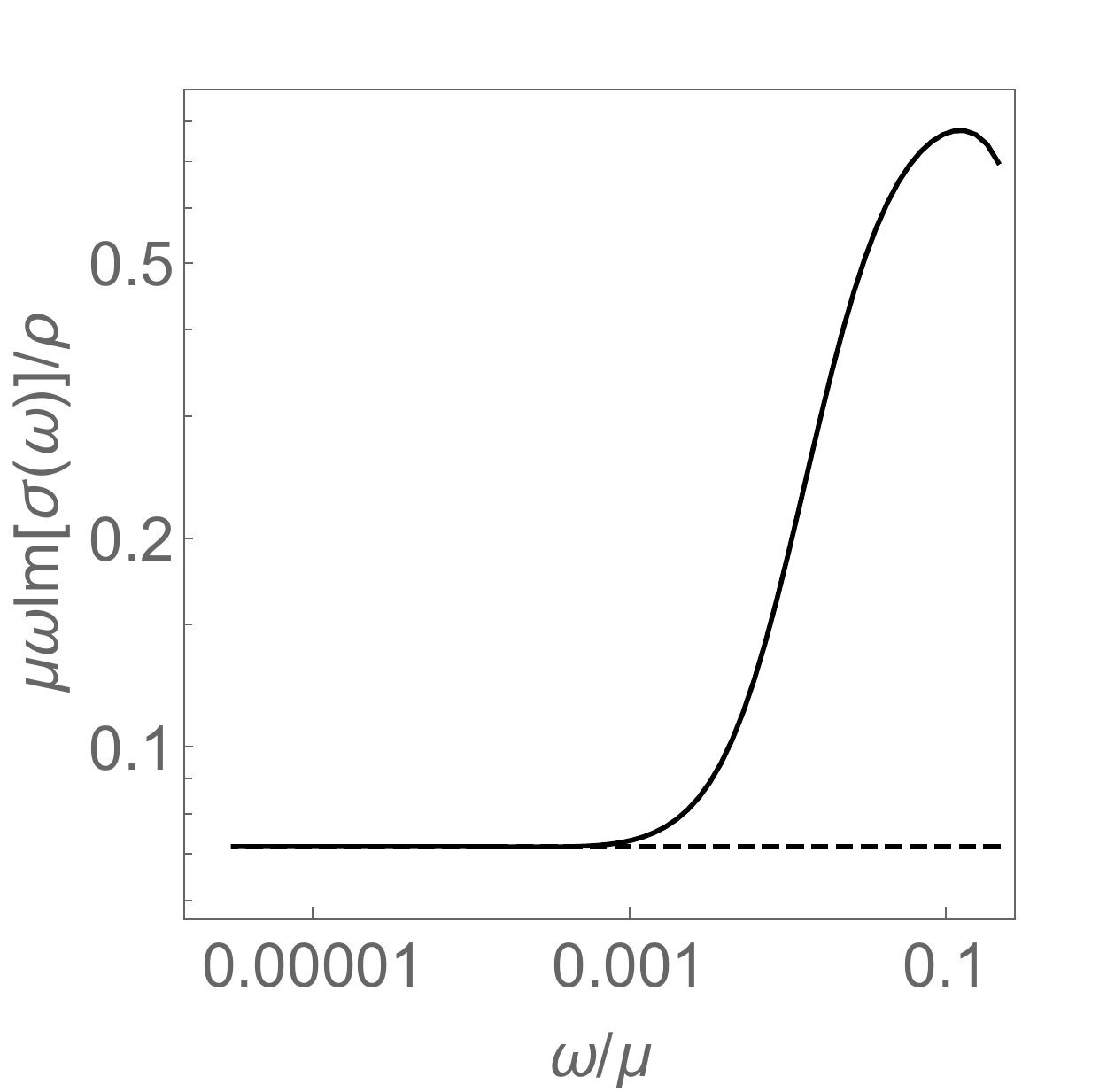}
\includegraphics[scale=.4]{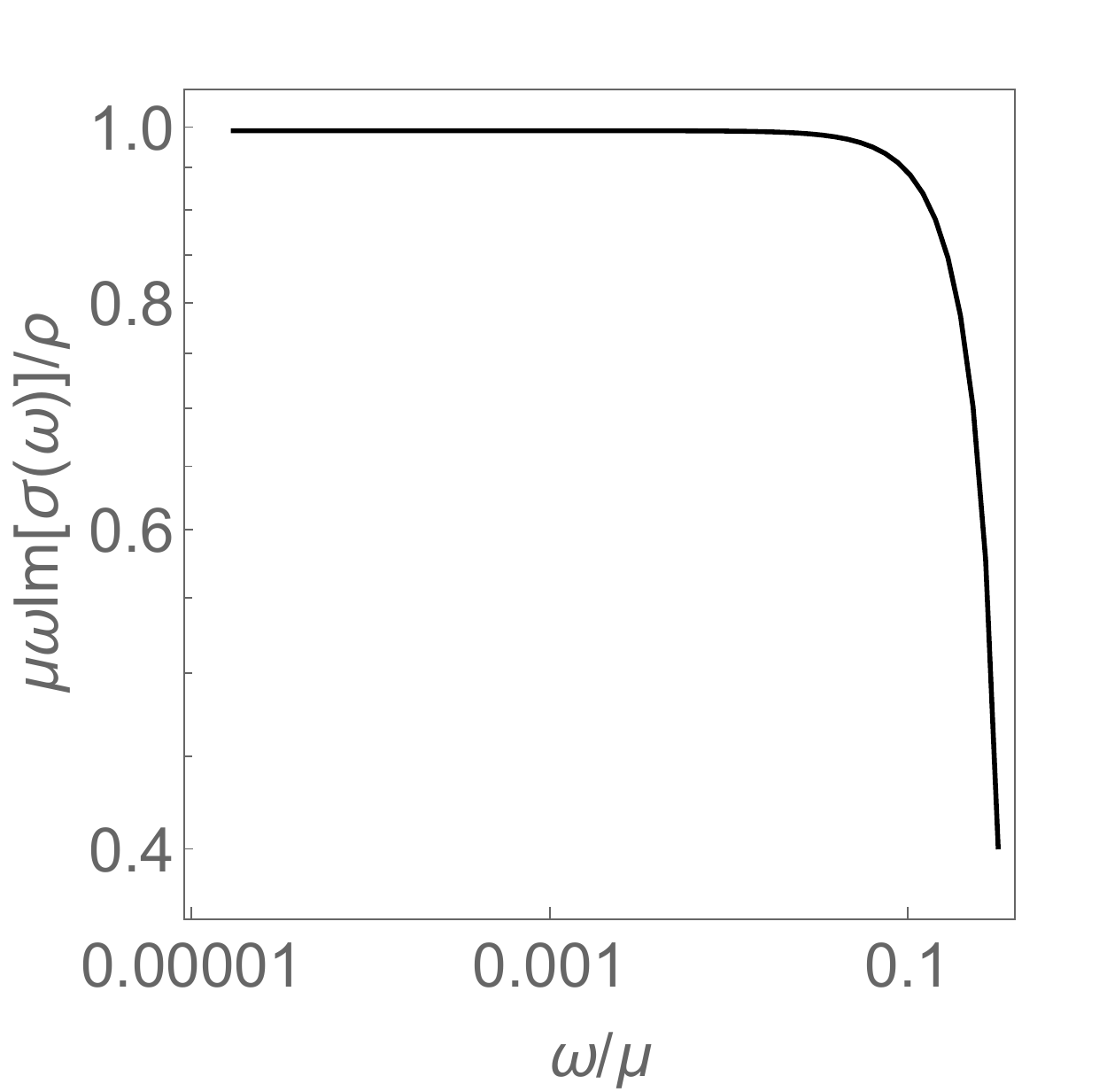}
\caption{\label{zeroTsemilocal}The normal density for $a=4$ semi-local quantum critical superfluids. (Left) Finite temperature $\rho_n$ from equation (\ref{rhonZdefinition}). (Center) Weight of the $T=0$ imaginary pole with $m/\mu=.315$ showing partial depletion of spectral weight and $\rho_n^{(0)}=\rho^{(0)}-\rho_s^{(0)}\neq 0$. The dashed line is the value we find as $T\to 0$ from equation (\ref{rhonZdefinition}). (Right) Weight of the $T=0$ imaginary pole with $m=0$ showing no depletion of spectral weight.}
\end{center}
\end{figure}

As a first step, we note that with no complex scalar, we may obtain a semi-local quantum critical geometry by investigating the zero temperature limit of an AdS-Reissner-Nordstrom black hole. If we break translations with neutral scalars, such a system was investigated in \cite{Andrade:2013gsa,Davison:2015bea}. Given the absence of a complex scalar, there is no spontaneous symmetry breaking and we expect breaking translations will completely remove the pole in the imaginary conductivity. In figure \ref{AdSRNresults}, we confirm this result. This is in contrast to the semi-local geometries with a superfluid. In figure \ref{zeroTsemilocal}, we confirm that there is residual spectral weight in the imaginary pole, and hence $\rho_n^{(0)}\neq 0$. In fact, we confirm that this $\rho_n^{(0)}$ agrees with the result we obtained with our translation invariant finite temperature solutions.

\section{Vanishing $\rho_n^{(0)}$ but finite $\rho_{in}^{(0)}$ \label{convergenceappendix}}
Up to this point, the necessary conditions for a non-vanishing $\rho_n^{(0)}$ are either a non-vanishing $\rho_{in}^{(0)}$ and any values of $z, \theta$ or a vanishing $\rho_{in}^{(0)}$ with $d+2-z-\theta > 0$. Recalling that $\rho_{in}^{(0)}$ captures, in some sense, the degree to which charged degrees of freedom are uncondensed at zero temperature, these criteria seem not to apply to systems like $^4$He, which have $\rho_{n}^{(0)}$ vanishing but not all $^4$He atoms are in the superfluid state at zero temperature. We would like to have a holographic model that can describe such a state. In order to do so, let's recall the basic criteria for the vanishing of $\rho_{n}^{(0)}$. 
$\\$

We can expand
\begin{align}
a_{\hat{x}}(r) = \frac{a_{\hat{x}}}{\mu}{A_t}(r)\left[1 + (sT) \mathcal{A}_1(r) + (sT)^2\mathcal{A}_2(r)+...\right]
\end{align}
so long as the functions $\mathcal{A}_i(r)$ are sufficiently well-behaved. The expansion is guaranteed to be well-behaved if
\begin{align}
\lim_{r_h\to\infty}\int_0^{r_h} dr\frac{2q^2M_F\sqrt{BD}C^{d/2-1}\eta^2}{R^2}
\end{align}
converges, at least up to order $(sT)^2$. This allows a non-vanishing $\rho_n^{(0)}$ for certain Lifshitz phases. Now, let's ask what happens if we naively impose $z=1$ and $\rho_{in}^{(0)}\neq 0$. Such a solution requires $\theta\neq 0$. Using our original action, we must conclude that since both criteria for convergence are met, $\rho_{n}^{(0)} \neq 0$. Hence $z=1$ does not imply $\rho_n^{(0)}=0$.
$\\$

A slight technical point is that for $z=1$ partially condensed phases, the gauge field completely decouples from the equations of motion at $T=0$ as can be seen from equation (\ref{marginaldeformations}). Nevertheless, one can solve the Maxwell equation in the IR background and find that $A_t$ is still given by equation (\ref{marginaldeformations}) except that $A_0$ is not fixed by the IR parameters and its value must be determined by the full RG flow. Fortunately, this leads to $\rho_{in}^{(0)}\neq 0$ as desired.
$\\$

To find a $\rho_n^{(0)} = 0$ we need the expansion for $a_x$ to break down. Unfortunately, there is not a nice way to affect $\mathcal{A}_1$. On the other hand, we can affect $\mathcal{A}_2$ by modifying the mass term for the $U(1)$ gauge field. Taking
\begin{align}
M_F(\psi) \to M_0 \exp(\frac{d}{2}\iota \delta \psi) \sim M_0\left(\frac{\hat{r}}{L}\right)^{\iota\theta}
\end{align}
in the IR, we find that
\begin{align}
\int^{r_h} dr \frac{2q^2M_F\sqrt{BD}C^{d/2-1}\eta^2}{R^2} \sim \# \frac{L_t\tilde{L}}{L_x^2} \frac{sq^2 \eta_h^2}{2\pi \rho_{in}^2} \left(\frac{r_h}{L}\right)^{2-z+\iota\theta} \sim \# T^{1-\frac{2-d+(1+\iota)\theta}{z}}
\end{align}
In our earlier language of (\ref{alphadef}), this means that
\begin{align}
\alpha = 2z-(1+\iota)\theta - 4
\end{align}
Hence, if $2-z-d+(1+\iota)\theta>0$, the integral diverges. If $|\iota|$ is too large, this can disrupt the IR. We will only present results where this does not happen. 
$\\$

\begin{figure}[h]
\begin{center}
\includegraphics[scale=.24]{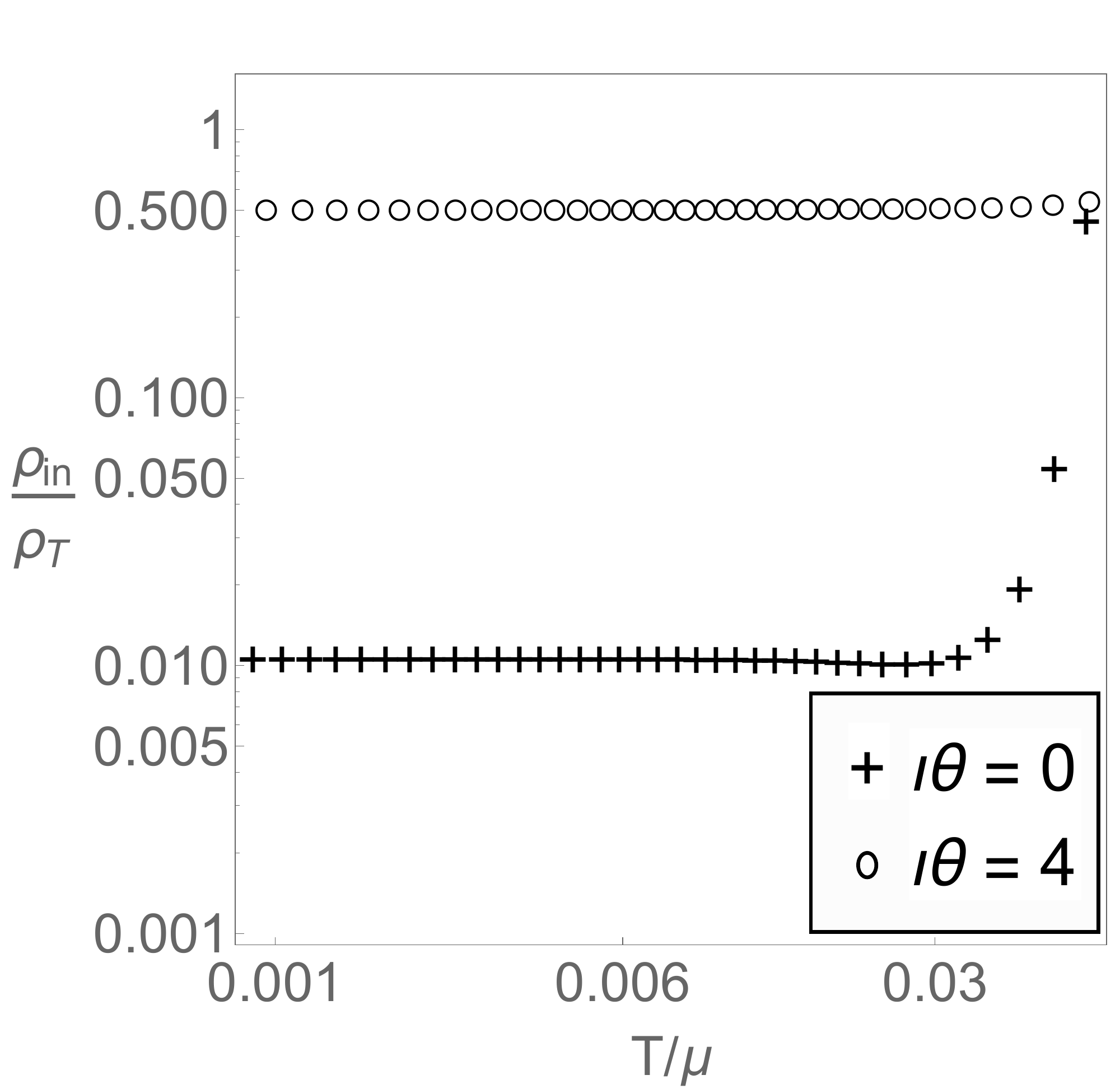}
\includegraphics[scale=.24]{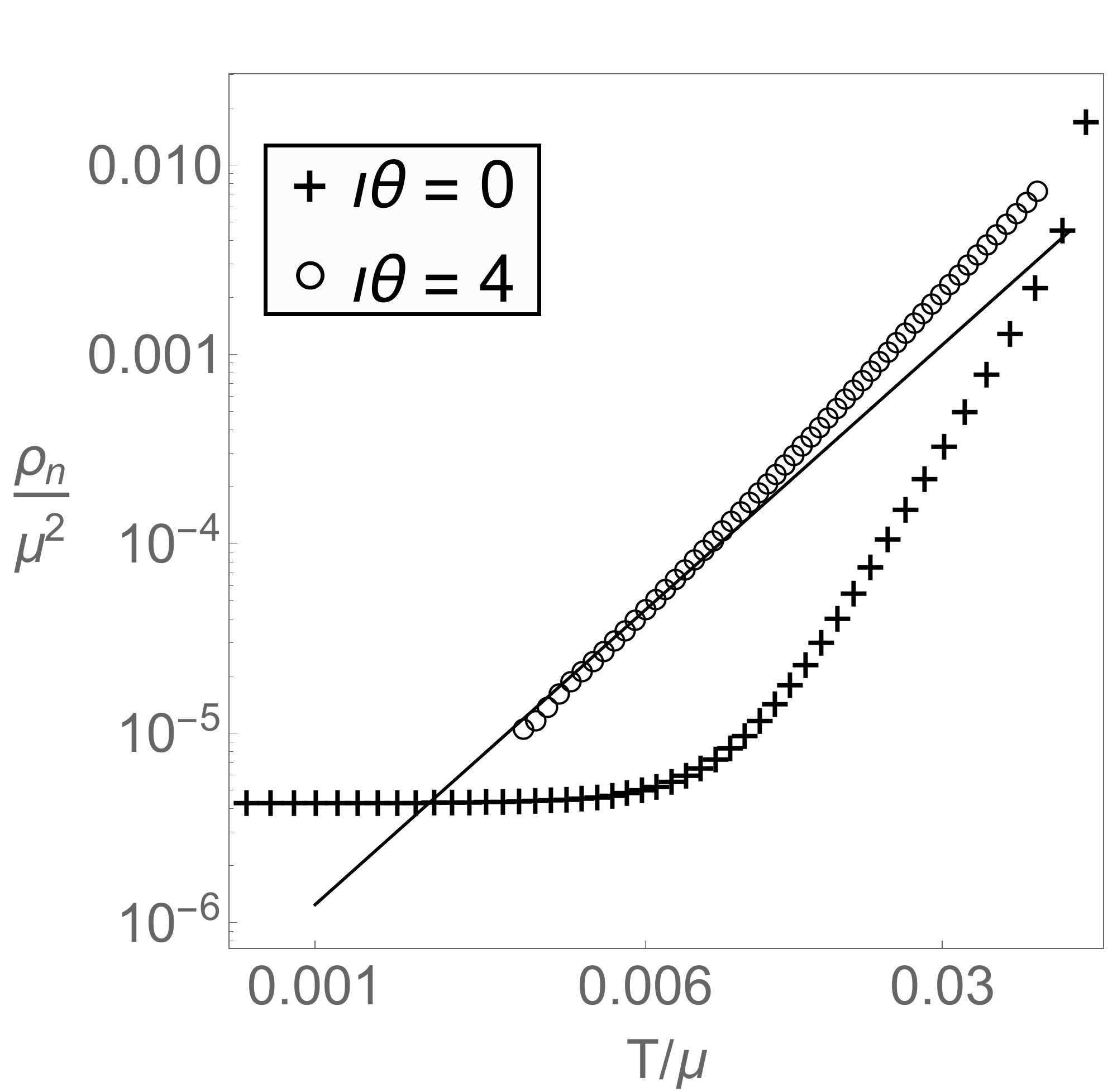}
\includegraphics[scale=.25]{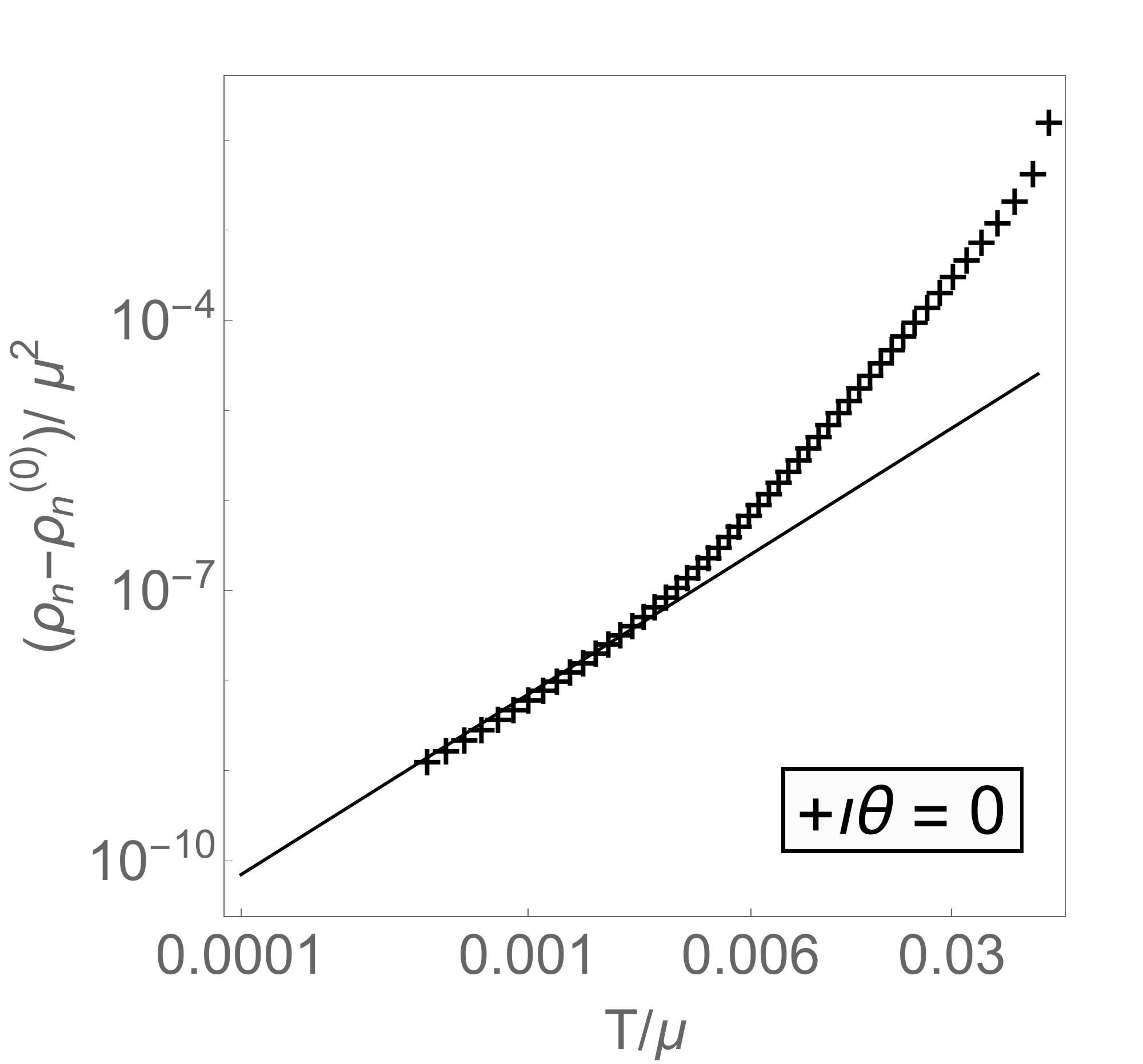}
\caption{\label{modifiedactionfigure} An example using the potential in \ref{fractionalizedexample} with $a=5$ so that $z=1, \theta=-1$ but $\rho_{in}^{(0)}\neq 0$ (left). For $\iota = -4$, we find that $\rho_{n} \sim \#T^2$ whereas for $\iota=0$, we find $\rho_n \sim \rho_n^{(0)} + \# T^2$ (middle and right).}
\end{center}
\end{figure}

Following our earlier discussion of (\ref{alphadef})
\begin{align}
\mu\rho_{n} \sim \begin{cases} 
\mu\rho_{n}^{(0)}+ \#(sT) & 2+\iota\theta<0\,,\\
\mu\rho_{n}^{(0)} + \# T^{\frac{z-2+d-(1+\iota)\theta}{z}} & 2-d-z+(\iota+1)\theta<0<2+\iota\theta\,,\\
\# T^{\frac{2-d+(1+\iota)\theta-z}{z}} &  2(1-d-z+\theta)+\iota\theta<0< 2-d-z +(\iota+1)\theta \,,\\
\#(sT) & 2(1-d-z+\theta)+\iota\theta = 0\,.
\end{cases}
\end{align}
We demonstrate this scaling in figure \ref{modifiedactionfigure}. In systems like $^4$He, $\mu\rho_n \sim sT$, and hence is satisfied by the last case. Of course, this analysis required an additional field, the dilaton, which is likely not present in systems like $^4$He. This simplified our analysis but is not necessary. We can modify the action in different ways, for instance by setting $\psi=0$ but letting $\eta$ run in the IR. The analysis for these cases is nearly identical to that with the dilaton since $\eta$ is irrelevant when $\psi\neq 0$, so we omit writing them here. In such a system, we also found solutions with $\rho^{(0)}_{in}\neq 0$ and $\mu\rho_n \sim sT$.
$\\$

We note that this behavior only depends on the behavior of potentials in the IR. Hence, from an RG perspective, $\rho_n$ can have an anomalous dimension from a non-trivial $\iota$. Nevertheless, since everything is determined from IR quantities, the analysis suggests that quantum critical superfluids with general critical exponents may be amenable to an effective field theory treatment. 

\bibliographystyle{JHEP}
\bibliography{local}

\providecommand{\href}[2]{#2}\begingroup\raggedright\begin{thebibliography}{10}

\bibitem{Landau}
L.~Landau, \emph{Theory of the superfluidity of helium ii},
  \href{https://doi.org/10.1103/PhysRev.60.356}{\emph{Phys. Rev.} {\bfseries
  60} (1941) 356}.

\bibitem{Tisza}
L.~Tisza, \emph{The theory of liquid helium},
  \href{https://doi.org/10.1103/PhysRev.72.838}{\emph{Phys. Rev.} {\bfseries
  72} (1947) 838}.

\bibitem{Khalatnikov1}
I.~Khalatnikov and V.~Lebedev, \emph{Relativistic hydrodynamics of a superfluid
  liquid},
  \href{https://doi.org/https://doi.org/10.1016/0375-9601(82)90268-7}{\emph{Physics
  Letters A} {\bfseries 91} (1982) 70 }.

\bibitem{Khalatnikov2}
V.~Lebedev and I.~Khalatnikov, \emph{Relativistic hydrodynamics of a superfluid
  liquid}, {\emph{Zh. Eksp. Teor. Fiz.} {\bfseries 56} (1982) }.

\bibitem{Khalatnikov3}
B.~Carter and I.~M. Khalatnikov, \emph{Equivalence of convective and potential
  variational derivations of covariant superfluid dynamics},
  \href{https://doi.org/10.1103/PhysRevD.45.4536}{\emph{Phys. Rev. D}
  {\bfseries 45} (1992) 4536}.

\bibitem{Khalatnikov4}
B.~Carter and I.~Khalatnikov, \emph{Momentum, vorticity, and helicity in
  covariant superfluid dynamics},
  \href{https://doi.org/https://doi.org/10.1016/0003-4916(92)90348-P}{\emph{Annals
  of Physics} {\bfseries 219} (1992) 243 }.

\bibitem{Israel1}
W.~Israel, \emph{Covariant superfluid mechanics},
  \href{https://doi.org/https://doi.org/10.1016/0375-9601(81)90169-9}{\emph{Physics
  Letters A} {\bfseries 86} (1981) 79 }.

\bibitem{Israel2}
W.~Israel, \emph{Equivalence of two theories of relativistic superfluid
  mechanics},
  \href{https://doi.org/https://doi.org/10.1016/0375-9601(82)90298-5}{\emph{Physics
  Letters A} {\bfseries 92} (1982) 77 }.

\bibitem{Son:2000ht}
D.~T. Son, \emph{{Hydrodynamics of relativistic systems with broken continuous
  symmetries}}, \href{https://doi.org/10.1142/S0217751X01009545}{\emph{Int. J.
  Mod. Phys.} {\bfseries A16S1C} (2001) 1284}
  [\href{https://arxiv.org/abs/hep-ph/0011246}{{\ttfamily hep-ph/0011246}}].

\bibitem{Bhattacharya:2011eea}
J.~Bhattacharya, S.~Bhattacharyya and S.~Minwalla, \emph{{Dissipative
  Superfluid dynamics from gravity}},
  \href{https://doi.org/10.1007/JHEP04(2011)125}{\emph{JHEP} {\bfseries 04}
  (2011) 125} [\href{https://arxiv.org/abs/1101.3332}{{\ttfamily 1101.3332}}].

\bibitem{Bhattacharya:2011tra}
J.~Bhattacharya, S.~Bhattacharyya, S.~Minwalla and A.~Yarom, \emph{{A Theory of
  first order dissipative superfluid dynamics}},
  \href{https://doi.org/10.1007/JHEP05(2014)147}{\emph{JHEP} {\bfseries 05}
  (2014) 147} [\href{https://arxiv.org/abs/1105.3733}{{\ttfamily 1105.3733}}].

\bibitem{Bhattacharyya:2012xi}
S.~Bhattacharyya, S.~Jain, S.~Minwalla and T.~Sharma, \emph{{Constraints on
  Superfluid Hydrodynamics from Equilibrium Partition Functions}},
  \href{https://doi.org/10.1007/JHEP01(2013)040}{\emph{JHEP} {\bfseries 01}
  (2013) 040} [\href{https://arxiv.org/abs/1206.6106}{{\ttfamily 1206.6106}}].

\bibitem{Leggett}
A.~J. Leggett, \emph{Quantum liquids: Bose condensation and Cooper pairing in
  condensed-matter systems}. Oxford university press, 2006,
  \href{https://doi.org/10.1093/acprof:oso/9780198526438.001.0001}{10.1093/acprof:oso/9780198526438.001.0001}.

\bibitem{Schmitt:2014eka}
A.~Schmitt, \emph{{Introduction to Superfluidity}},
  \href{https://doi.org/10.1007/978-3-319-07947-9}{\emph{Lect. Notes Phys.}
  {\bfseries 888} (2015) pp.1}
  [\href{https://arxiv.org/abs/1404.1284}{{\ttfamily 1404.1284}}].

\bibitem{Son:2002zn}
D.~T. Son, \emph{{Low-energy quantum effective action for relativistic
  superfluids}},  \href{https://arxiv.org/abs/hep-ph/0204199}{{\ttfamily
  hep-ph/0204199}}.

\bibitem{Nicolis:2011cs}
A.~Nicolis, \emph{{Low-energy effective field theory for finite-temperature
  relativistic superfluids}},
  \href{https://arxiv.org/abs/1108.2513}{{\ttfamily 1108.2513}}.

\bibitem{Delacretaz:2019brr}
L.~V. Delacrétaz, D.~M. Hofman and G.~Mathys, \emph{{Superfluids as
  Higher-form Anomalies}},
  \href{https://doi.org/10.21468/SciPostPhys.8.3.047}{\emph{SciPost Phys.}
  {\bfseries 8} (2020) 047} [\href{https://arxiv.org/abs/1908.06977}{{\ttfamily
  1908.06977}}].

\bibitem{Leggett2}
A.~J. Leggett, \emph{{On the Superfluid Fraction of an Arbitrary Many-Body
  System at T = 0}}, {\emph{Journal of Statistical Physics} {\bfseries 93}
  (1998) 927}.

\bibitem{Gouteraux:2019kuy}
B.~Goutéraux and E.~Mefford, \emph{{Normal charge densities in quantum
  critical superfluids}},
  \href{https://doi.org/10.1103/PhysRevLett.124.161604}{\emph{Phys. Rev. Lett.}
  {\bfseries 124} (2020) 161604}
  [\href{https://arxiv.org/abs/1912.08849}{{\ttfamily 1912.08849}}].

\bibitem{Bozovic1}
I.~Bo{\v{z}}ovi{\'{c}}, X.~He, J.~Wu and A.~T. Bollinger, \emph{{Dependence of
  the critical temperature in overdoped copper oxides on superfluid density}},
  \href{https://doi.org/10.1038/nature19061}{\emph{Nature} {\bfseries 536}
  (2016) 309}.

\bibitem{Bozovic2}
F.~Mahmood, X.~He, I.~Bo\ifmmode \check{z}\else
  \v{z}\fi{}ovi\ifmmode~\acute{c}\else \'{c}\fi{} and N.~P. Armitage,
  \emph{Locating the missing superconducting electrons in the overdoped
  cuprates
  ${\mathrm{la}}_{2\ensuremath{-}x}{\mathrm{sr}}_{x}{\mathrm{cuo}}_{4}$},
  \href{https://doi.org/10.1103/PhysRevLett.122.027003}{\emph{Phys. Rev. Lett.}
  {\bfseries 122} (2019) 027003}
  [\href{https://arxiv.org/abs/1802.02101}{{\ttfamily 1802.02101}}].

\bibitem{LeeHone1}
N.~R. Lee-Hone, J.~S. Dodge and D.~M. Broun, \emph{Disorder and superfluid
  density in overdoped cuprate superconductors},
  \href{https://doi.org/10.1103/PhysRevB.96.024501}{\emph{Phys. Rev. B}
  {\bfseries 96} (2017) 024501}
  [\href{https://arxiv.org/abs/1704.04803}{{\ttfamily 1704.04803}}].

\bibitem{LeeHone2}
N.~R. Lee-Hone, V.~Mishra, D.~M. Broun and P.~J. Hirschfeld, \emph{Optical
  conductivity of overdoped cuprate superconductors: Application to
  ${\mathrm{la}}_{2\ensuremath{-}x}{\mathrm{sr}}_{x}{\mathrm{cuo}}_{4}$},
  \href{https://doi.org/10.1103/PhysRevB.98.054506}{\emph{Phys. Rev. B}
  {\bfseries 98} (2018) 054506}
  [\href{https://arxiv.org/abs/1802.10198}{{\ttfamily 1802.10198}}].

\bibitem{LeeHone3}
N.~R. {Lee-Hone}, H.~U. {{\"O}zdemir}, V.~{Mishra}, D.~M. {Broun} and P.~J.
  {Hirschfeld}, \emph{Low energy phenomenology of the overdoped cuprates:
  Viability of the landau-bcs paradigm},
  \href{https://doi.org/10.1103/physrevresearch.2.013228}{\emph{Physical Review
  Research} {\bfseries 2} (2020) }
  [\href{https://arxiv.org/abs/1902.08286}{{\ttfamily 1902.08286}}].

\bibitem{Gubser:2008px}
S.~S. Gubser, \emph{{Breaking an Abelian gauge symmetry near a black hole
  horizon}}, \href{https://doi.org/10.1103/PhysRevD.78.065034}{\emph{Phys.
  Rev.} {\bfseries D78} (2008) 065034}
  [\href{https://arxiv.org/abs/0801.2977}{{\ttfamily 0801.2977}}].

\bibitem{Hartnoll:2008vx}
S.~A. Hartnoll, C.~P. Herzog and G.~T. Horowitz, \emph{{Building a Holographic
  Superconductor}},
  \href{https://doi.org/10.1103/PhysRevLett.101.031601}{\emph{Phys. Rev. Lett.}
  {\bfseries 101} (2008) 031601}
  [\href{https://arxiv.org/abs/0803.3295}{{\ttfamily 0803.3295}}].

\bibitem{Hartnoll:2008kx}
S.~A. Hartnoll, C.~P. Herzog and G.~T. Horowitz, \emph{{Holographic
  Superconductors}},
  \href{https://doi.org/10.1088/1126-6708/2008/12/015}{\emph{JHEP} {\bfseries
  12} (2008) 015} [\href{https://arxiv.org/abs/0810.1563}{{\ttfamily
  0810.1563}}].

\bibitem{Ammon:2008fc}
M.~Ammon, J.~Erdmenger, M.~Kaminski and P.~Kerner, \emph{{Superconductivity
  from gauge/gravity duality with flavor}},
  \href{https://doi.org/10.1016/j.physletb.2009.09.029}{\emph{Phys. Lett. B}
  {\bfseries 680} (2009) 516}
  [\href{https://arxiv.org/abs/0810.2316}{{\ttfamily 0810.2316}}].

\bibitem{Ammon:2009fe}
M.~Ammon, J.~Erdmenger, M.~Kaminski and P.~Kerner, \emph{{Flavor
  Superconductivity from Gauge/Gravity Duality}},
  \href{https://doi.org/10.1088/1126-6708/2009/10/067}{\emph{JHEP} {\bfseries
  10} (2009) 067} [\href{https://arxiv.org/abs/0903.1864}{{\ttfamily
  0903.1864}}].

\bibitem{Gauntlett:2009dn}
J.~P. Gauntlett, J.~Sonner and T.~Wiseman, \emph{{Holographic superconductivity
  in M-Theory}},
  \href{https://doi.org/10.1103/PhysRevLett.103.151601}{\emph{Phys. Rev. Lett.}
  {\bfseries 103} (2009) 151601}
  [\href{https://arxiv.org/abs/0907.3796}{{\ttfamily 0907.3796}}].

\bibitem{Gauntlett:2009bh}
J.~P. Gauntlett, J.~Sonner and T.~Wiseman, \emph{{Quantum Criticality and
  Holographic Superconductors in M-theory}},
  \href{https://doi.org/10.1007/JHEP02(2010)060}{\emph{JHEP} {\bfseries 02}
  (2010) 060} [\href{https://arxiv.org/abs/0912.0512}{{\ttfamily 0912.0512}}].

\bibitem{Arean:2010wu}
D.~Arean, M.~Bertolini, C.~Krishnan and T.~Prochazka, \emph{{Type IIB
  Holographic Superfluid Flows}},
  \href{https://doi.org/10.1007/JHEP03(2011)008}{\emph{JHEP} {\bfseries 03}
  (2011) 008} [\href{https://arxiv.org/abs/1010.5777}{{\ttfamily 1010.5777}}].

\bibitem{Arean:2011gz}
D.~Arean, M.~Bertolini, C.~Krishnan and T.~Prochazka, \emph{{Quantum Critical
  Superfluid Flows and Anisotropic Domain Walls}},
  \href{https://doi.org/10.1007/JHEP09(2011)131}{\emph{JHEP} {\bfseries 09}
  (2011) 131} [\href{https://arxiv.org/abs/1106.1053}{{\ttfamily 1106.1053}}].

\bibitem{Bobev:2011rv}
N.~Bobev, A.~Kundu, K.~Pilch and N.~P. Warner, \emph{{Minimal Holographic
  Superconductors from Maximal Supergravity}},
  \href{https://doi.org/10.1007/JHEP03(2012)064}{\emph{JHEP} {\bfseries 03}
  (2012) 064} [\href{https://arxiv.org/abs/1110.3454}{{\ttfamily 1110.3454}}].

\bibitem{Donos:2012yu}
A.~Donos, J.~P. Gauntlett, J.~Sonner and B.~Withers, \emph{{Competing orders in
  M-theory: superfluids, stripes and metamagnetism}},
  \href{https://doi.org/10.1007/JHEP03(2013)108}{\emph{JHEP} {\bfseries 03}
  (2013) 108} [\href{https://arxiv.org/abs/1212.0871}{{\ttfamily 1212.0871}}].

\bibitem{Adam:2012mw}
A.~Adam, B.~Crampton, J.~Sonner and B.~Withers, \emph{{Bosonic
  Fractionalisation Transitions}},
  \href{https://doi.org/10.1007/JHEP01(2013)127}{\emph{JHEP} {\bfseries 01}
  (2013) 127} [\href{https://arxiv.org/abs/1208.3199}{{\ttfamily 1208.3199}}].

\bibitem{Herzog:2008he}
C.~P. Herzog, P.~K. Kovtun and D.~T. Son, \emph{{Holographic model of
  superfluidity}},
  \href{https://doi.org/10.1103/PhysRevD.79.066002}{\emph{Phys. Rev.}
  {\bfseries D79} (2009) 066002}
  [\href{https://arxiv.org/abs/0809.4870}{{\ttfamily 0809.4870}}].

\bibitem{Herzog:2009md}
C.~P. Herzog and A.~Yarom, \emph{{Sound modes in holographic superfluids}},
  \href{https://doi.org/10.1103/PhysRevD.80.106002}{\emph{Phys. Rev.}
  {\bfseries D80} (2009) 106002}
  [\href{https://arxiv.org/abs/0906.4810}{{\ttfamily 0906.4810}}].

\bibitem{Sonner:2010yx}
J.~Sonner and B.~Withers, \emph{{A gravity derivation of the Tisza-Landau Model
  in AdS/CFT}}, \href{https://doi.org/10.1103/PhysRevD.82.026001}{\emph{Phys.
  Rev.} {\bfseries D82} (2010) 026001}
  [\href{https://arxiv.org/abs/1004.2707}{{\ttfamily 1004.2707}}].

\bibitem{Herzog:2011ec}
C.~P. Herzog, N.~Lisker, P.~Surowka and A.~Yarom, \emph{{Transport in
  holographic superfluids}},
  \href{https://doi.org/10.1007/JHEP08(2011)052}{\emph{JHEP} {\bfseries 08}
  (2011) 052} [\href{https://arxiv.org/abs/1101.3330}{{\ttfamily 1101.3330}}].

\bibitem{Gubser:2009cg}
S.~S. Gubser and A.~Nellore, \emph{{Ground states of holographic
  superconductors}},
  \href{https://doi.org/10.1103/PhysRevD.80.105007}{\emph{Phys. Rev.}
  {\bfseries D80} (2009) 105007}
  [\href{https://arxiv.org/abs/0908.1972}{{\ttfamily 0908.1972}}].

\bibitem{Horowitz:2009ij}
G.~T. Horowitz and M.~M. Roberts, \emph{{Zero Temperature Limit of Holographic
  Superconductors}},
  \href{https://doi.org/10.1088/1126-6708/2009/11/015}{\emph{JHEP} {\bfseries
  11} (2009) 015} [\href{https://arxiv.org/abs/0908.3677}{{\ttfamily
  0908.3677}}].

\bibitem{Gouteraux:2012yr}
B.~Gout\'eraux and E.~Kiritsis, \emph{{Quantum critical lines in holographic
  phases with (un)broken symmetry}},
  \href{https://doi.org/10.1007/JHEP04(2013)053}{\emph{JHEP} {\bfseries 04}
  (2013) 053} [\href{https://arxiv.org/abs/1212.2625}{{\ttfamily 1212.2625}}].

\bibitem{Gouteraux:2013oca}
B.~Goutéraux, \emph{{Universal scaling properties of extremal cohesive
  holographic phases}},
  \href{https://doi.org/10.1007/JHEP01(2014)080}{\emph{JHEP} {\bfseries 01}
  (2014) 080} [\href{https://arxiv.org/abs/1308.2084}{{\ttfamily 1308.2084}}].

\bibitem{Ammon:2012je}
M.~Ammon, M.~Kaminski and A.~Karch, \emph{{Hyperscaling-Violation on Probe
  D-Branes}}, \href{https://doi.org/10.1007/JHEP11(2012)028}{\emph{JHEP}
  {\bfseries 11} (2012) 028} [\href{https://arxiv.org/abs/1207.1726}{{\ttfamily
  1207.1726}}].

\bibitem{Delacretaz:2020nit}
L.~V. Delacretaz, \emph{{Heavy Operators and Hydrodynamic Tails}},
  \href{https://arxiv.org/abs/2006.01139}{{\ttfamily 2006.01139}}.

\bibitem{Hartnoll:2011pp}
S.~A. Hartnoll and L.~Huijse, \emph{{Fractionalization of holographic Fermi
  surfaces}},
  \href{https://doi.org/10.1088/0264-9381/29/19/194001}{\emph{Class. Quant.
  Grav.} {\bfseries 29} (2012) 194001}
  [\href{https://arxiv.org/abs/1111.2606}{{\ttfamily 1111.2606}}].

\bibitem{Hartnoll:2011fn}
S.~A. Hartnoll, \emph{{Horizons, holography and condensed matter}},  in
  \emph{Black holes in higher dimensions}, G.~T. Horowitz, ed., pp.~387--419,
  (2012), \href{https://arxiv.org/abs/1106.4324}{{\ttfamily 1106.4324}}.

\bibitem{Hartnoll:2012wm}
S.~A. Hartnoll and E.~Shaghoulian, \emph{{Spectral weight in holographic
  scaling geometries}},
  \href{https://doi.org/10.1007/JHEP07(2012)078}{\emph{JHEP} {\bfseries 07}
  (2012) 078} [\href{https://arxiv.org/abs/1203.4236}{{\ttfamily 1203.4236}}].

\bibitem{Hartnoll:2012ux}
S.~A. Hartnoll and D.~Radicevic, \emph{{Holographic order parameter for charge
  fractionalization}},
  \href{https://doi.org/10.1103/PhysRevD.86.066001}{\emph{Phys. Rev. D}
  {\bfseries 86} (2012) 066001}
  [\href{https://arxiv.org/abs/1205.5291}{{\ttfamily 1205.5291}}].

\bibitem{Anantua:2012nj}
R.~J. Anantua, S.~A. Hartnoll, V.~L. Martin and D.~M. Ramirez, \emph{{The Pauli
  exclusion principle at strong coupling: Holographic matter and momentum
  space}}, \href{https://doi.org/10.1007/JHEP03(2013)104}{\emph{JHEP}
  {\bfseries 03} (2013) 104} [\href{https://arxiv.org/abs/1210.1590}{{\ttfamily
  1210.1590}}].

\bibitem{Gouteraux:2016arz}
B.~Goutéraux and V.~L. Martin, \emph{{Spectral weight and spatially modulated
  instabilities in holographic superfluids}},
  \href{https://doi.org/10.1007/JHEP05(2017)005}{\emph{JHEP} {\bfseries 05}
  (2017) 005} [\href{https://arxiv.org/abs/1612.03466}{{\ttfamily
  1612.03466}}].

\bibitem{Martin:2019sxc}
V.~L. Martin and N.~Monga, \emph{{Spectral weight in Chern-Simons theory with
  symmetry breaking}},
  \href{https://doi.org/10.1007/JHEP10(2019)116}{\emph{JHEP} {\bfseries 10}
  (2019) 116} [\href{https://arxiv.org/abs/1905.07417}{{\ttfamily
  1905.07417}}].

\bibitem{Martin:2019ulo}
V.~L. Martin, \emph{{Spectral weight in holography with momentum relaxation}},
  \href{https://arxiv.org/abs/1908.04312}{{\ttfamily 1908.04312}}.

\bibitem{Autti_2020}
S.~Autti, J.~T. Mäkinen, J.~Rysti, G.~E. Volovik, V.~V. Zavjalov and V.~B.
  Eltsov, \emph{Exceeding the landau speed limit with topological bogoliubov
  fermi surfaces},
  \href{https://doi.org/10.1103/physrevresearch.2.033013}{\emph{Physical Review
  Research} {\bfseries 2} (2020) }.

\bibitem{Gouteraux:2018wfe}
B.~Gout\'eraux, N.~Jokela and A.~P\"onni, \emph{{Incoherent conductivity of
  holographic charge density waves}},
  \href{https://doi.org/10.1007/JHEP07(2018)004}{\emph{JHEP} {\bfseries 07}
  (2018) 004} [\href{https://arxiv.org/abs/1803.03089}{{\ttfamily
  1803.03089}}].

\bibitem{Iqbal:2011in}
N.~Iqbal, H.~Liu and M.~Mezei, \emph{{Semi-local quantum liquids}},
  \href{https://doi.org/10.1007/JHEP04(2012)086}{\emph{JHEP} {\bfseries 04}
  (2012) 086} [\href{https://arxiv.org/abs/1105.4621}{{\ttfamily 1105.4621}}].

\bibitem{Davison:2015bea}
R.~A. Davison and B.~Goutéraux, \emph{{Dissecting holographic
  conductivities}}, \href{https://doi.org/10.1007/JHEP09(2015)090}{\emph{JHEP}
  {\bfseries 09} (2015) 090}
  [\href{https://arxiv.org/abs/1505.05092}{{\ttfamily 1505.05092}}].

\bibitem{Davison:2015taa}
R.~A. Davison, B.~Goutéraux and S.~A. Hartnoll, \emph{{Incoherent transport in
  clean quantum critical metals}},
  \href{https://doi.org/10.1007/JHEP10(2015)112}{\emph{JHEP} {\bfseries 10}
  (2015) 112} [\href{https://arxiv.org/abs/1507.07137}{{\ttfamily
  1507.07137}}].

\bibitem{Davison:2018ofp}
R.~A. Davison, S.~A. Gentle and B.~Goutéraux, \emph{{Slow relaxation and
  diffusion in holographic quantum critical phases}},
  \href{https://doi.org/10.1103/PhysRevLett.123.141601}{\emph{Phys. Rev. Lett.}
  {\bfseries 123} (2019) 141601}
  [\href{https://arxiv.org/abs/1808.05659}{{\ttfamily 1808.05659}}].

\bibitem{Davison:2018nxm}
R.~A. Davison, S.~A. Gentle and B.~Goutéraux, \emph{{Impact of irrelevant
  deformations on thermodynamics and transport in holographic quantum critical
  states}}, \href{https://doi.org/10.1103/PhysRevD.100.086020}{\emph{Phys.
  Rev.} {\bfseries D100} (2019) 086020}
  [\href{https://arxiv.org/abs/1812.11060}{{\ttfamily 1812.11060}}].

\bibitem{Lucas:2015vna}
A.~Lucas, \emph{{Conductivity of a strange metal: from holography to memory
  functions}}, \href{https://doi.org/10.1007/JHEP03(2015)071}{\emph{JHEP}
  {\bfseries 03} (2015) 071}
  [\href{https://arxiv.org/abs/1501.05656}{{\ttfamily 1501.05656}}].

\bibitem{Son:2002sd}
D.~T. Son and A.~O. Starinets, \emph{{Minkowski space correlators in AdS / CFT
  correspondence: Recipe and applications}},
  \href{https://doi.org/10.1088/1126-6708/2002/09/042}{\emph{JHEP} {\bfseries
  09} (2002) 042} [\href{https://arxiv.org/abs/hep-th/0205051}{{\ttfamily
  hep-th/0205051}}].

\bibitem{Carter:1995if}
B.~Carter and D.~Langlois, \emph{{The Equation of state for cool relativistic
  two constituent superfluid dynamics}},
  \href{https://doi.org/10.1103/PhysRevD.51.5855}{\emph{Phys. Rev.} {\bfseries
  D51} (1995) 5855} [\href{https://arxiv.org/abs/hep-th/9507058}{{\ttfamily
  hep-th/9507058}}].

\bibitem{Raghu_2015}
S.~Raghu, G.~Torroba and H.~Wang, \emph{Metallic quantum critical points with
  finite bcs couplings},
  \href{https://doi.org/10.1103/physrevb.92.205104}{\emph{Physical Review B}
  {\bfseries 92} (2015) } [\href{https://arxiv.org/abs/1507.06652}{{\ttfamily
  1507.06652}}].

\bibitem{Wang_2017}
H.~Wang, S.~Raghu and G.~Torroba, \emph{Non-fermi-liquid superconductivity:
  Eliashberg approach versus the renormalization group},
  \href{https://doi.org/10.1103/physrevb.95.165137}{\emph{Physical Review B}
  {\bfseries 95} (2017) } [\href{https://arxiv.org/abs/1612.01971}{{\ttfamily
  1612.01971}}].

\bibitem{Wang_2018}
H.~Wang, Y.~Wang and G.~Torroba, \emph{Superconductivity versus quantum
  criticality: Effects of thermal fluctuations},
  \href{https://doi.org/10.1103/physrevb.97.054502}{\emph{Physical Review B}
  {\bfseries 97} (2018) } [\href{https://arxiv.org/abs/1708.04624}{{\ttfamily
  1708.04624}}].

\bibitem{damia2020thermal}
J.~A. Damia, M.~Solis and G.~Torroba, \emph{Thermal effects in non-fermi liquid
  superconductivity},  2020.

\bibitem{Harrison:2012vy}
S.~Harrison, S.~Kachru and H.~Wang, \emph{{Resolving Lifshitz Horizons}},
  \href{https://doi.org/10.1007/JHEP02(2014)085}{\emph{JHEP} {\bfseries 02}
  (2014) 085} [\href{https://arxiv.org/abs/1202.6635}{{\ttfamily 1202.6635}}].

\bibitem{Horowitz:2011gh}
G.~T. Horowitz and B.~Way, \emph{{Lifshitz Singularities}},
  \href{https://doi.org/10.1103/PhysRevD.85.046008}{\emph{Phys. Rev. D}
  {\bfseries 85} (2012) 046008}
  [\href{https://arxiv.org/abs/1111.1243}{{\ttfamily 1111.1243}}].

\bibitem{Bao:2012yt}
N.~Bao, X.~Dong, S.~Harrison and E.~Silverstein, \emph{{The Benefits of Stress:
  Resolution of the Lifshitz Singularity}},
  \href{https://doi.org/10.1103/PhysRevD.86.106008}{\emph{Phys. Rev. D}
  {\bfseries 86} (2012) 106008}
  [\href{https://arxiv.org/abs/1207.0171}{{\ttfamily 1207.0171}}].

\bibitem{Valle:2007xx}
M.~A. Valle, \emph{{Hydrodynamic fluctuations in relativistic superfluids}},
  \href{https://doi.org/10.1103/PhysRevD.77.025004}{\emph{Phys. Rev.}
  {\bfseries D77} (2008) 025004}
  [\href{https://arxiv.org/abs/0707.2665}{{\ttfamily 0707.2665}}].

\bibitem{Kovtun:2012rj}
P.~Kovtun, \emph{{Lectures on hydrodynamic fluctuations in relativistic
  theories}}, \href{https://doi.org/10.1088/1751-8113/45/47/473001}{\emph{J.
  Phys.} {\bfseries A45} (2012) 473001}
  [\href{https://arxiv.org/abs/1205.5040}{{\ttfamily 1205.5040}}].

\bibitem{Andrade:2013gsa}
T.~Andrade and B.~Withers, \emph{{A simple holographic model of momentum
  relaxation}}, \href{https://doi.org/10.1007/JHEP05(2014)101}{\emph{JHEP}
  {\bfseries 05} (2014) 101} [\href{https://arxiv.org/abs/1311.5157}{{\ttfamily
  1311.5157}}].

\end{thebibliography}\endgroup

\end{document}